\documentclass[twoside,11pt]{article} 
\usepackage{jmlr2e}

\ShortHeadings{Learning Attribute Patterns in SLAMs}{Gu and Xu}
\firstpageno{1}

\usepackage{amsfonts}
\usepackage{amsmath,amssymb,color,xcolor}
\usepackage{graphicx}
\usepackage{subcaption}
\usepackage{array,multirow}
\usepackage{float}
\usepackage{afterpage}
\usepackage{natbib}
\usepackage{makecell}
\usepackage[ruled,vlined]{algorithm2e}
\usepackage{blkarray}
\usepackage{arydshln}
\usepackage{booktabs}

\allowdisplaybreaks

\newcommand{\bo}{\boldsymbol }
\newcommand{\mt}{\mathcal }
\def\tilde{\widetilde}
\newcommand{\reals}{\mathbb R}
\DeclareMathOperator{\inp}{\normalfont{\text{input}}}
\newcommand{\RR}{\mathbf R}

\newcommand*{\qed}{\hfill\ensuremath{\square}}%
\newcommand{\aaa}{\boldsymbol \alpha}

\newcommand{\ttt}{\boldsymbol \theta}
\newcommand{\wt}{\widetilde}
\newcommand{\one}{\mathbf 1}
\newcommand{\zero}{\mathbf 0}
\newcommand{\norm}[1]{\lVert#1\rVert}

\newcommand{\mca}{\mathcal A}
\newcommand{\mc}{\mathcal C}
\newcommand{\mo}{\mathbf 1}

\newcommand{\rr}{\boldsymbol r}
\newcommand{\mz}{\boldsymbol 0}
\newcommand{\eeta}{\boldsymbol \eta}
\newcommand{\ma}{\mathbf A}
\newcommand{\II}{\mathbf I}

\newcommand{\ee}{\boldsymbol e}
\newcommand{\pp}{{\boldsymbol p}}

\newcommand{\qq}{\boldsymbol q}
\newcommand*{\Cdot}{{\scalebox{1.5}{$\cdot$}}}
\newcommand{\bcdot}{\boldsymbol\cdot}

\newcommand{\TT}{\boldsymbol \Theta}
\newcommand{\nnu}{\boldsymbol \nu}
\newcommand{\uu}{\boldsymbol u}
\newcommand{\vv}{\boldsymbol v}

\usepackage{tikz}
\usetikzlibrary{bayesnet}
\usepackage{tkz-graph}
\usepackage{tikz-qtree}
\usetikzlibrary{arrows}
\tikzset{
  LabelStyle/.style = { rectangle, rounded corners, draw,
                        minimum width = 2em, fill = yellow!50,
                        text = red, font = \bfseries },
  VertexStyle/.append style = { inner sep=0pt,
                                minimum size=1pt
                                },
  EdgeStyle/.append style = {->,  left} }

\usetikzlibrary{shapes, calc, shapes, arrows, positioning}
\tikzstyle{neuron}=[draw,circle,minimum size=20pt,inner sep=0pt, fill=white]
\tikzstyle{stateTransition}=[thick]
\tikzstyle{learned}=[text=red]

\tikzstyle{place}=[circle,draw=black,fill=white,inner sep=0pt,minimum size=6mm]

\tikzstyle{group1}=[rounded corners,fill=red!10,inner sep=1ex]

\tikzstyle{group2}=[rounded corners,fill=blue!10,inner sep=1ex]

\tikzstyle{group3}=[rounded corners,fill=orange!10,inner sep=1ex]

\tikzstyle{group4}=[rounded corners,fill=lime!10,inner sep=1ex]

\tikzstyle{group5}=[rounded corners,fill=cyan!10,inner sep=1ex]

\usetikzlibrary{arrows.meta}

\tikzset{>={Latex[width=2mm,length=2mm]}}

\newcommand{\eff}{\text{\normalfont{eff}}}
\newcommand{\screen}{\text{\normalfont{screen}}}
\newcommand{\ave}{\text{\normalfont{ave}}}

\begin{document}

\title{Learning Attribute Patterns in High-Dimensional Structured Latent Attribute Models}

\author{\name Yuqi Gu \email yuqigu@umich.edu\\ 
\name Gongjun Xu \email gongjun@umich.edu\\ 
\addr Department of Statistics\\ 
University of Michigan\\ 
Ann Arbor, MI 48109, USA}

\editor{}

\maketitle

\begin{abstract}
	{Structured latent attribute models (SLAMs) are a special family of discrete  latent variable models widely used in social and biological sciences. 
	This paper considers the problem of learning significant attribute patterns from a SLAM with potentially high-dimensional configurations of the latent attributes. 
	We address the theoretical identifiability issue, propose a penalized likelihood method for the selection of the attribute patterns, and further establish the selection consistency in such an overfitted SLAM  with  diverging number of latent patterns.
	 The good performance of the proposed methodology is illustrated by simulation studies and two real datasets  in educational assessment. 
	}
\end{abstract}

\section{Introduction}
\textit{Structured Latent Attribute Models} (SLAMs) are widely used statistical and machine learning tools in modern social and biological sciences.
 SLAMs offer a framework to  achieve fine-grained inference on individuals' latent attributes based on their observed multivariate responses, and also obtain the latent subgroups of a population based on the inferred attribute patterns.
In practice, each latent attribute is often assumed to be discrete and has  particular scientific interpretations,  such as   mastery or deficiency of some targeted skill  in educational assessment \citep{Junker, dela2011}, 
presence or absence of some underlying mental disorder  in  psychiatric diagnosis \citep{Templin, dela2018}, the existence or nonexistence of some disease pathogen  in subjects' biological samples \citep{wu2017}. In these scenarios, the framework of SLAMs enables one to simultaneously achieve the machine learning task of clustering, and the scientific purpose of diagnostic inference.

Different from the exploratory nature of traditional latent variable  models, SLAMs often have some   additional  scientific information for model fitting.
In particular, the observed variables are assumed to have certain structured dependence on the unobserved latent attributes, where the dependence is introduced through a  binary design matrix to respect the scientific context.
  The rich structure and nice interpretability of SLAMs make them popular in many scientific disciplines, such as cognitive diagnosis in educational assessment \citep{Junker,davier2008general,HensonTemplin09, rupp2010diagnostic,dela2011},   psychological and psychiatric measurement  for diagnosis of mental disorders \citep{Templin,dela2018}, and
  epidemiological and medical studies for   scientifically constrained  clustering  \citep{wu2017, wu2018}.
 
One challenge in modern applications of SLAMs is that  the number of discrete latent attributes could be large, leading to a high-dimensional space for all the possible configurations of the attributes, i.e., high-dimensional
 latent attribute patterns. In many applications, the number of potential attribute patterns is  much larger than the sample size. 
For scientific interpretability and practical use, it is often assumed that not all the possible   attribute patterns exist in the population.
Examples with large number of potential latent attribute patterns and moderate sample sizes can be found in   educational assessment   \citep{lee2011cognitive, choi2015cdm, yamaguchi} and the medical diagnosis of disease etiology \citep{wu2017}. For instance, Example \ref{exp-timss0} in Section 2 presents a dataset from Trends in   International Mathematics and Science Study (TIMSS), which has   $13$ binary latent attributes (i.e., $2^{13}=8192$ possible latent attribute patterns) while only 757 students' responses are observed. In cognitive diagnosis, it is  of interest to select the significant  attribute patterns among these $2^{13}=8192$. 
In such high-dimensional scenarios, existing  estimation methods often tend to over select the number of latent attribute patterns, and may not scale to   datasets with large number of latent attribute patterns. 
Moreover,   theoretical questions remain open on whether and when the ``sparse"    latent attribute patterns are identifiable  and can be consistently learned from the data.

 Identifiability of SLAMs has long been an issue in the literature \citep[e.g.,][]{davier2008general, decarlo2011analysis, MarisBechger, davier2014dina, Xu15}.  
 SLAMs can be viewed as a special class of restricted latent class models and their identifiability has a close connection  with the study of tensor  decompositions, by noting that the probability distribution of SLAMs can be viewed as a mixture of specially structured tensor products.
 In the  literature, it is known that  unrestricted latent class models   are not identifiable \citep{gyllenberg1994non}. Nonetheless, \cite{practicalID} showed through extensive simulations that they are almost always identifiable, which the authors termed as practical identifiability.   \cite{allman2009} further established \textit{generic} identifiability of various latent variable models, including latent class models. Generic identifiability is weaker than strict identifiability, and it implies that the model parameters are almost surely identifiable with respect to the Lebesgue measure of the parameter space. 
 The study of \cite{allman2009} is based on an identifiability result of the three-way tensor decomposition in \cite{kruskal1977three}.
Other analysis of tensor decomposition has also been developed to study the identifiability of various latent variable models \citep[e.g.,][]{drton2007,hsu2013,anand2014,bhaskara2014,anand2015topic,jaffe2018}.
 However, the structural constraints imposed by the   design matrix make these results not directly applicable to SLAMs.

 With the aid of the structural constraints, strict identifiability of SLAMs has been  obtained under certain conditions on the design matrix \citep{xu2017,xu2018,id-dina,partial}. 
 However,  these   works either make the strong  assumption that all the possible combinations of the attributes exist in the population with positive probabilities \citep{xu2017, xu2018}, or assume these significant  attribute patterns are known a priori \citep{partial}. These assumptions are difficult to meet in practice for SLAMs with high-dimensional attributes patterns and
the fundamental learnability issue of the sparse attribute patterns in  SLAMs remains
unaddressed.

In terms of estimation, learning sparse attribute patterns from a high-dimensional space of  latent attribute patterns is related to learning the significant mixture components in a  highly overfitted mixture model. Researchers have shown that the estimation of the mixing distributions in overfitted mixture models is technically challenging  and it usually leads to nonstandard convergence rate \citep[e.g.,][]{chen1995,ho2016convergence,heinrich}.
Estimating the number of components in the mixture model goes beyond only estimating the parameters of a mixture, by learning at least the order of the mixing distribution \citep{heinrich}. 
    This problem was also studied  in \cite{rousseau2011} from a Bayesian perspective; however, the  Bayesian estimator  in \cite{rousseau2011} may not guarantee the frequentist selection consistency, as to be shown in Section 3.
In the setting of SLAMs with the    structural constraints and a large number (larger than sample size)  of potential latent attribute patterns, it is not clear how to consistently select the significant attribute patterns.

Our contributions in this paper contain the following aspects. First, we characterize the identifiability requirement needed for a SLAM with an arbitrary subset of attribute patterns to be learnable, and establish mild identifiability conditions. 
Our new identifiability conditions   significantly extends the results of previous works   \citep{xu2017, xu2018} to more general and practical settings. 
Second, we propose a statistically consistent  method to perform attribute pattern  selection. 
In particular, we establish theoretical guarantee for   selection consistency in the setting of high dimensional latent attribute patterns, where both the sample size and the number of latent attribute patterns can go to infinity.  
Our analysis also shows that imposing the popular Dirichlet prior on the population proportions would fail to select the true model consistently, when the convergence rate of the SLAM is slower than the usual root-$N$ rate.
As for computation, we develop two approximation algorithms to maximize the penalized likelihood for pattern selection.
In addition, we propose a fast screening strategy for SLAMs as a preprocessing step that can scale to huge number of potential latent attribute patterns with a moderate sample size, and establish its sure screening property.

The rest of the paper is organized as follows. Section 2 introduces the general setup of structured latent attribute models and motivates our study. Section 3 investigates the learnability requirement and propose mild sufficient conditions for learnability. Section 4 proposes the estimation methodology and establishes theoretical guarantee for the proposed methods. Section 5 and Section 6 include simulations and real data analysis respectively. The proofs of the theoretical results are deferred to the Appendix.

\section{Model Setup and Motivation}
 \subsection{  Structured Latent Attribute Models and Examples}\label{sec-setup}
  We first introduce the general setup of SLAMs. 
Consider a SLAM with $J$ items which depend on the $K$ latent attributes of interest.  There are two types of subject-specific variables in the model,  the observed responses to items $\RR=(R_1,\ldots,R_J)$ and the latent attribute pattern $\aaa=(\alpha_1,\ldots,\alpha_K)$, both assumed to be binary vectors in this work. 
The $J$-dimensional vector  $\RR\in\{0,1\}^J$ denotes the observed binary responses to the set of $J$ items. 
The $K$-dimensional vector $\aaa\in\{0,1\}^K$ denotes a profile of existence or non-existence of the $K$ attributes. 

A key structure that specifies how the items depend on the latent attributes is called the $Q$-matrix, which is a $J\times K$ matrix with binary entries. We denote $Q=(q_{j,k})$ and $q_{j,k}\in\{1,0\}$ reflects whether or not item $j$ requires (i.e., depends on) attribute $k$.
We denote the $j$th row vector of $Q$ by $\qq_j$, then the $K$-dimensional binary vector $\qq_j$ reflects the full attribute requirements of item $j$. For an attribute pattern $\aaa$, we say $\aaa$ possesses all the required attributes of item $j$, if $\aaa\succeq\qq_j$, where $\aaa\succeq\qq_j$ denotes $\alpha_k\geq q_{j,k}$ for all $k=1,\ldots,K$. 
   Example \ref{exp-timss0} below gives an example of the $Q$-matrix.

\begin{example}\label{exp-timss0}
{
Trends in   International Mathematics and Science Study (TIMSS) is a large scale cross-country educational assessment. TIMSS evaluates the mathematics and science abilities of fourth and eighth graders  every four years since 1995. 
Researchers have used 
SLAMs   to analyze the TIMSS data \citep[e.g.,][]{lee2011cognitive, choi2015cdm, yamaguchi}.
For example, a $23\times 13$ $Q$-matrix constructed by mathematics educators was specified for the TIMSS 2003 eighth grade mathematics assessment \citep{choi2015cdm}. Thirteen attributes ($K=13$) are identified, which fall in five big categories of skill domains measured by the eighth grade exam, Number, Algebra, Geometry, Measurement, and Data. Table \ref{tab-tim-43} shows the first and last three rows of the $Q$-matrix, i.e., $\{\qq_j: j=1,2,3,21,22,23\}$.
\begin{table}[h!]
\centering
\caption{$Q$-matrix, TIMSS 2003 $8$th Grade Data}
\begin{tabular}{l|ccccccccccccc}
\hline
Item  &  $\alpha_1$ & $\alpha_2$ & $\alpha_3$ & $\alpha_4$ & $\alpha_5$ & $\alpha_6$ & $\alpha_7$ & $\alpha_8$ & $\alpha_9$ & $\alpha_{10}$ & $\alpha_{11}$ & $\alpha_{12}$ & $\alpha_{13}$ \\
\hline
1 & 1  &   0  &   0    & 0    & 0   &  0   &  0  &   0  &   0 &    0   &  1    & 0  &   1 \\
2 &     0   &  0  &   0   &  0  &   0 &    1  &   0  &   0  &   0  &   0  &   0   &  0   &  0 \\
3 &     0  &   1    & 0  &   0  &   0   &  0   &  1  &   0  &   0 &    0   &  0  &   0    & 0\\
\vdots & \vdots & \vdots & \vdots  & \vdots & \vdots  & \vdots & \vdots  & \vdots & \vdots  & \vdots & \vdots  & \vdots & \vdots \\
21 & 0  &   0   &  0  &   0  &   1  &   0   &  0  &   0  &   0  &   0    & 0  &   0   &  0 \\
22 &     0   &  1   &  0   &  0  &   0  &   0   &  0  &   0   &  0  &   0  &   0  &   0  &   0 \\
23 &   0  &   0   &  0   &  1  &   0   &  0    & 0  &   0 &    1 &    0   &  0  &   0   &  0 \\
\hline
\end{tabular}
\label{tab-tim-43}
\end{table}
}
\end{example}

The $Q$-matrix constrains the model parameters in a certain way to reflect the scientific assumptions. 
We next introduce the model parameters and how the $Q$-matrix impose constraints on them in general.
Conditional on a subject's latent attribute pattern  $\aaa\in \{0,1\}^K$, his/her responses to the $J$ items are assumed to be independent Bernoulli random variables with parameters $\theta_{1,\aaa},\ldots,\theta_{J,\aaa}$. Specifically, $\theta_{j,\aaa}=\mathbb P(R_j=1\mid  \aaa)$ denotes the positive response probability, and is also called an item parameter of item $j$. We collect all the item parameters in the matrix $\TT=(\theta_{j,\aaa})$, which has size $J\times 2^K$ with rows indexed by the $J$ items and columns by the $2^K$ attribute patterns. 
For pattern $\aaa\in\{0,1\}^K$, we denote its corresponding column vector in $\TT$ by $\TT_{\Cdot,\aaa}$.

One key assumption in SLAMs is that for a latent attribute pattern $\aaa=(\alpha_1,\ldots,\alpha_K)$ and item $j$, the parameter $\theta_{j,\aaa}$ is only determined by 
whether $\aaa$ possesses the attributes  in the set 
$
\mt K_{j}=\{k\in\{1,\ldots,K\}:  q_{j,k}=1\};
$ that is, those attributes related to item $j$ as specified in the $Q$-matrix. We will sometimes call the attributes in $\mt K_{j}$ the \textit{required attributes of item $j$}.
Under this assumption, all latent attribute patterns in the set  
\begin{equation}\label{eq-cdm-cj}
\mc_j = \{\aaa\in\{0,1\}^K:\,\aaa\succeq\qq_j\}
\end{equation}
  share the same value of $\theta_{j,\aaa}$; namely, 
\begin{equation}\label{eq-cons}
\quad\max_{\aaa\in \mc_j}\theta_{j,\aaa} = \min_{\aaa\in \mc_j}\theta_{j,\aaa}  \mbox{ for any }  j\in\{1,\ldots,J\}.
\end{equation}
We will call the set $\mc_j$ a \textit{constraint set}.
Thus, the $Q$-matrix  puts constraints on $\TT$ by forcing certain entries of it to be the same; specifically, for an item $j$, those attribute patterns that only differ in the attributes in $\{1,\ldots,K\}\setminus\mt K_j$ are constrained to  the same level of $\theta_{j,\aaa}$'s.  
Different SLAMs model the dependence of $\theta_{j,\aaa}$ on 
the required attributes differently to reflect the underlying scientific assumptions. Please see Examples \ref{exp-dina} and \ref{exp-gdina} for examples.

In addition to \eqref{eq-cons}, another key assumption in SLAMs is the monotonicity assumption 
that 
\begin{equation}\label{eq-cons2}
 \theta_{j,\aaa} >  \theta_{j,\aaa'} \mbox{ for any } \aaa\in\mc_j, \,\aaa'\not\in\mc_j.
\end{equation}
Constraint  \eqref{eq-cons2} is commonly used in our motivating applications of cognitive diagnosis in educational assessment, where \eqref{eq-cons2} indicates subjects mastering  all required attributes of an item are more ``capable" of giving a positive response to   it (i.e., with higher $\theta_{j,\aaa}$),  than those who lack some required attributes.
Nonetheless, our theoretical results of model learnability in Section \ref{sec-id} also applies if  \eqref{eq-cons2} is relaxed  to
 \begin{equation}\label{eq-cons3}
 \theta_{j,\aaa} \neq   \theta_{j,\aaa'} \mbox{ for any } \aaa\in\mc_j,\,\aaa'\not\in\mc_j.
\end{equation}
This allows more flexibility in the model assumptions of SLAMs used in other applications.

Next we introduce some popular  SLAMs in educational and psychological applications. These models are also called Cognitive Diagnosis Models in the psychometrics literature.
The first type of SLAMs have exactly two item parameters associated with each item.
\begin{example}[two-parameter SLAM]\label{exp-dina}
The two-parameter SLAM specify exactly two levels of item parameters for each item $j$, which we denote by $\theta_j^+$ and $\theta_j^-$ with $\theta_j^+>\theta_j^-$. The popular Deterministic Input Noisy output ``And" gate (DINA) model introduced in \cite{Junker} is a two-parameter SLAM.
It assumes 
the general form of $\theta_{j,\aaa}$ can be rewritten as
$$
\theta^{\text{two-para}}_{j,\aaa} =
\begin{cases}
\theta_j^+, & \text{if}~~
\aaa\in\mc_j,\\
\theta_j^-,   & \text{if}~~
\aaa\not\in\mc_j.\\
\end{cases}
$$
In the application of the  two-parameter SLAM in educational assessment, the item parameters $\theta_j^+$ and $\theta_j^-$ have the following   interpretations. The  $1-\theta_j^+$ is called the slipping parameter,  denoting the probability of a ``capable'' subject slips the correct response, despite mastering all the required attributes of item $j$; and $\theta_j^-$ is called the guessing parameter,  denoting the probability of a ``non-capable'' subject coincidentally giving the correct response by guessing, despite lacking some required attributes of item $j$.
In this case, the unique item parameters in matrix $\TT$   reduce to $(\ttt^+,\ttt^-)$, where $\ttt^+=(\theta_1^+,\ldots,\theta_J^+)^\top$ and $\ttt^-=(\theta_1^-,\ldots,\theta_J^-)^\top$.
Under the two-parameter SLAM, the constraint set of each item $j$ takes the form of \eqref{eq-cdm-cj} and satisfies \eqref{eq-cons} and \eqref{eq-cons2}.

\end{example}

Another family of SLAMs are the multi-parameter models, which allow each item to have multiple levels of item parameters.
\begin{example}[multi-parameter SLAMs]\label{exp-gdina}
Multi-parameter SLAMs can be categorized into two general types, the main-effect models and the all-effect models.
The main-effect models assume the main effect of the required attributes play a role in distinguishing the positive response probabilities. The item parameters can be written as
\begin{equation}\label{maineff}
\theta^{\text{main-eff}}_{j,\aaa}   =
f\Big(\beta_{j,0}+ {\sum}_{k\in \mathcal K_{j}}\beta_{j,k}\alpha_k\Big),
\end{equation}
where 
$f(\cdot)$ is a link function. Different link functions $f(\cdot)$ lead to different models, including the popular reduced Reparameterized Unified Model \citep[reduced-RUM;][]{dibello1995unified} with $f(\cdot)$ being  the exponential function, the Linear Logistic Model \citep[LLM;][]{maris1999estimating} with $f(\cdot)$ being  the sigmoid function, and the Additive Cognitive Diagnosis Model \citep[ACDM;][]{dela2011} with $f(\cdot)$   the identity function.

Another type of multi-parameter SLAMs are the all-effect models. The item parameter of an all-effect model can be written as
\begin{equation}\label{alleff}
\theta^{\text{all-eff}}_{j,\aaa} =
f\Big({\sum}_{S\subseteq \mathcal K_{j}}\beta_{j,S}{\prod}_{k\in S}\alpha_k\Big),
\end{equation}
When $f(\cdot)$ is the  identity function, \eqref{alleff} is the Generalized DINA (GDINA) model proposed by \cite{dela2011}; and when $f(\cdot)$ is the sigmoid function, \eqref{alleff} is the Log-linear Cognitive Diagnosis Models (LCDMs) proposed by \cite{HensonTemplin09}; see also the General Diagnostic Models (GDMs) proposed in \cite{davier2008general}. 
\end{example}

Under the multi-parameter SLAMs, the constraint set of each item $j$ also takes the form of \eqref{eq-cdm-cj}.
  Those attribute patterns in $\mc_j$ still share the same value of item parameters by the definition; and what is different from the two-parameter counterpart is that those $\aaa$ not in $\mc_j$ can have different levels of item parameters.
We next give another example of multi-parameter SLAMs.
\begin{example}[Deep Boltzmann Machines]
The Restricted Boltzmann Machine (RBM) \citep{smo1986,DLbook} is a popular neural network model. RBM is an undirected probabilistic graphical model, with one layer of latent (hidden) binary variables, one layer of observed (visible) binary variables, and a bipartite graph structure between the two layers. 
We denote variables in the observed layer by $\RR$ and variables in the latent layer by $\aaa$, with their   lengths  $J$ and $K$, respectively.
Under a RBM,  the probability  mass function of $\RR$  and $\aaa$ is
 $\mathbb P(\RR, \aaa)
\propto\exp( 
  - \RR^\top \bo W^Q \aaa   - \bo f^\top \RR - \bo b^\top \aaa ),$
 where $\bo f$, $\bo b$, and $\bo W^Q=(w_{j,k})$ are the parameters. The binary $Q$-matrix then specifies the sparsity structure in $\bo W^Q$, by constraining $w_{j,k}\neq 0$ only if $q_{j,k}\neq 0$. 
The Deep Boltzmann Machine (DBM) is a generalization of RBM by allowing multiple latent layers. 
Consider a DBM with two latent layers $\aaa^{(1)}$ and $\aaa^{(2)}$ of length $K_1$ and $K_2$, respectively. 
The probability mass function of $(\RR, \aaa^{(1)},\aaa^{(2)})$ in this  DBM can be written as
\begin{align}\label{eq-dbm}
\mathbb P(\RR, \aaa^{(1)},\aaa^{(2)})\propto \exp\Big( 
&- \RR^\top \bo W^Q \aaa^{(1)}  - (\aaa^{(1)})^\top \bo U \aaa^{(2)}  
 -  \bo f^\top \RR - \bo b_1^\top \aaa^{(1)} -  \bo b_2^\top \aaa^{(2)}
\Big),
\end{align}
where $\bo f\in\reals^J$, $\bo b_i\in\reals^{K_i}$ for $i=1,2$, and $\bo W^Q=(w_{j,k})\in\reals^{J\times K_1}$, $\bo U\in\reals^{K_1\times K_2}$ are model parameters;  Figure \ref{fig-rbm} gives an example of a DBM with a $5\times 4$ $Q$-matrix. 
 For $\bo f=(f_1,\ldots,f_J)^\top$  and   $\aaa^{(1)}=(\alpha^{(1)}_1,\ldots,\alpha^{(1)}_{K_1})$,
 the conditional distribution of an observed variable $R_j$ given the latent variables is
\begin{align}\label{eq-cond}
\mathbb P(R_j=1\mid\aaa^{(1)},\aaa^{(2)},\cdots) 
=    \mathbb P(R_j=1\mid\aaa^{(1)})=~ \frac{\exp\Big(\sum_{k=1}^{K_1} w_{j,k}\alpha_k^{(1)} + f_j\Big)}{1+\exp\Big(\sum_{k=1}^{K_1} w_{j,k}\alpha_k^{(1)} + f_j\Big)}, 
\end{align}
where 
``~$\cdots$" represents deeper latent layers that potentially exist in a DBM.
Moreover, from \eqref{eq-dbm} we have
$
\mathbb P(\RR\mid \aaa^{(1)}) = \prod_{j=1}^J\mathbb P(R_j\mid \aaa^{(1)}),
$
so a DBM satisfies the local independence assumption that  the $R_j$'s are conditionally independent given the $\aaa^{(1)}$.
Therefore, a DBM can be viewed as a multi-parameter main-effect SLAM in \eqref{maineff} with a sigmoid link function.
 Viewing a DBM in this way, \eqref{eq-cond} gives the item parameter $\theta_{j,\aaa^{(1)}}$, and the constraint set of each item $j$ also takes the form 
 $\mc_j = \{\aaa^{(1)}\in\{0,1\}^{K_1}:\,\aaa^{(1)}\succeq\qq_j\}$.

\begin{figure}[h!]
\centering
\begin{minipage}{0.4\textwidth}
$$
Q=\begin{pmatrix}
1 & 0 & 1 & 0 \\
1 & 1 & 0 & 0 \\
0 & 1 & 1 & 0 \\
0 & 0 & 1 & 1 \\
0 & 1 & 0 & 1 \\
\end{pmatrix};
$$
\end{minipage}
\begin{minipage}{0.6\textwidth}
\centering
    \begin{tikzpicture}[scale=1.3]

    \node (v1)[neuron] at (0, -0.2) {$R_1$};
    \node (v2)[neuron] at (1, -0.2) {$R_2$};
    \node (v3)[neuron] at (2, -0.2) {$R_3$};
    \node (v4)[neuron] at (3, -0.2) {$R_4$};
    \node (v5)[neuron] at (4, -0.2) {$R_5$};
    \node[right=0.1cm of v5] (v) {\quad $\RR \in \{0, 1\}^5$};

    \node (h1)[neuron] at (0.5, 1) {$\alpha^{(1)}_1$};
    \node (h2)[neuron] at (1.5, 1) {$\alpha^{(1)}_2$};
    \node (h3)[neuron] at (2.5, 1) {$\alpha^{(1)}_3$};
    \node (h4)[neuron] at (3.5, 1) {$\alpha^{(1)}_4$};
    \node[right=0.1cm of h4] (h) {\quad $\aaa^{(1)} \in \{0, 1\}^4$};

    \node[] (W) at (4.8, 0.4) {\quad\quad\quad\quad $W^Q \in \mathbb{R}^{5 \times 4}$};

    \draw[stateTransition] (v1) -- (h1) node [midway,above=-0.06cm,sloped] {$w_{1,1}$};
    \draw[stateTransition] (v1) -- (h3) node [midway,above=-0.06cm,sloped] {};

    \draw[stateTransition] (v2) -- (h1) node [midway,above=-0.06cm,sloped] {};
    \draw[stateTransition] (v2) -- (h2) node [midway,above=-0.06cm,sloped] {};

    \draw[stateTransition] (v3) -- (h2) node [midway,above=-0.06cm,sloped] {};
    \draw[stateTransition] (v3) -- (h3) node [midway,above=-0.06cm,sloped] {};

    \draw[learned,stateTransition] (v4) -- (h3) node [midway,above=-0.06cm,sloped] {};
    \draw[stateTransition] (v4) -- (h4) node [midway,above=-0.06cm,sloped] {};
    
    \draw[stateTransition] (v5) -- (h2) node [midway,above=-0.06cm,sloped] {};
    \draw[stateTransition] (v5) -- (h4) node [midway,above=-0.06cm,sloped] {$w_{5,4}$};

    \node (h5)[neuron] at (0, 2.2) {$\alpha^{(2)}_{1}$};
    \node (h6)[neuron] at (1.33, 2.2) {$\alpha^{(2)}_{2}$};
    \node (h7)[neuron] at (2.67, 2.2) {$\alpha^{(2)}_{3}$};
    \node (h8)[neuron] at (4, 2.2) {$\alpha^{(2)}_{4}$};
    \node[right=0.1cm of h8] (h) {\quad $\aaa^{(2)} \in \{0, 1\}^4$};

        \draw[learned,stateTransition] (h1) -- (h5) node [midway,above=-0.06cm,sloped] {};
    \draw[stateTransition] (h1) -- (h6) node [midway,above=-0.06cm,sloped] {};

    \draw[stateTransition] (h2) -- (h5) node [midway,above=-0.06cm,sloped] {};
    \draw[stateTransition] (h2) -- (h6) node [midway,above=-0.06cm,sloped] {};
    
            \draw[learned,stateTransition] (h3) -- (h5) node [midway,above=-0.06cm,sloped] {};
    \draw[stateTransition] (h3) -- (h6) node [midway,above=-0.06cm,sloped] {};

    \draw[stateTransition] (h4) -- (h5) node [midway,above=-0.06cm,sloped] {};
    \draw[stateTransition] (h4) -- (h6) node [midway,above=-0.06cm,sloped] {};

    \draw[learned,stateTransition] (h1) -- (h7) node [midway,above=-0.06cm,sloped] {};
    \draw[stateTransition] (h1) -- (h8) node [midway,above=-0.06cm,sloped] {};

    \draw[stateTransition] (h2) -- (h7) node [midway,above=-0.06cm,sloped] {};
    \draw[stateTransition] (h2) -- (h8) node [midway,above=-0.06cm,sloped] {};
    
    \draw[learned,stateTransition] (h3) -- (h7) node [midway,above=-0.06cm,sloped] {};
    \draw[stateTransition] (h3) -- (h8) node [midway,above=-0.06cm,sloped] {};

    \draw[stateTransition] (h4) -- (h7) node [midway,above=-0.06cm,sloped] {};
    \draw[stateTransition] (h4) -- (h8) node [midway,above=-0.06cm,sloped] {};
    
\end{tikzpicture}

\end{minipage}

\caption{Deep Boltzmann Machine}
\label{fig-rbm}
\end{figure}
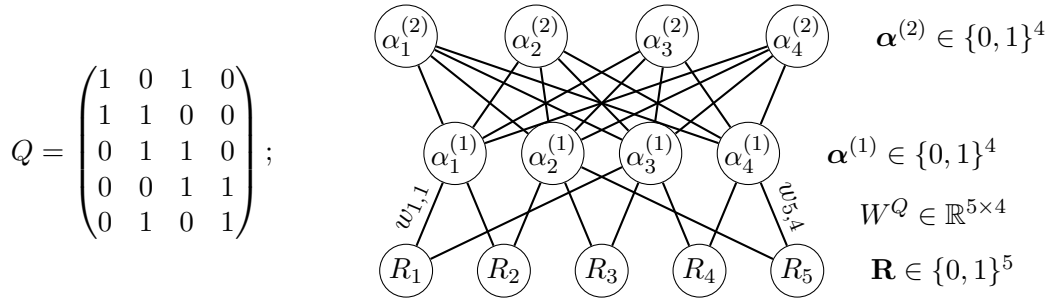
 \end{example}

\subsection{Motivation and Problem}

One challenge in modern applications of SLAMs is that the number of potential latent attribute patterns $2^K$ increases exponentially with $K$ and could be much larger than the sample size $N$. It is often assumed that a relatively small portion of   attribute patterns exist in the population.
For instance, Example \ref{exp-timss0} has   $2^K=2^{13}=8192$ different configurations of attribute patterns.
Given the limited sample size $757$, it is desirable to   learn the potentially small set of significant  attribute patterns from data.

Another motivation for assuming   a small number of attribute patterns  exist in the population, results from the possible hierarchical structure among the targeted attributes.  
{
For instance, in  educational assessment of a set of underlying latent skill attributes, some attributes often serve as prerequisites for some others \citep{leighton2004attribute, templin2014hierarchical}. Specifically, the prerequisite relationship depicts the different level of difficulty of the skill attributes, and also reveals the order in which these skills are learned in the population of students. 
For instance, if attribute $\alpha_1$ is a prerequisite for attribute $\alpha_2$, then the attribute pattern $(\alpha_1=0,\alpha_2=1)$ does not exists in the population, naturally resulting in a sparsity structure of the existence of attribute patterns.
When the number of attributes is large and the underlying hierarchy structure is complex and  unknown, it is   desirable to learn the hierarchy of attributes directly from data. 
In such cases with attribute hierarchy, the number of patterns respecting the hierarchy could be far fewer than $2^K$.

The problem of interest is that, given a moderate sample size, how to consistently estimate the small set of latent attribute patterns among all the possible $2^K$ patterns. 
As discussed in the introduction, in the high-dimensional case when the total number of attribute patterns is   large or even larger than the sample size, the questions of when the true model with the significant attribute patterns are learnable from data, and how to perform consistent pattern selection, remain open in the literature.

  This problem is equivalent to selecting the nonzero elements of  the population proportion parameters $\pp=(p_{\aaa}:\,\aaa\in\{0,1\}^K)$, where we use $p_{\aaa}$ to denote the proportion of the subjects with attribute pattern $\aaa$ in the population. The $\pp$ satisfies  $p_{\aaa}\in [0,1]$ for $\aaa\in\{0,1\}^K$ and $\sum_{\aaa\in\{0,1\}^K} p_{\aaa}=1$. 
In this work, we will treat  the latent attribute patterns $\aaa$ as random variables (random effects).
For any subject, his/her attribute pattern is a random vector $\ma\in\{0,1\}^K$ that (marginally) follows a categorical distribution with population proportion parameters $\pp=(p_{\aaa}:\,\aaa\in\{0,1\}^K)$.  
 One main reason for this random effect assumption is that, when the number of observed variables per subject (i.e., $J$) does not increase with the sample size $N$ asymptotically, the counterpart fixed effect model can not consistently estimate the model parameters. As a consequence, the fixed effect approach can not give consistent selection of significant attribute patterns. This scenario with relatively small $J$ but larger $N$ and $2^K$ is commonly seen in the motivating applications in educational and psychological assessments.

We would like to point out that we give the joint distribution of the attributes  full flexibility by modeling it as a categorical distribution with $2^K-1$ free proportion parameters $p_{\aaa}$'s. Modeling in this way allows those ``sparse" significant attribute patterns to have arbitrary structures among the $2^K$ possibilities.
On the contrary, any simpler parametric modeling of the distribution of $\aaa$ with fewer   parameters would fail to capture all the possibilities of the attributes' dependency.

Under the introduced notations, the probability mass function of a subject's response vector $\RR=(R_1,\ldots,R_J)^\top$ can be written as
$\mathbb P(\RR=\rr\mid \TT, \,\pp) = \sum_{\aaa\in \{0,1\}^K}p_{\aaa} \prod_{j=1}^J \theta_{j,\aaa}^{r_j} (1-\theta_{j,\aaa})^{1-r_j},$ for
$\rr\in\{0,1\}^J.$
Alternatively, the responses  can be viewed as a $J$-th order tensor  and  the probability mass function of $\RR$ can be written as a probability tensor
\begin{equation}\label{eq-tensor}
\mathbb P(\RR\mid \TT, \,\pp) = \sum_{l=1}^{2^K} p_{\aaa_l} \, 
\begin{pmatrix}
\theta_{1,\aaa_l} \\
1-\theta_{1,\aaa_l}
\end{pmatrix} \circ 
\begin{pmatrix}
\theta_{2,\aaa_l} \\
1-\theta_{2,\aaa_l}
\end{pmatrix} \circ \cdots \circ 
\begin{pmatrix}
\theta_{J,\aaa_l} \\
1-\theta_{J,\aaa_l}
\end{pmatrix},
\end{equation}
where ``$\circ$" denotes the tensor outer product and $\theta$'s are constrained by \eqref{eq-cons} and \eqref{eq-cons2}.

 In the following sections, we first  investigate the learnability requirement of learning a SLAM with an arbitrary set of true attribute patterns, and provide   identifiability conditions in Section \ref{sec-id}. 
Then in Section \ref{sec-reg}, we propose a penalized likelihood method to select the attribute patterns, and we establish theoretical guarantee for the proposed method.

\section{Learnability Requirement and Conditions}\label{sec-id}

To facilitate the discussion on identifiability of SLAMs, we need to introduce a new notation, the $\Gamma$-matrix. 
We first introduce the $J\times 2^K$ constraint matrix $\Gamma^{\text{all}}$ that is entirely determined by the $Q$-matrix. The rows of $\Gamma^{\text{all}}$  are indexed by the $J$ items, and columns by the $2^K$ latent attribute patterns in $\{0,1\}^K$. 
The $(j,\aaa)$th entry of $\Gamma^{\text{all}}_{j,\aaa}$ is defined as 
\begin{equation}\label{eq-gall}
\Gamma^{\text{all}}_{j,\aaa}=I(\aaa\succeq\qq_j)=I(\aaa\in\mc_j),\quad j\in\{1,\ldots,J\},~\aaa\in\{0,1\}^K,
\end{equation}
 which is a binary indicator of whether attribute pattern $\aaa$ possess all the required attributes of item $j$. 
We will also call $\Gamma^{\text{all}}$ the \textit{constraint matrix}, since its entries indicate  what latent attribute patterns are constrained to have the highest level of parameters for each item. For example, consider the $2\times 2$ $Q$-matrix in the following \eqref{eq-q24}.  
Then its corresponding $\Gamma$-matrix $\Gamma^{\text{all}}$ with a saturated set of attribute patterns takes the following form. 
\begingroup
\setlength{\belowdisplayskip}{0pt}
\begin{align}\label{eq-q24}
Q = 
\begin{pmatrix}
0 & 1\\
1 & 1\\
\end{pmatrix}
~\Longrightarrow~
\Gamma^{\text{all}} = 
\begin{blockarray}{cccc}
\aaa_1 & \aaa_2 & \aaa_3 & \aaa_4 \\
(0,0) & (0,1) & (1,0) & (1,1)\\
\begin{block}{(cccc)}
0 & 1 & 0 & 1\\
0 & 0 & 0 & 1\\
\end{block}
\end{blockarray}~.
\end{align}
\endgroup
More generally, we generalize the definition of the constraint 
 matrix $\Gamma^{\text{all}}$ in \eqref{eq-gall} to an arbitrary subset $\mca$ of the entire attribute pattern space $\{0,1\}^K$, and an arbitrary set of items $S\subseteq[J]$. For $S\subseteq[J]$ and $\mca\subseteq\{0,1\}^K$, we simply denote by $\Gamma^{(S,\mca)}$ the $|S|\times |\mathcal A|$ submatrix of $\Gamma^{\text{all}}$ with column indices from $\mca$. When $S=\{1,\ldots,J\}$, we will sometimes just denote $\Gamma^{(S,\mca)}$ by $\Gamma^{\mca}$ for simplicity. Then $\Gamma^{\mca}$ itself can be viewed as the constraint matrix  for a SLAM with attribute pattern space $\mca$, and $\Gamma^{\mca}$  directly characterizes how the items constrain the positive response probabilities of latent attribute patterns in $\mca$.

Given the $Q$-matrix, we denote by $\mca_0\subseteq\{0,1\}^K$ the set of true attribute patterns existing in the population, i.e., $\mca_0 = \{\aaa\in \{0,1\}^K: p_{\aaa} >0 \}$.
In knowledge space theory \citep{duntsch1995}, the set $\mca_0$ of patterns corresponds to the \textit{knowledge structure} of the population.
We further denote by $\TT^{\mca_0}$ the item parameter matrix respecting the constraints imposed by $\Gamma^{\mca_0}$; specifically, $\TT^{\mca_0}=(\theta_{j,\aaa})$ has the same size as $\Gamma^{\mca_0}$, with rows and columns indexed by the $J$ items and the attribute patterns in $\mca_0$, respectively.
For any positive integer $k\leq 2^K$, we let $\mathcal T^{k-1}$ be the $k$-dimensional simplex, i.e., $\mathcal T^{k-1} = \{(x_1,x_2,\ldots,x_k):\, x_i\geq 0,~ \sum_{i=1}^k x_k=1\}$.
We denote the true proportion parameters by $\pp^{\mca_0} = (p_{\aaa},\aaa\in\mca_0)\in\mathcal T^{|\mca_0|-1}$, then $\pp^{\mca_0}\succ \zero$ by the definition of $\mca_0$.

The following toy example illustrates why we need to establish identifiability guarantee for pattern selection.

\begin{example}\label{exp-id}
Consider the $2\times 2$ $Q$-matrix together with its corresponding $2\times 4$ $\Gamma$-matrix in Equation \eqref{eq-q24}.
Consider two attribute pattern sets, the true set $\mca_0 = \{\aaa_1=(0,0),\aaa_2=(0,1)\}$ and an alternative set $\mca_1 = \{\aaa_2=(0,1),\aaa_3=(1,0)\}$. 
Under the two-parameter SLAM, for any valid item parameters $\TT$ restricted by $\Gamma$ and any proportion parameters $\pp = (p_{\aaa_1},p_{\aaa_2},p_{\aaa_3},p_{\aaa_4})$ such that $p_{\aaa_1}=p_{\aaa_3}$, we have
 $\mathbb P(\RR = \rr\mid \TT^{\mca_0},\, (p_{\aaa_1}, p_{\aaa_2}))=\mathbb P(\RR = \rr\mid \TT^{\mca_1},\,(p_{\aaa_3}, p_{\aaa_2})).$  
This is   because $\Gamma^{\mca_0}=\Gamma^{\mca_1}$ from \eqref{eq-q24} and hence $\TT^{\mca_0}=\TT^{\mca_1}$; and also $(p_{\aaa_1}, p_{\aaa_2})=(p_{\aaa_3}, p_{\aaa_2})$ by our construction that $p_{\aaa_1}=p_{\aaa_3}$.
This implies 
even if one knows exactly there are  two latent attribute patterns in the population, one can never tell which two patterns those are based on the likelihood function. 
In this sense, $\mca_0$ is not identifiable, due to the fact that $\Gamma^{\mca_0}$ and $\Gamma^{\mca_1}$ do not lead to distinguishable distributions of responses under the two-parameter SLAM.
\end{example}

From the above example, to make sure the set of true attribute patterns $\mca_0$ is learnable from the observed multivariate responses, we need the $\Gamma^{\mca_0}$-matrix to have certain structures.
We   state  the formal definition of (strict) learnability of $\mca_0$.

\begin{definition}[strict learnability of $\mca_0$]\label{def-id}
Given $Q$, $\mca_0$ is said to be (strictly) learnable,
if for any constraint matrix $\Gamma^{\mca}$ of size $J\times |\mca|$ with $|\mca|\leq |\mca_0|$, any valid item parameters $\TT^{\mca}$ respecting constraints given by $\Gamma^{\mca}$, and any proportion parameters $\pp^{\mca}\in\mathcal T^{|\mca|-1}$, $\pp^{\mca}\succ \zero$, the following equality
\begin{equation}\label{eq-id-def}
\mathbb P(\RR\mid \TT^{\mca_0},\, \pp^{\mca_0})
= \mathbb P(\RR\mid \TT^{\mca},\, \pp^{\mca})
\end{equation}
implies $\mca=\mca_0$. Moreover, if \eqref{eq-id-def} implies $(\TT^{\mca},\,\pp^{\mca})=(\TT^{\mca_0},\,\pp^{\mca_0})$, then we say the model parameters $(\TT^{\mca_0},\,\pp^{\mca_0})$ are (strictly) identifiable.
\end{definition}

Next we further introduce some  notations and definitions about the constraint matrix $\Gamma$ and then present the needed identifiability result.
Consider an arbitrary subset of items $S\subseteq\{1,\ldots,J\}$.
For $\aaa,\aaa'\in\mca$, we denote $\aaa\succeq_{S}\aaa'$ under $\Gamma^{\mca}$, if
for each $j\in S$ 
there is $\Gamma^{\mca}_{j,\aaa}\geq\Gamma^{\mca}_{j,\aaa'}$.
If viewing $\Gamma_{j,\aaa}=1$ as $\aaa$ being ``capable'' of item $j$, then $\aaa\succeq_{S}\aaa'$ would mean $\aaa$ is at least as capable as $\aaa'$ of items in set $S$. 
Then under $\Gamma$, any subset of items $S$ defines a partial order ``$\succeq_{S}$" on the set of latent attribute patterns $\mca$.
For two item sets $S_1$ and $S_2$, we say $``\succeq_{S_1}" = ``\succeq_{S_2}"$ under $\Gamma^{\mca}$,
 if for any $\aaa'$, $\aaa\in\mca$, we have
$\aaa\succeq_{S_1}\aaa'$ under $\Gamma^{\mca}$ if and only if  $\aaa\succeq_{S_2}\aaa'$ under $\Gamma^{\mca}$.
The next theorem gives conditions that ensure the constraint matrix $\Gamma$ as well as the $\Gamma$-constrained model parameters are jointly identifiable.

\begin{theorem}[conditions for strict learnability]\label{thm-id}
Consider a SLAM with an arbitrary set of true attribute patterns $\mca_0\subseteq\{0,1\}^K$, and a corresponding constraint matrix $\Gamma^{\mca_0}$. If this true $\Gamma^{\mca_0}$ satisfies the following conditions, then $\mca_0$ is identifiable.\begin{enumerate}
\item[A.] There exist two disjoint item sets $S_1$ and $S_2$, such that $\Gamma^{(S_i,\mca_0)}$ has distinct column vectors for $i=1,2$ and ``\,$\succeq_{S_1} = \succeq_{S_2}$" under $\Gamma^{\mca_0}$.

\item[B.] For any $\aaa$, $\aaa'\in\mca_0$ where $\aaa'\succeq_{S_i} \aaa$ under $\Gamma^{\mca_0}$ for $i=1$ or $2$, there exists some $j\in(S_1\cup S_2)^c$ such that $\Gamma_{j,\aaa}^{\mca_0}\neq \Gamma_{j,\aaa'}^{\mca_0}$.
\item[C.] Any column vector of $\Gamma^{\mca_0}$ is different from any column vector of $\Gamma^{\mca_0^c}$, where $\mca_0^c = \{0,1\}^K\setminus\mca_0$.
\end{enumerate}
\end{theorem}

 Recall that each column in the $\Gamma$-matrix corresponds to a latent attribute pattern, then
 Conditions $A$ and $B$ help ensure the $\Gamma$-matrix of the true patterns $\Gamma^{\mca_0}$ contains enough information to distinguish between these true patterns.
  Specifically,  Condition $A$ requires $\Gamma^{\mca_0}$ to contain two vertically stacked submatrices corresponding to item sets $S_1$ and $S_2$, each having distinct columns, i.e., each being able to distinguish between the true patterns; and Condition $B$ requires the remaining submatrix of $\Gamma^{\mca_0}$ to distinguish those pairs of true patterns that have some order ($\aaa'\succeq_{S_i} \aaa$) based on the first two item sets $S_1$ or $S_2$. 
  Condition $C$ is necessary for identifiability of $\mca_0$ by ensuring that any true pattern would have a different column vector in $\Gamma^{\text{all}}$ from that of any false pattern. Condition $C$ is satisfied for any $\mca_0\subseteq\{0,1\}^K$ if the $Q$-matrix contains an identity submatrix $I_K$, because such a $Q$-matrix will give a $\Gamma^{\text{all}}$ that has all the $2^K$ columns distinct.

We would  like to point out that  our identifiability conditions in Theorem \ref{thm-id} do not depend on the unknown parameters (e.g., $\TT$ and $\pp$), but only rely on the structure of the constraint matrix $\Gamma$.
{The $\Gamma$-matrix with respect to the true set of patterns $\mca_0$ is the key quantity that defines the latent structure of a SLAM. Generally, it is hard to establish identifiability conditions that only depend on the cardinality of $\mca_0$ but not on $\Gamma^{\mca_0}$. For instance, in Example \ref{exp-id}, the two sets $\mca_0$ and $\mca_1$ have the same cardinality but can not be distinguished under the conditions there; indeed further conditions on $Q$ (and the resulting $\Gamma$) are needed to guarantee identifiability.}

The developed identifiability conditions generally apply to any SLAM satisfying the constraints \eqref{eq-cons} and \eqref{eq-cons2} introduced in Section \ref{sec-setup}. If one makes further assumptions on $\TT$, such as assuming each item $j\in[J]$ has exactly two item parameters to make it a two-parameter model, then the conditions in Theorem \ref{thm-id} may be further relaxed. For example, in the saturated case with $\mca_0=\{0,1\}^K$, the sufficient identifiability conditions developed in \cite{xu2017} for a general SLAM require $Q$ to contain two copies of $I_K$ as submatrices, while the necessary and sufficient conditions established in \cite{id-dina} for the two-parameter SLAM require $Q$ to have just one submatrix $I_K$. We expect that in the current case with an arbitrary $\mca_0\subseteq\{0,1\}^K$, the conditions in Theorem \ref{thm-id} can also be relaxed under the two-parameter model in a technically nontrivial way. For the reason of generality, we focus on SLAMs under the general constraints \eqref{eq-cons} and \eqref{eq-cons2} in this work. 
\color{black}
 
When the conditions in Theorem \ref{thm-id} are satisfied, $\mca_0$ is identifiable; and from Theorem 4.1 in \cite{partial}, the model parameters $(\TT^{\mca_0},\pp^{\mca_0})$ associated with $\mca_0$ are also identifiable.
\begin{corollary}\label{cor1}
	Under the conditions in Theorem \ref{thm-id},  the model parameters $(\TT^{\mca_0},\pp^{\mca_0})$ associated with $\mca_0$ are   identifiable.
\end{corollary}
Note that the result of Theorem \ref{thm-id} differs from the existing works \cite{xu2017}, \cite{xu2018} and \cite{partial} in that those works assume $\mca_0$ is known a priori  and study the   identifiability of $(\TT^{\mca_0},\pp^{\mca_0})$, while  in the current work  $\mca_0$ is unknown and we focus on the identifiability of $\mca_0$ itself.
 This is crucially needed in order to guarantee that we can learn the set of true attribute patterns.
 
\begin{remark}
The identifiability result in Theorem \ref{thm-id} and Corollary \ref{cor1} is  related to the uniqueness of tensor decomposition. As shown in \eqref{eq-tensor}, the probability mass function of the multivariate responses of each subject can be viewed as a higher order tensor with constraints on entries of the tensor, and unique decomposition of the tensor correspond to identification of the constraint matrix as well as the model parameters.
The identifiability conditions in Theorem \ref{thm-id} are weaker than the general conditions for uniqueness of three-way tensor decomposition  in \cite{kruskal1977three}, which is a celebrated result in the literature. Kruskal's conditions require the  tensor can be decomposed as a Khatri-Rao product of three matrices, two having full-rank and the other having Kruskal rank at least two (Kruskal rank of a matrix is the largest number $T$ such that every set of $T$ columns of it are linearly independent). 
Consider an example with $J=5$, $K=2$, $\mca_0 = \{\aaa_2 = (0,1), ~\aaa_3 = (1,0)\}$, and the corresponding $\Gamma^{\mca_0}$ in the  form of \eqref{gamma-52}. Then we can set $S_1=\{1,2\}$, $S_2=\{3,4\}$ and Condition $A$ in Theorem \ref{thm-id} is satisfied. 
Further, Condition $B$ is also satisfied since $\aaa_2\nsucceq_{S_i} \aaa_3$ and $\aaa_3\nsucceq_{S_i} \aaa_2$ under $\Gamma^{\mca_0}$.
Therefore, Theorem 1 guarantees the set $\mca_0$ is identifiable, and further guarantees the parameters $(\TT^{\mca_0}, \pp^{\mca_0})$ are identifiable. On the contrary,  results based on Kruskal's conditions for unique three-way tensor decomposition can not guarantee identifiability, because other than two full rank structures given by the items in $S_1$ and $S_2$, the remaining item 5 in $(S_1\cup S_2)^c$ corresponds to a structure with Kruskal rank only one.
\begingroup
\setlength{\belowdisplayskip}{0pt}
\begin{equation}\label{gamma-52}
Q = 
\begin{blockarray}{cc}
\begin{block}{(cc)}
 1 & 0 \\
 0 & 1 \\
\cmidrule{1-2}
 1 & 0 \\
 0 & 1 \\
\cmidrule{1-2}
 1 & 1 \\
\end{block}
\end{blockarray}
\quad\Longrightarrow\quad
\Gamma^{\mca_0} = 
\begin{blockarray}{cc}
 \aaa_2 & \aaa_3 \\
 (0,1) & (1,0) \\
\begin{block}{(cc)}
 1 & 0 \\
 0 & 1 \\
\cmidrule{1-2}
 1 & 0 \\
 0 & 1 \\
\cmidrule{1-2}
 0 & 0 \\
\end{block}
\end{blockarray}~.
\end{equation}
\endgroup
\end{remark}

We next discuss two extensions of the developed identifiability theory. 
First, Theorem \ref{thm-id} guarantees the strict learnability of ${\mca_0}$.
Under a multi-parameter SLAM, these conditions can be relaxed if  the aim is to obtain the so-called generic joint identifiability of ${\mca_0}$,
which means that ${\mca_0}$ is learnable 
with the true model parameters ranging almost everywhere in the restricted parameter space except a  set of Lebesgue measure zero.
  Specifically, we have the following definition.

\begin{definition}[generic learnability of the true model]\label{def-gid}
Denote the parameter space of $(\TT^{\mca_0},\pp^{\mca_0})$ constrained by $\Gamma^{\mca_0}$ by $\Omega$.
We say $\mca_0$ is generically identifiable, if there exists a subset $\mathcal V$ of $\Omega$ that has   Lebesgue measure zero, such that
for any $(\TT^{\mca_0},\pp^{\mca_0})\in\Omega\setminus \mathcal V$, Equation \eqref{eq-id-def}  
implies $\mca=\mca_0$. 
Moreover, if for any $(\TT^{\mca_0},\pp^{\mca_0})\in\Omega\setminus \mathcal V$, Equation \eqref{eq-id-def} implies $(\TT^{\mca},\,\pp^{\mca})=(\TT^{\mca_0},\,\pp^{\mca_0})$, we say the model parameters $(\TT^{\mca_0},\,\pp^{\mca_0})$ are generically identifiable.
\end{definition}

The generic learnability result is presented in the next theorem.
 
\begin{theorem}[conditions for generic learnability]\label{prop-genid}
Consider a multi-parameter SLAM with the set of true attribute patterns $\mca_0$ and the $J\times |\mca_0|$ constraint matrix $\Gamma^{\mca_0}$. If this true $\Gamma^{\mca_0}$ satisfies Condition $C$ and also the following conditions, then $\mca_0$ is generically identifiable.
\begin{enumerate}
\item[$A^\star$.] There exist two disjoint item sets $S_1$ and $S_2$, such that altering some entries from 0 to 1 in $\Gamma^{(S_1\cup S_2,\,\mca_0)}$  can yield a $\widetilde \Gamma^{(S_1\cup S_2,\,\mca_0)}$ satisfying Conditions $A$. That is, $\widetilde\Gamma^{(S_i,\,\mca_0)}$ has distinct columns for $i=1,2$ and $``\succeq_{S_1}" = ``\succeq_{S_2}"$ under $\widetilde\Gamma^{\mca_0}$.

\item[$B^\star$.] 
For any $\aaa$, $\aaa'\in\mca_0$ where $\aaa'\succeq_{S_i} \aaa$ under $\Gamma^{\mca_0}$ for $i=1$ or $2$, there exists some $j\in(S_1\cup S_2)^c$ such that $\tilde\Gamma_{j,\aaa}^{\mca_0}\neq \tilde\Gamma_{j,\aaa'}^{\mca_0}$.\\
\end{enumerate}
\end{theorem}

We also have the following corollary, where the identifiability requirements are directly characterized by the structure of the design $Q$-matrix.
\begin{corollary}\label{cor-gid}
If the $Q$-matrix satisfies the following conditions, then for any true set of attribute patterns $\mca_0\subseteq\{0,1\}^K$ such that $\Gamma^{\mca_0}$ satisfies Condition $C$, the set $\mca_0$ is generically identifiable. 
\begin{enumerate}
\item[($A^{\star\star}$)] The $Q$ contains two $K\times K$ sub-matrices $Q_1$, $Q_2$, such that for $i=1, 2$, 
\begin{equation}\label{eq-diag}
Q = \begin{pmatrix}
Q_1 \\
Q_2 \\
Q'
\end{pmatrix}_{J\times K};\quad
Q_i =
\begin{pmatrix}
    1 & * & \dots  & * \\
    * & 1 & \dots  & * \\
    \vdots & \vdots & \ddots & \vdots \\
    * & * & \dots  & 1
\end{pmatrix}_{K\times K},\quad i=1,2,
\end{equation}
where each `$*$'   can be either zero or one.
\item[($B^{\star\star}$)] 
With $Q$ in the form of \eqref{eq-diag}, there is $\sum_{j=2K+1}^J q_{j,k}\geq 1$ for each $k\in\{1,\ldots,K\}$.
\end{enumerate}
\end{corollary}

\begin{remark}
When the conditions in Theorem \ref{cor-gid} are satisfied, $\mca_0$ is generically identifiable and from Theorem 4.3 in \cite{partial}, the model parameters $(\TT^{\mca_0},\pp^{\mca_0})$ are also generically identifiable.
Corollary \ref{cor-gid} differs from Theorem 4.3 in \cite{partial} in that, here we allow the true set of attribute patterns $\mca_0$ to be unknown and arbitrary, and study its identifiability, while 
\cite{partial} assumes  $\mca_0$  is pre-specified and studies  the identifiability of the model parameters $(\TT^{\mca_0},\pp^{\mca_0})$. 
\end{remark}

\begin{remark}\label{rmk-gid}
Under the conditions for generic identifiability in Theorem \ref{prop-genid} or Corollary \ref{cor-gid}, we can obtain the explicit forms of the measure zero set $\mathcal V$ ($\mathcal V\subseteq\Omega$) where the non-identifiability may occur. Under either Theorem \ref{prop-genid} or Corollary \ref{cor-gid}, the set $\mathcal V$ is characterized by the solution set of certain polynomials about the parameters $(\TT,\pp)$ (see the proofs for details). The solution set of these polynomials indeed defines a lower-dimensional manifold in the parameter space. Therefore, Theorem \ref{prop-genid} and Corollary \ref{cor-gid} supplement Theorem \ref{thm-id} by relaxing the original conditions and establish identifiability  when $(\TT,\pp)$ satisfy certain shape constraints, i.e., $(\TT,\pp)$ do not fall on that manifold $\mathcal V$ in the parameter space.
\end{remark}
\color{black}

The above generic identifiability of $\mca_0$ ensures the nonidentifiability case happens only in a zero  measure set. 
The second extension of Theorem \ref{thm-id} regards a case when the nonidentifiability case lies in a positive measure set. 
This happens when certain latent attribute patterns always have the same positive response probabilities to all the items, i.e., $\TT_{\Cdot,\aaa} = \TT_{\Cdot,\aaa'}$ for some $\aaa\neq\aaa'$. 
We define $\aaa$ and $\aaa'$ to be in the same equivalence class if $\TT_{\Cdot,\aaa} = \TT_{\Cdot,\aaa'}$.
For instance, still consider the following $2\times 2$   $Q$-matrix under the two-parameter SLAM introduced in Example \ref{exp-dina},
\begin{equation}\label{exam}
Q=\begin{pmatrix}
0 & 1 \\
1 & 1
\end{pmatrix},
\end{equation}
then attribute patterns $\aaa_1=(0,0)$ and $\aaa_3=(1,0)$ are equivalent under the two-parameter SLAM, as can be seen from the $\Gamma^{\text{all}}$ in \eqref{eq-q24}.
Therefore the two latent patterns $\aaa_1$ and  $\aaa_3$ are not identifiable, no matter which values  the true model parameters take.

In this case when both strict and generic identifiability do not hold, we study the $\pp$-partial identifiability, a concept introduced in \cite{partial}.
Specifically, when some attribute patterns have the same positive response probabilities across all items, we define the set of these attribute patterns as  an  equivalence class, and aim to identify the proportion of this equivalence class, instead of the separate proportions of these equivalent patterns in the population. 
For instance, in the above example in \eqref{exam}, because $\aaa_1$ and $\aaa_3$ are equivalent, there are three equivalence classes: $\{\aaa_1=(0,0),\aaa_3=(1,0)\}$, $\{\aaa_2=(0,1)\}$, and $\{\aaa_4=(1,1)\}$. We denote these three equivalence classes by $[\aaa_1]$ (or $[\aaa_3]$, since $[\aaa_1]=[\aaa_3]$), $[\aaa_2]$ and $[\aaa_4]$, since $\aaa_1$, $\aaa_2$ and $\aaa_4$ form a complete set of representatives of the equivalence classes. 
For any $Q$, we denote the induced  set of equivalence classes by $\mca^{\text{equiv}}=\{[\aaa_{1}],\ldots,[\aaa_{C}]\}$ of latent patterns, where $\aaa_{1},\ldots,\aaa_{C}$ form a complete set of representatives of the equivalence classes. 
In this case, the pattern selection problem of interest is to learn which equivalence classes in $\mca^{\text{equiv}}$ are significant.

	For the two-parameter SLAM introduced in Example \ref{exp-dina}, two attribute patterns $\aaa_1,\aaa_2$ are in the same equivalence class if and only if $\Gamma^{\mca}_{\bcdot, \aaa_1}= \Gamma^{\mca}_{\bcdot, \aaa_2}$. This is because under the two-parameter SLAM, the $\Gamma$-matrix determined by the $Q$-matrix with $\Gamma_{j,\aaa}=I(\aaa\succeq\qq_j)$ fully captures the model structure in the sense that $\theta_{j,\aaa}=\theta_j^+ \Gamma_{j,\aaa} + \theta_j^- (1-\Gamma_{j,\aaa})$. In this case, we can obtain a complete set of representatives of the 
equivalence classes directly from the $\qq$-vectors, which are
\begin{equation}\label{eq-equiv}
\mca_Q = \{\vee_{j\in S}\,\qq_j:\,S\subseteq\{1,\ldots,J\}\},
\end{equation}
where $\vee_{j\in S}\,\qq_j = (\max_{j\in S}\,q_{j,1},\ldots,\max_{j\in S}\,q_{j,K})$.
For $S=\varnothing$, we define the vector $\vee_{j\in S}\,\qq_j$  to be $\zero_K$, the all-zero attribute pattern. 
The reasons for $\mca_Q$ being a complete set of  representatives are that, first, $\Gamma^{\mca_Q}$ has distinct columns and contains all the unique column vectors in $\Gamma^{\text{all}}$; and second, for any other pattern  not in $\mca_Q$, there is some pattern in $\mca_Q$ such that the two patterns have identical column vectors in $\Gamma^{\text{all}}$.
 It is not hard to see that $\mca_Q = \{0,1\}^K$ if and only if the $Q$-matrix contains a submatrix $I_K$.

 For multi-parameter SLAMs introduced in Example \ref{exp-gdina}, two attribute patterns $\aaa_1,\aaa_2$ are in the same equivalence class if   $\Gamma_{\bcdot, \aaa_1}= \Gamma_{\bcdot, \aaa_2} =\mathbf 1.$ This can be seen by considering
$\Gamma_{\bcdot, \aaa_1}= \Gamma_{\bcdot, \aaa_1} \neq\mathbf 1$, i.e., $\Gamma_{j,\aaa_1}=\Gamma_{j,\aaa_2}=0$ for some item $j$. Then different from the two-parameter SLAMs,  for such item $j$, the $\theta_{j,\aaa_1}$ and  $\theta_{j,\aaa_2}$ are not always the same by the modeling assumptions of multi-parameter SLAMs. Indeed, under a multi-parameter SLAM, for item $j$, patterns in the set $\mca_0\setminus\mc_j$ can have multiple levels of item parameters.

We have the following corollary of Theorem \ref{thm-id} on the identifiability,  when certain attribute patterns are not distinguishable. 
We denote the set of significant equivalence classes by $\mca^{\normalfont{\text{equiv}}}_{0}=\{[\aaa_{\ell_1}],\ldots,[\aaa_{\ell_m}]\}$, which is a subset of the saturated set $\mca^{\normalfont{\text{equiv}}}=\{[\aaa_{1}],\ldots,[\aaa_{C}]\}$.
Denote the set of representative patterns of the significant equivalence classes by $\{\aaa_{\ell_1},\ldots,\aaa_{\ell_m}\}=\mca^{\normalfont{\text{rep}}}$.

\begin{corollary}\label{cor-parid}
If the matrix $\Gamma^{\mca^{\normalfont{\text{rep}}}}$ satisfies Conditions $A$, $B$ and $C$,    $\mca^{\normalfont{\text{equiv}}}_{0}$ is  identifiable.
\end{corollary}

\begin{remark}
Under the two-parameter SLAM with $\mca^{\text{equiv}}=\{[\aaa_{1}],\ldots,[\aaa_{C}]\}$, the $\Gamma$-matrix $\Gamma^{\{\aaa_1,\ldots,\aaa_C\}}$ by definition would have distinct column vectors. Therefore any column vector of $\Gamma^{\mca^{\normalfont{\text{rep}}}}$ in Corollary \ref{cor-parid}  must be different form any column vector of $\Gamma^{\{\aaa_1,\ldots,\aaa_C\}\setminus\mca^{\normalfont{\text{rep}}}}$. In this case, Condition $C$ is automatically satisfied. And in order to identify $\mca^{\normalfont{\text{equiv}}}_{0}$, one only needs to check if $\Gamma^{\mca^{\normalfont{\text{rep}}}}$ satisfies Conditions $A$ and $B$.
\end{remark}

\section{Penalized Likelihood approach to pattern selection}\label{sec-reg}

\subsection{Shrinkage Estimation}\label{sec-shrink}
The developed identifiability conditions guarantee that the true set of patterns can be distinguished from any alternative set that has not more than $|\mca_0|$  patterns, since they would lead to different probability mass functions of the responses.
As $\mca_0 = \{\aaa\in \{0,1\}^K: p_{\aaa} >0 \}$, we know that learning the significant attribute patterns is equivalent to the selection of nonzero elements of $\pp$.
In practice, if we directly overfit the data with all the $2^K$ possible attribute patterns, the corresponding maximum likelihood estimator (MLE) can not correctly recover the sparsity structure of the proportion parameters $\pp$.
In this case, we propose to impose some regularization on the proportion parameters $\pp$, and perform pattern selection through maximizing a penalized likelihood function.

In general, we denote by $\mca_{\text{input}}$ the set of candidate attribute patterns  given to the shrinkage estimation method as input.
If the saturated space of all the possible attribute patterns are considered,  $\mca_{\text{input}}=\{0,1\}^K$ and it contains all the $2^K$ possible configurations of attributes.
When $2^K\gg N,$ we propose to use a preprocessing step that returns a proper subset $\mca_{\text{input}}$ of the saturated set $\{0,1\}^K$ as candidate attribute patterns, and then perform the shrinkage estimation (please see Section \ref{sec-screen} for the preprocessing procedure).

We first introduce the general data likelihood of a structured latent attribute model.
Given a sample of size $N$, we denote the $i$th subject's response by  $\RR_i=(R_{i,1},\ldots,R_{i,J})^\top$, $i=1,\ldots, N$. We further use $\mathcal R$ to denote the $N\times J$ data matrix $(\RR_1^\top, \ldots, \RR_N^\top)^\top.$
The marginal likelihood can be written as
\begin{align}\label{eq-lk}
L(\TT,\pp\mid\mathcal R) = \prod_{i=1}^N\Big[\sum_{\aaa\in\mca_{\text{input}}} p_{\aaa}\prod_{j=1}^J \theta_{j,\aaa}^{R_{i,j}} (1-\theta_{j,\aaa})^{1-R_{i,j}}  \Big],
\end{align}
where the constraints on $\TT$ imposed by $Q$ are made implicit. 
We denote the corresponding log likelihood by $\ell(\TT,\pp )=\log L(\TT,\pp\mid \mathcal R)$.

As the proportion parameters $\pp$ belongs to a simplex, in order to encourage sparsity of $\pp$, we propose to use a $\log$-type penalty   with a tuning parameter $\lambda<0$. 
 Specifically, we use the following penalized likelihood  as the objective function,
  
\begin{align}\label{eq-orig}
\ell^{\lambda}(\TT,\pp ) \
= &~ \ell(\TT,\pp ) + \lambda\sum_{\aaa\in\mca_{\text{input}}} \log_{\rho_N} (p_{\aaa}),\quad \lambda\in(-\infty,0),
\end{align} 
where  $ \log_{\rho_N}(p_{\aaa})   =  \log(p_{\aaa}) \cdot I(p_{\aaa}>\rho_N)+  \log(\rho_N)\cdot I(p_{\aaa}\leq \rho_N)$
 and 
$\rho_N$ is a small threshold parameter that is introduced to   circumvent the singularity issue of the $\log$ function at zero.
Specifically, we take
\begin{equation}\label{eq-rho-N}
\rho_N \asymp N^{-d},
\end{equation}
for some constant $d\geq 1$, where for two sequences  $\{a_N\}$ and $\{b_N\}$, we denote $a_N\lesssim b_N$ if $a_N = O(b_N)$ and $a_N\asymp b_N$ if  $a_N\lesssim b_N$ and $b_N\lesssim a_N$. 
Any attribute pattern $\aaa$ whose estimated $p_{\aaa}<\rho_N$ will be considered as 0, and hence not selected.
  The tuning parameter $\lambda\in(-\infty,0)$ controls the sparsity level of the estimated proportion vector $\pp$, and a smaller $\lambda$ leads to a sparser   solution (with more estimated $p_{\aaa}$ falling below $\rho_N$). Given a $\lambda\in(-\infty,0)$, we denote the estimated set of patterns by $\widehat\mca^{\lambda}=\{\aaa\in\mca_{\inp}:\,\widehat p_{\aaa}>\rho_N,\, (\widehat\TT,\widehat\pp)=\arg\max_{\TT,\pp}\ell^{\lambda}(\TT,\pp)\}$.

\begin{remark} \label{rmk5}
 	In the literature, \cite{jhchen2001} and \cite{jhchen2004} 
	used a similar form of penalty as the summation term in our \eqref{eq-orig}, but instead imposed $\lambda>0$ to avoid sparse solutions of the proportion parameters. These works used that penalty in order to avoid singularity when performing restricted likelihood ratio test. 	While our goal here is to encourage sparsity of $\pp$ so that significant attribute patterns can be selected.

 The formulation of \eqref{eq-orig} can also be interpreted in a Bayesian way, where the penalty term regarding the proportions $\pp$ is the logarithm of the Dirichlet prior density with hyperparameter $\beta=\lambda+1$ over the proportions.
	 But note that when $\beta<0$, the penalty term is not a proper prior density.
	 Our later Proposition \ref{prop-rate} reveals that, under nonstandard convergence rate of the mixture model, the traditional Bayesian way of imposing a proper Dirichlet prior over proportions is not sufficient for selecting significant attribute patterns consistently. Instead, this classical procedure will yield too many false patterns being selected. Therefore, our novelty of allowing $\lambda$ in \eqref{eq-orig} to be negative with arbitrarily large magnitude is crucial to selection consistency.
 
Other than the nice connection to the Dirichlet prior density in the Bayesian literature,   the log-type penalty in \eqref{eq-orig} also facilitates the computation based on modified EM and variational EM algorithms, as shown in our Algorithms \ref{algo-pem} and \ref{algo-fpvem}. For such reasons, this work uses the log-type penalty.
 There are also alternative ways of imposing penalty on the proportion parameters $\pp$ that would lead to selection consistency, such as the truncated $L_1$ penalty used in \cite{shen2012} for high-dimensional feature selection. 
 \end{remark}

We denote the MLE obtained from directly maximizing $L(\TT,\pp \mid \mathcal R)$ in \eqref{eq-lk} by  $\widehat\TT$ and $\widehat\pp$, 
and denote the ``oracle" MLE of the parameters obtained by maximizing the likelihood constrained to the true set of attribute patterns by $(\widehat\TT^{\mca_0},\widehat\pp^{\mca_0})$.
We denote the rate of convergence of $\ell(\widehat\TT,\widehat\pp)$ to $\ell(\widehat\TT^{\mca_0},\widehat\pp^{\mca_0})$ by $\delta\in(0,1]$, that is,  
\begin{equation}\label{eq-rate}
\big[\ell(\widehat\TT,\widehat\pp) - \ell(\widehat\TT^{\mca_0},\widehat\pp^{\mca_0})\big]/{N} = O_{P}(N^{-\delta}).
\end{equation} 
When $\delta=1$, \eqref{eq-rate} implies $\ell(\widehat\TT,\widehat\pp)$ converges with the usual root-$N$ rate, and $\delta<1$ would imply a slower convergence rate. In the literature, \cite{ho2016convergence} and \cite{heinrich} have studied the technically involved problem of convergence rate of the mixing distribution of certain   mixture models,  and showed these models may not have the standard  root-$N$ rate. As implied by these works, for complicated models like SLAMs, the convergence rate of the mixing distribution is likely to be slower than root-$N$, so as the convergence rate of $\ell(\widehat\TT,\widehat\pp)$.
For a set $\mca$, denote its cardinality by $|\mca|$. We have the following theorem.

\begin{theorem}[selection consistency]\label{thm-select}
Suppose  the true constraint matrix $\Gamma^{\mca_0}$ associated with $\mca_0$ satisfies conditions $A$ and $B$ stated in Theorem \ref{thm-id}. 
The true parameters satisfy
\begin{align}\label{eq-min}
\min_{\aaa\in \mca_0} p_{\aaa} > c_0; \quad
\theta_{j,\aaa^\star} - \max_{\aaa:\,\Gamma_{j,\aaa}=0}\theta_{j,\aaa} \geq c_1, ~\forall~j=1,\ldots,J\text{ and }\aaa^\star\in\mc_j,
\end{align}
where $c_0, c_1>0$ are some constants.
Assume $\log|\mca_{\normalfont{\text{input}}}|=o(N)$ and $|\mca_{\normalfont{\text{input}}}|\cdot \rho_N = O(N^{-\delta})$.
Then there exist a sequence of tuning parameters $\{\lambda_N\}$ satisfying $N^{1-\delta}/|\log \rho_N| \lesssim -\lambda_N \lesssim N/|\log \rho_N|$ such that
$\mathbb P(\widehat \mca^{\lambda_N} = \mca_0) \to 1$ as $N\to \infty.$ 
\end{theorem}

\begin{remark}\label{rmk-pmin}
Together with our identifiability result in Theorem \ref{thm-id}, the assumption \eqref{eq-min}  helps 
distinguish the true patterns from any alternative set of patterns with no larger cardinality,
and further helps establish selection consistency.
It is possible to further extend the current result and relax the constant lower bound assumption, though identifiability conditions would   need to be adapted carefully to the case with a growing number of significant patterns and a shrinking magnitude of the proportions; we leave this for future work.
\end{remark}

The proof of Theorem \ref{thm-select} also reveals that if the convergence rate of $U_N$ are slower than $\sqrt{N}$ with $\delta<1$ in \eqref{eq-rate}, then the tuning parameter $\lambda$ in \eqref{eq-orig} has to satisfy $\lambda<-1$ in order to have pattern selection consistency; otherwise the issue of over selecting exists.  Under the Bayesian interpretation as discussed in Remark \ref{rmk5}, this result implies that imposing the popular Dirichlet prior with a proper parameter  $\beta = \lambda+1 \in(0,1)$ is not sufficient for consistent selection of the significant mixture components (i.e., latent attribute patterns). Therefore, the approach proposed by \cite{rousseau2011} would not yield frequentist selection consistency in this considered scenario.
We state this in the following proposition. \begin{proposition}[selection inconsistency of Dirichlet prior]\label{prop-rate}
Suppose $\delta<1$ in \eqref{eq-rate}, i.e., the rate of convergence of $\ell(\widehat\TT,\widehat\pp)$ is slower than the usual $\sqrt{N}$-rate. 
Then there {does not} exist a sequence of $\{\lambda_N, ~N=1,2,...\}\subseteq[-1,0)$ such that $\mathbb P(\widehat{\mathcal A}^{\lambda_N}=\mathcal A_0)\to 1$ as $N\to\infty$.
\end{proposition}

\begin{example}\label{exp-fpvem}
	To visualize how the numbers of selected patterns differ for our proposed method based on maximizing \eqref{eq-orig} with $\beta=\lambda+1\in(-\infty,1)$, and the   variational EM algorithm resulting from imposing a proper Dirichlet prior over the proportions, we conduct a simulation study. 
	In a simulation setting of $K=10$ and $J=30$, for each sample size $N=500$ and $1000$, we carry out $200$ independent simulations and in each run record the number of selected attribute patterns from the proposed method, and that from the variational EM algorithm.
We plot the histogram corresponding to the proposed method 
 (FP-VEM, see Section 4 for details), together with that corresponding to Variational EM (VEM) with a small Dirichlet parameter $\beta=0.01$. 
 For both algorithms, we use the same threshold $\rho_N=1/(2N)$ for selecting attribute patterns in the end of the algorithm, by selecting patterns whose posterior means exceeds $\rho_N$.
Here we did not plot the results corresponding to VEM with  $\beta$ smaller than $0.01$, because we found the VEM algorithm with smaller $\beta$ values generally has convergence issues and in many cases it fails to converge but just jumps between several solutions.
One can see from Figure \ref{fig-hist} that the proposed method selects 10 patterns for most of datasets, which are indeed the 10 true patterns; while VEM over selects the attribute patterns. 
\end{example}
\begin{figure}[h!]
\begin{subfigure}{0.5\textwidth}
\includegraphics[width=\linewidth]{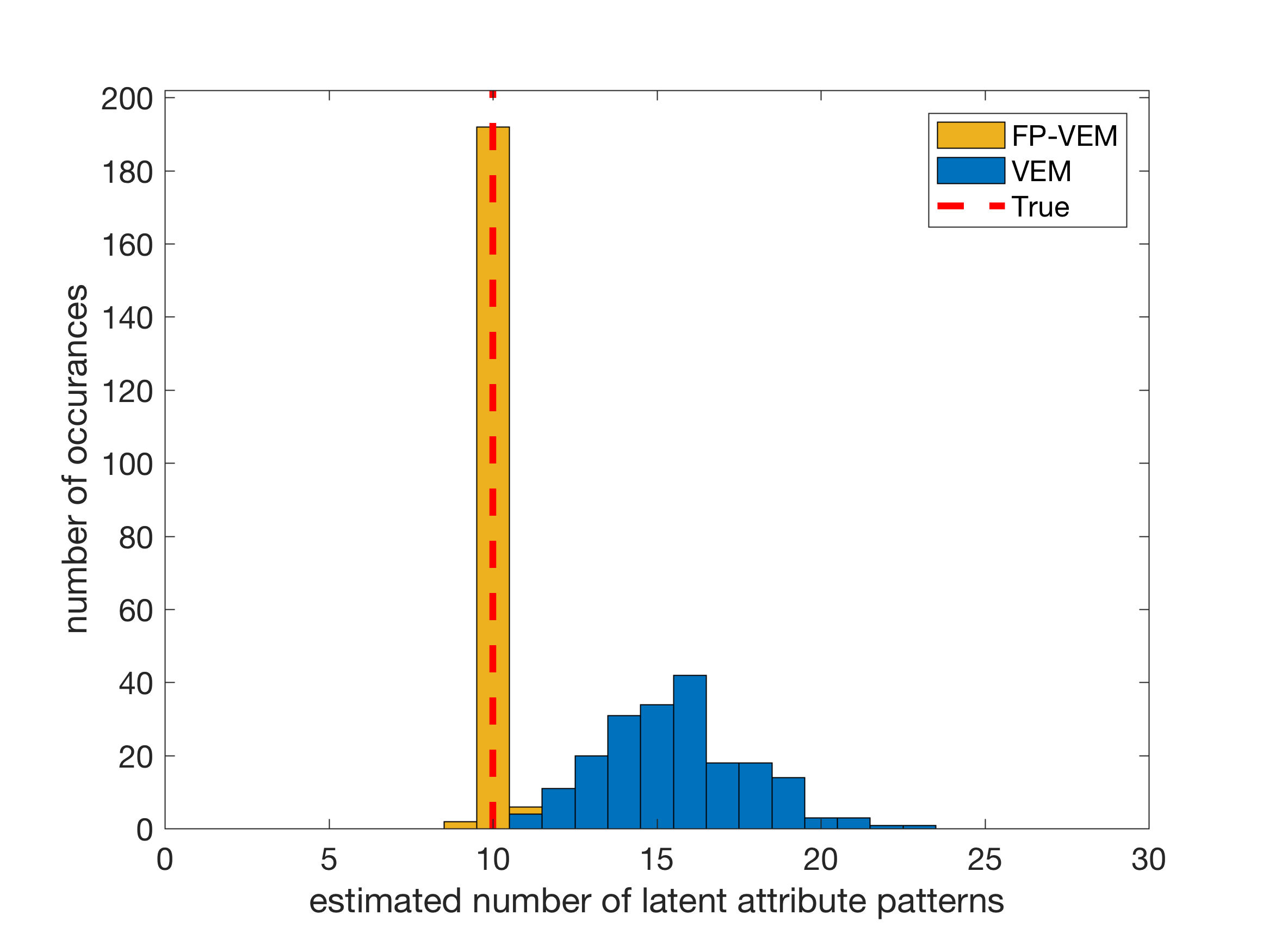}
\caption{$K=10,~N=500$}
\end{subfigure}\hspace*{\fill}
\begin{subfigure}{0.5\textwidth}
\includegraphics[width=\linewidth]{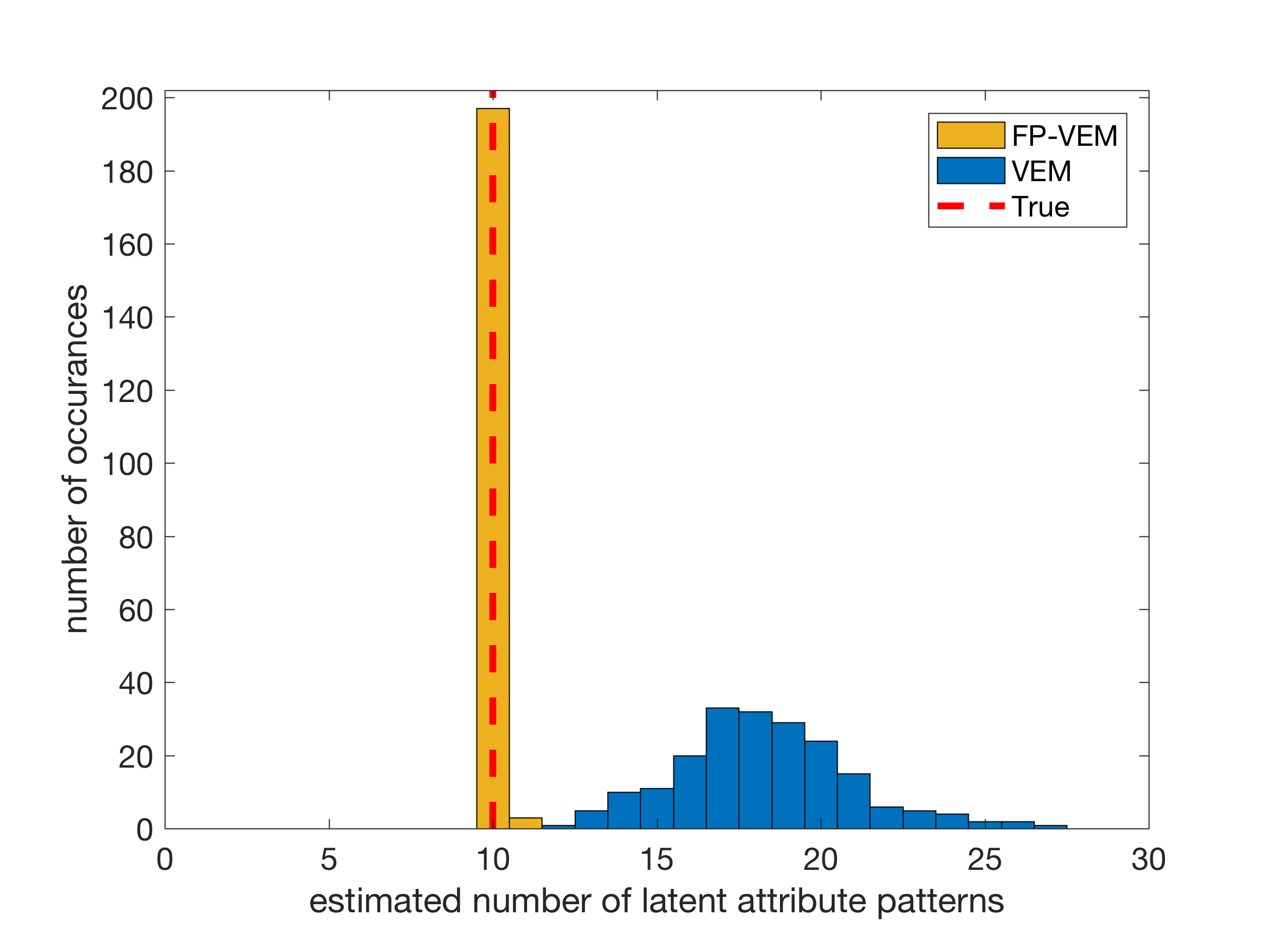}
\caption{$K=10,~N=1000$} 
\end{subfigure}
\caption[]{Histograms of estimated number of latent attribute patterns. VEM represents Variational EM with $\beta=\lambda+1=0.01$, and FP-VEM represents the proposed Algorithm 2 in Section 4.
The true number of latent attribute patterns is $|\mca_0|=10$. } 
\label{fig-hist}
\end{figure}

We next propose two algorithms to perform pattern selection, one being a modification of an EM algorithm, and the other being a variational EM algorithm resulting from an alternative   formulation of the   problem.

\subsubsection{Modified EM algorithm.}
We first consider using an EM algorithm with a slight modification in the E step to maximize \eqref{eq-orig}. 
For each subject $i=1,\ldots,N$, denote his/her latent attribute pattern by $\ma_i = (A_{i,1},\ldots,A_{i,K})$, then $\ma_i \in\{0,1\}^K$.
The complete log likelihood corresponding to \eqref{eq-orig} is
\begin{align}\label{eq-complete}
& \ell^{\lambda}_{\text{comp}}(\TT,\pp\mid\mathcal R, \ma)
=  \sum_{\aaa_l\in\mca_{\text{input}}}\Big(\sum_{i} I(\ma_i=\aaa_l) +\lambda\Big)\log _{\rho_N}(p_{\aaa_l})
\\ \notag 
&\qquad + \sum_{\aaa_l\in\mca_{\text{input}}}\sum_{i} I(\ma_i=\aaa_l)\sum_{j} \Big[R_{i,j}\log(\theta_{j,\aaa_l})
 + (1-R_{i,j})\log(1-\theta_{j,\aaa_l})\Big].
\end{align}
Following the standard formulation of the EM algorithm \citep{EM}, in the E step of the $(t+1)$-th iteration, conditional expectations of $\ell^{\lambda}_{\text{comp}}(\TT,\pp\mid\mathcal R, \ma)$ is evaluated with respect to  the posterior distribution of latent variables $\ma_i$'s given the current iterates of  parameters $\TT^{(t)}$ and $\pp^{(t)}$. Specifically, in the E step we replace the indicator $I(\ma_i=\aaa_l)$ in \eqref{eq-complete} by the probability $\varphi_{i,l} = \mathbb P(\ma_i=\aaa_l\mid\TT^{(t)},\pp^{(t)})$; and this is equivalent to updating 
$$Q(\TT,\pp\mid\TT^{(t)},\pp^{(t)}) 
:=\mathbb E\Big[\ell^{\lambda}_{\text{comp}}(\TT,\pp\mid\mathcal R, \ma) ~\Big\vert~ \mathcal \TT^{(t)},\pp^{(t)}\Big].$$
 In the M step, 
 we update
  $(\TT^{(t+1)},\pp^{(t+1)}) = \arg\max Q(\TT,\pp\mid \TT^{(t)},\pp^{(t)})$.
 Note that directly using a negative $\lambda$ in the EM algorithm may  yield an invalid E step, due to   potentially negative updates for some proportion parameters (e.g., $p_{\aaa}$'s).
When this happens, we do a thresholding in the E step as an approximation by replacing the probably negative class potential ($\Delta_l$ in Algorithm \ref{algo-pem}) with a pre-specified small constant $c>0$. {In practice, Algorithm \ref{algo-pem}'s performance appears not  sensitive to   small values of $c$, and we take   $c=0.01$ in our numerical experiments; see   Appendix A.2 for a sensitivity study of  the parameter $c$.}

 \begin{algorithm}[h!]
\caption{PEM: Penalized EM for $\log$-penalty with $\lambda\in(-\infty,0)$}
\label{algo-pem}
\SetKwInOut{Input}{Input}
\SetKwInOut{Output}{Output}

\KwData{$Q$, responses $\mathcal R$, and candidate attribute patterns $\mca_{\text{input}}$.
}

Initialize $\bo\Delta = (\Delta^{(0)}_1,\ldots,\Delta^{(0)}_{|\mca_{\text{input}}|})$. 

 \While{not converged}{
 In the $(t+1)$th iteration,
 
    \For{$(i,l)\in[N]\times[|\mca_{\normalfont{\text{input}}}|]$}{
\begin{align*}
\varphi^{(t+1)}_{i,\aaa_l}~=~ 
 \frac{\Delta^{(t)}_l\cdot \exp \Big\{ \sum_{j} \Big[ R_{i,j}\log(\theta^{(t)}_{j,\aaa_l}) + (1-R_{i,j})\log(1-\theta^{(t)}_{j,\aaa_l}) \Big] \Big\} }{\sum_m \Delta^{(t)}_m\cdot 
 \exp \Big\{ \sum_{j} \Big[ R_{i,j}\log(\theta^{(t)}_{j,\aaa_m}) + (1-R_{i,j})\log(1-\theta^{(t)}_{j,\aaa_m}) \Big] \Big\}};
\end{align*}
    }
    \For{$l\in[|\mca_{\normalfont{\text{input}}}|]$}{
     $\Delta^{(t+1)}_l = \max\{c,~\lambda + \sum_{i=1}^N \varphi^{(t+1)}_{i,\aaa_l}\};$ ~ ($c>0$ is pre-specified);\\
    }
        $\pp^{(t+1)}\leftarrow \bo\Delta^{(t+1)}/(\sum_{l}\Delta^{(t+1)}_l);$
        
    \For{$j\in[J]$}{
    $
    \TT^{(t+1)}
    = {\arg\max}_{\TT} ~ 
    \Big\{\sum_{\aaa_l}\sum_{i} \varphi^{(t+1)}_{i,\aaa_l}\sum_{j} \Big[R_{i,j}\log (\theta_{j,\aaa_l})
 + (1-R_{i,j})\log (1-\theta_{j,\aaa_l})\Big]\Big\};
    $
    } 
  }
  After the total $T$ iterations,

 \Output{$\{\aaa_l\in\mca_{\text{input}}:~ p^{(T)}_{\aaa_l}>\rho_N \}$.}

\end{algorithm}

\begin{remark}\label{rmk-algo-dina}
Under the two-parameter SLAM, the DINA model, and the identity-link multi-parameter SLAM, the GDINA model in Examples \ref{exp-dina} and \ref{exp-gdina}, the M-step of updating the item parameters $\{\theta_{j,\aaa}\}$'s in Algorithm \ref{algo-pem} has closed forms. Specifically, under DINA, for any item $j$ the update for the unique parameters $(\theta_j^+,\theta_j^-)$  takes the form
    $$(\theta_j^+)^{(t+1)} =  \frac{\sum_{i} \sum_{\aaa} R_{i,j} \Gamma_{j,\aaa} \varphi^{(t+1)}_{i,\aaa}}{\sum_{i} \sum_{\aaa} \Gamma_{j,\aaa} \varphi^{(t+1)}_{i,\aaa}}, \quad
(\theta_j^-)^{(t+1)} = \frac{\sum_{i} \sum_{\aaa} R_{i,j} (1-\Gamma_{j,\aaa}) \varphi^{(t+1)}_{i,\aaa}}{\sum_{i} \sum_{\aaa} (1-\Gamma_{j,\aaa}) \varphi^{(t+1)}_{i,\aaa}}.$$
Under GDINA, for item $j$,
the update for the unique parameters $\theta_{j,\{k_1,\ldots,k_l\}}$ with $\{k_1,\ldots,k_l\}\subseteq\mathcal K_{j}$  takes the following form,
$$
\theta^{(t+1)}_{j,\,\{k_1,\ldots,k_l\}} = \frac{{\sum_{i}\sum_{\aaa}I(\{k\in\mathcal K_{j}:\, \alpha_k=1\} =\{k_1,\ldots,k_l\})}  R_{i,j}\varphi^{(t+1)}_{i,\aaa}}{{\sum_{i}\sum_{\aaa}I(\{k\in\mathcal K_{j}:\, \alpha_k=1\} =\{k_1,\ldots,k_l)\}}\varphi^{(t+1)}_{i,\aaa}}.
$$
In addition, when certain latent patterns are not distinguishable as discussed earlier in Corollary \ref{cor-parid}, we can  easily  modify Algorithm  1  from selecting attribute patterns to selecting equivalence classes of attribute patterns. For instance, under a two-parameter SLAM, given the row vectors $\{\qq_j,j\in[J]\}$ of $Q$, we first obtain the representatives of the $Q$-induced equivalence classes:
$
\mca_Q = \{\vee_{j\in S}\,\qq_j:\,S\subseteq\{1,\ldots,J\}\},
$
then get the ideal response matrix of $\mca_Q$, namely
$ \Gamma(\bcdot,\mca_Q) = (\gamma_{j,l})_{J\times |\mca_Q|}$ where $\gamma_{j,l} = I(\aaa_l\succeq \qq_j)$
for $\aaa_l\in\mca_Q$ and $j\in[J]$. After initializing $\boldsymbol\Delta=(\Delta_1,\ldots,\Delta_{|\mca_Q|})$, we just follow the same iterative procedure as that of Algorithm 1  for the two-parameter SLAM. In the end of the algorithm, after calculating $\nu_{[\aaa_l]} = \Delta_l/(\sum_m \Delta_m)$, we select those $[\aaa_l]$ with proportion $\nu_{[\aaa_l]}$ above a pre-specified threshold.
From the selected equivalence classes of attribute profiles,  we can go back to obtain their representatives which are combinations of the $\qq$-vectors from $\mca_Q$. 

\end{remark}

In practice when applying the PEM algorithm, we recommend using a sequential procedure with a range of $\lambda$ values $\lambda_1>\lambda_2>\cdots>\lambda_B$, where $\lambda_1>-1$ is close to 0 and $\lambda_B$ should be less than $-1$. Specifically, we start with the relatively large $\lambda_1$ and use the estimated parameters from PEM with $\lambda_1$ as initial values for the next round of PEM with $\lambda_2$. We do this sequentially with estimates from PEM with $\lambda_b$ serve as initializations for PEM with $\lambda_{b+1}$. When this sequential procedure ends, we choose the final model from   the total number of $B$ estimated ones using  certain information criterion.

Given the large  model space, we propose to use the Extended  Bayesian Information Criterion (EBIC) introduced in \cite{EBIC} to select the tuning parameter.
 Recall that we denote by $\mca^{\lambda}$ the selected set of attribute patterns obtained by maximizing the penalized likelihood function \eqref{eq-orig} with the specific tuning parameter $\lambda$. And we denote the item parameters and proportion parameters defined on this $\mca^{\lambda}$ by $\TT^{\mca^{\lambda}}$ and $\pp^{\mca^{\lambda}}$, respectively.
The EBIC family have the following information criterion
\begin{equation}\label{eq-ebic}
\text{BIC}_{\gamma}(\mca^{\lambda}) = -2\ell(\TT^{\mca^{\lambda}}, \pp^{\mca^{\lambda}}) + \lvert \mca^{\lambda}\rvert\log N + 2\gamma \log {|\mca_{\text{input}}|\choose \lvert \mca^{\lambda}\rvert},
\end{equation}
where the EBIC parameter $\gamma\in[0,1]$. A smaller EBIC value implies a more favorable model.
 Selection consistency of the EBIC for high-dimensional model is established in Theorem 1 of \cite{EBIC} for $\gamma$ greater than a certain threshold. 
 When $\gamma=0$, EBIC becomes the the classical BIC. Generally, larger $\gamma$ yields a more parsimonious model. Here we choose $\gamma = 1$, for which the conditions in Theorem 1  for selection consistency in \cite{EBIC} is satisfied. 

\begin{example}\label{exp-path}
	Figure \ref{fig-path} presents an illustration of the solution paths of the estimated proportions versus $\lambda$ based on a simulated dataset with  $N=150$, $K=10$, and $J=30$. The $Q$-matrix  $Q=(Q_1^\top,Q_2^\top,Q_3^\top)^\top$ with $Q_i$ in the following form,
{\small
\begin{equation}\label{eq-3q}
Q_1 = \begin{pmatrix}
1 &  & & 0\\
 & \ddots &  & \\
&  & \ddots & \\
0 & &  & 1\\
\end{pmatrix},\quad
Q_2 = \begin{pmatrix}
1 & 1 & & 0\\
 & \ddots & \ddots & \\
&  & \ddots & 1 \\
0 & &  & 1\\
\end{pmatrix},\quad
Q_3 = \begin{pmatrix}
1 & 1 & & 0\\
1 & \ddots & \ddots & \\
& \ddots & \ddots & 1 \\
0 & & 1 & 1
\end{pmatrix}.
\end{equation}}
When generating the data, 10 attribute patterns are randomly selected from the $2^{10}=1024$ possible ones as true patterns, and the proportion of each of them is set to be $0.1$. The item parameters are set as $1-\theta^+_j=\theta^-_j=0.2$ for each $j$ under a two-parameter SLAM.
{In the current setting with $K=10$, we take the set of patterns as input to the PEM algorithm to be $\mca_{\inp}=\{0,1\}^{K}$.}
Figure \ref{fig-path}(a) plots the solution paths of the estimated proportions of all the $2^{10}=1024$ attribute patterns as $\lambda$ varies in  $\{-0.2,-0.4,\cdots,-4.8, -5.0\}$. The 10 true attribute patterns are plotted with  colored lines with circles while the remaining $2^{10}-10$ attribute patterns are plotted with black solid lines. Figure \ref{fig-path}(b) plots the estimated support size of $\pp$ versus $\lambda$, and the EBIC value versus $\lambda$. For definition of EBIC see the end of this subsection. We observe that when $\lambda\in[-4.4,-1.4]$, Algorithm \ref{algo-pem} selects the correct model with 10 true attribute patterns. This interval of $\lambda$ corresponds to a ``stable window" of the estimation algorithm that gives the correct selection and also has the smallest EBIC value.
For this specific dataset, the proposed method along with EBIC succeeds in selecting the true model. Please see Section \ref{sec-simu} for more simulation results which show that the proposed methods combined with EBIC indeed have good performance in general.
\end{example}

\begin{figure}[h!]
\begin{subfigure}{0.5\textwidth}
\includegraphics[width=\linewidth]{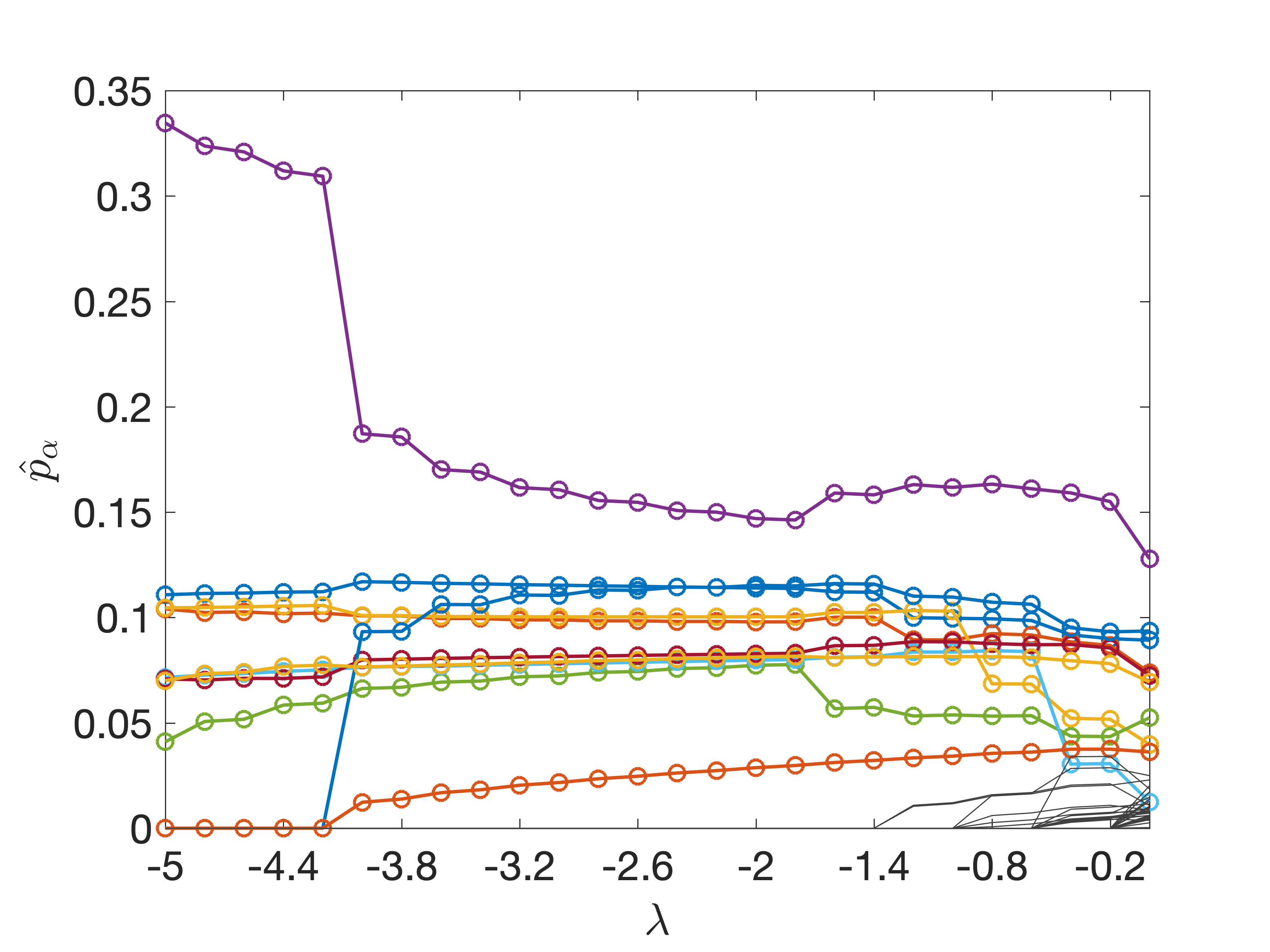}
\caption{solution paths versus $\lambda$}
\end{subfigure}\hspace*{\fill}
\begin{subfigure}{0.5\textwidth}
\includegraphics[width=\linewidth]{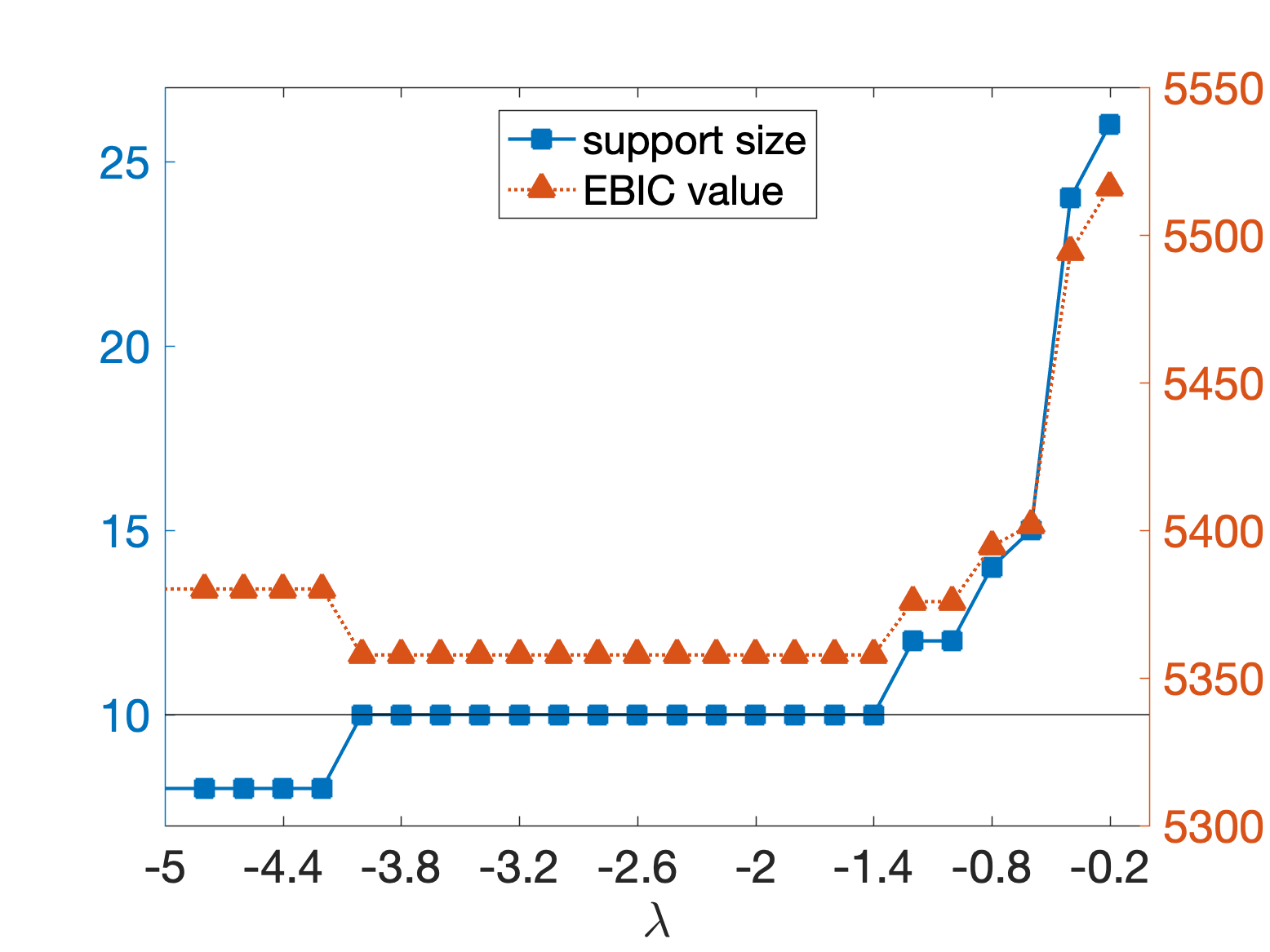}
\caption{EBIC values and support sizes versus $\lambda$}
\end{subfigure}
\caption[]{PEM solution paths and EBIC values in one trial, $N=150$.} 
\label{fig-path}
\end{figure}

\subsubsection{Variational EM algorithm from an alternative formulation.}
In the following, we discuss an alternative formulation of the   objective function \eqref{eq-orig} and propose a variational EM algorithm for estimation, by treating   the proportion parameters $\pp$ as latent variables.  
As discussed in Remark \ref{rmk5}, for the objective function \eqref{eq-orig} with $\lambda\in(-\infty,-1]$, the penalty term $\prod_{l=1}^{2^K} p_{\aaa_l}^{\lambda}$ does not correspond to  a proper Dirichlet distribution density.
However, for  any arbitrarily small $\lambda$ value,   the objective function \eqref{eq-orig}  can   be replaced by the following alternative formulation:
 \begin{align}\label{eq-pseudo}
 \ell_{\text{pseudo}}^{\lambda, \Upsilon}(\TT,\pp) 
=&~ \Upsilon\cdot  \ell(\TT,\pp) +  (\beta-1)\sum_{\aaa\in\mca_{\text{input}}} \log_{\rho_N}
(p_{\aaa})  \quad\mbox{ for }\beta\in(0,1),~\Upsilon\in(0,1].
\end{align}
where we introduce a new parameter $\Upsilon\in(0,1]$ and replace $\lambda$ with $\beta-1$ to respect the convectional notation of a Dirichlet  distribution with hyperparameter $\beta\in(0,1)$ to encourage sparsity.
With $\beta\in(0,1)$ and $\Upsilon\in(0,1]$, the ratio $(1-\beta)/\Upsilon$ can be arbitrarily large when $\Upsilon$ is arbitrarily close to zero,   therefore making \eqref{eq-pseudo} equivalent to \eqref{eq-orig}.

In the new objective function \eqref{eq-pseudo}, the penalty term $\prod_{l=1}^{2^K} p_{\aaa_l}^{\beta-1}$,  $\beta\in(0,1),$ can be viewed as a well-defined     Dirichlet density function for the latent variables $\pp$. 
One intuition behind \eqref{eq-pseudo} is that given moderate sample size and large number of potential latent attribute patterns, one needs to downweight the influence of data likelihood and magnify the  prior information encoded by the Dirichlet density, in order to have the sufficient extent of shrinkage.
 The fractional powered likelihood multiplied by the Dirichlet density can then be treated as a loss function to minimize.
 The idea of assigning a fractional power to the likelihood was also used in the Bayesian literature, such as \cite{bissiri2016} and \cite{holmes2017power} for Bayesian learning under model misspecification, and \cite{yang2017alpha} and \cite{cherief2018} for variational Bayesian inference. 
Different from these works, here we use the alternative formulation \eqref{eq-pseudo} of the original objective function \eqref{eq-orig} in order to consistently select the significant    latent attribute patterns. 

The formulation \eqref{eq-pseudo}   allows a variational EM algorithm for obtaining the item parameters $\TT$ and the posterior means of the latent variables $\pp$. 
Here we treat $\TT$ still as model parameters, then we
follow the general derivation of variational algorithms in  \cite{blei2017vari} to derive Algorithm 2. We denote the digamma function by $\Psi(x) = \frac{d}{dx} \log \Gamma(x)$ for $x\in(0,\infty)$. 
In particular, the complete likelihood is
\begin{align}\label{eq-complete-vem}
& \ell^{\lambda,\Upsilon}_{\text{comp}}(\TT\mid\mathcal R, \ma,\pp) = 
 \sum_{\aaa\in\mca_{\text{input}}}
\Big\{\Upsilon\cdot\Big[\sum_{i} I(\ma_i=\aaa) \Big] +\beta-1\Big\}\log _{\rho_N}(p_{\aaa})
\\ \notag 
& \qquad  + \Upsilon\cdot \Big\{\sum_{\aaa\in\mca_{\text{input}}}\sum_{i} I(\ma_i=\aaa)\sum_{j} \Big[R_{i,j}\log (\theta_{j,\aaa})
 + (1-R_{i,j})\log (1-\theta_{j,\aaa})\Big]\Big\}.
\end{align}
In the variational E step, we first obtain the conditional probability of $I(\ma_i=\aaa_l)$ for each individual $i$ and each input attribute pattern $\aaa_l$, which we denote by  $\varphi_{i,\aaa_l}$. In updating this $\varphi_{i,\aaa_l}$, the variational posterior distribution of the $p_{\aaa}$'s are used, which is still a Dirichlet distribution with mean parameters $(\Delta_1,\ldots,\Delta_{|\mca_{\text{input}}|})$ updated in the previous E step (or from initializations if in the first iteration).
Then   we update the mean parameters for the variational posterior distribution of $p_{\aaa_l}$'s based on the obtained $\varphi_{i,\aaa_l}$.
After finishing this E step, in the M step we maximize the complete likelihood with respect to $\TT$, by substituting the $I(\ma_i=\aaa_l)$'s with  $\varphi_{i,\aaa_l}$'s. Note that when taking the derivatives of \eqref{eq-complete-vem} with respect to $\theta_{j,\aaa_l}$'s does not involve either terms of $p_{\aaa_l}$ or terms of $\Upsilon$ and $\beta$, so only $\varphi_{i,\aaa_l}$ are used in the M step for updating $\TT$. Indeed, the M step of updating $\TT$ in the current Algorithm \ref{algo-fpvem} takes the same form as that of Algorithm \ref{algo-pem}.

\begin{algorithm}[h!]
\caption{FP-VEM: Fractional Power Variational EM for $\Upsilon\in(0,1]$}
\label{algo-fpvem}
\SetKwInOut{Input}{input}
\SetKwInOut{Output}{output}

\KwData{$Q$, $\mathcal R$, and candidate attribute patterns $\mca_{\text{input}}$.}

Initialize $\boldsymbol\Delta =  (\Delta^{(0)}_1,\ldots,\Delta^{(0)}_{|\mca_{\text{input}}|})
=(\beta,\ldots,\beta)$. 

 \While{not converged}{
 In the $(t+1)$th iteration,
 
    \For{$(i,l)\in[N]\times[|\mca_{\normalfont{\text{input}}}|]$}{
$$\varphi^{(t+1)}_{i,\aaa_l}~=~ 
 \frac{\exp \Big\{ \Psi(\Delta^{(t)}_l)+\Upsilon\cdot\sum_{j} \Big[ R_{i,j}\log (\theta^{(t)}_{j,\aaa_l}) + (1-R_{i,j})\log (1-\theta^{(t)}_{j,\aaa_l}) \Big] \Big\} }{\sum_m 
 \exp \Big\{ \Psi(\Delta^{(t)}_m)+\Upsilon\cdot\sum_{j} \Big[ R_{i,j}\log (\theta^{(t)}_{j,\aaa_m}) + (1-R_{i,j})\log (1-\theta^{(t)}_{j,\aaa_m}) \Big] \Big\}};$$
    }
    \For{$l\in[|\mca_{\normalfont{\text{input}}}|]$}{
     $\Delta^{(t+1)}_l \leftarrow \beta + {\Upsilon}\times\sum_{i=1}^N \varphi^{(t+1)}_{i,l};$\\
    }
    
    \For{$j\in[J]$}{
    $
    \TT^{(t+1)}
    = {\arg\max}_{\TT} ~ 
    \Big\{\sum_{\aaa_l}\sum_{i} \varphi^{(t+1)}_{i,\aaa_l}\sum_{j} \Big[R_{i,j}\log (\theta_{j,\aaa_l})
 + (1-R_{i,j})\log (1-\theta_{j,\aaa_l})\Big]\Big\}
    $ 
    } 
  }

  After the total $T$ iterations,
 
 \For{$\aaa_l\in\mca_{\normalfont{\text{input}}}$}{
  $p_{\aaa_l} \leftarrow \Delta^{(T)}_l/(\sum_{m}\Delta^{(T)}_m)$.\\
} 
 \Output{$\{\aaa_l\in\mca_{\text{input}}:~ p_{\aaa_l}>\rho_N \}$.}

\end{algorithm}

Similar to Algorithm \ref{algo-pem}, in the practical use of Algorithm \ref{algo-fpvem} for pattern selection, we recommend using a sequential fitting procedure. For a small fixed $\lambda>0$, we choose a sequence of $\Upsilon$ values $1>\Upsilon_1>\Upsilon_2>\cdots>\Upsilon_B>0$ where $\Upsilon_1$ should be close to 1 and $\Upsilon_B$ should be relatively small. In our simulation studies, we found a $\Upsilon_B=0.3$ is sufficient in most of cases. Then we sequentially run Algorithm \ref{algo-fpvem} for $B$ times with fractional powers $\Upsilon_1,\ldots,\Upsilon_B$ respectively and use estimated parameters from FP-VEM with $\Upsilon_b$ as initial values for FP-VEM with $\Upsilon_{b+1}$. In the end, we also use EBIC to select the best $\Upsilon$.
Since $\beta$ and $\Upsilon$ can be viewed as  acting together through the term $(1-\beta)/\Upsilon$, in terms of parameter tuning, in practice we recommend fixing $\beta$ to a relatively small value, say $\beta=0.01$, and let the fractional power $\Upsilon\in(0,1]$ vary to control the sparsity level of the proportion parameters.

\subsection{Screening as a preprocessing step when $2^K\gg N$}\label{sec-screen}
In many applications of SLAMs,  the number of   attribute patterns $2^K$ could be much larger than $N$. 
This is especially the case in the application of SLAMs in epidemiological and medical diagnosis \citep{wu2017, wu2018}.
In such scenarios, given a sample with size of several thousands or hundreds, it is desirable to develop an efficient screening procedure to bring down the  number of candidate attribute patterns, and then perform the estimation.

We next describe our screening approach. Recall that for each subject $i=1,\ldots,N$, we denote his/her latent attribute pattern by $\ma_i = (A_{i,1},\ldots,A_{i,K}) \in\{0,1\}^K$.
In the screening stage we jointly estimate the item parameters $\TT$ and the $\{\ma_i,i\in[N]\}$
 to get a rough estimation of each subject $i$'s attribute pattern, and gather all the $N$ estimated attribute profiles as candidate patterns. The estimation of $\pp$ is postponed to the estimation stage.
Under the basic two-parameter SLAM, the complete log likelihood involving the latent variables takes the form
\begin{align}\label{eq-screen}
\ell_{\text{complete}}(\TT,\ma)
=& \sum_{i=1}^N \sum_{j=1}^J\Big[ R_{i,j}\Big(  \prod_{k}A_{i,k}^{q_{j,k}} \log \theta^+_j + (1-\prod_{k}A_{i,k}^{q_{j,k}}) \log \theta^-_j \Big) \Big]  \\ \notag
&  + (1-R_{i,j}) \Big(  \prod_{k}A_{i,k}^{q_{j,k}} \log(1- \theta^+_j) + (1-\prod_{k}A_{i,k}^{q_{j,k}}) \log(1- \theta^-_j) \Big)\Big].
\end{align}
We next derive an algorithm with a stochastic EM flavor to estimate the posterior mean  of each latent variable $A_{i,k}$, denoted by a matrix $(\widehat a_{i,k})$ of size $N\times K$ where  $\widehat a_{i,k}=\mathbb E[A_{i,k}\mid\Cdot]$. 
In the end of the algorithm, we obtain the binary matrix $W$ containing the candidate attribute patterns by defining $W = (w_{i,k})_{N\times K}$ with $w_{i,k} = I(\widehat a_{i,k}> {1}/{2}).$
In such a screening procedure, we first utilize the dependency among the $K$ attributes in iterative updates, then partly ignore the dependency in the last step through applying Bayes' rule to each single attribute. This results in fast and valid screening   of attribute patterns.
Viewing the $i$th row vector of $W$ as the estimated attribute pattern of subject $i$,  the unique row vectors in $W$ give the selected attribute patterns in the screening stage. We denote this set of candidate patterns by  $\widehat\mca_{\text{screen}}$.
As long as the screening has the nice property of ``no false exclusion", meaning the rows in $W$ contain all the true attribute patterns, then the screening stage is considered successful. The selected candidate patterns are passed along to the shrinkage estimation stage as input patterns.

We say the screening procedure has the \textit{sure screening property} if as $N$ goes to infinity, the probability of all the true attribute patterns included in $\widehat\mca_{\text{screen}}$ goes to one.
The next theorem establishes the sure screening property of the proposed screening procedure.
 \begin{theorem}[sure screening property]
\label{prop-ra}
Suppose the identifiability conditions in Theorem \ref{thm-id} and the constraints \eqref{eq-min} are satisfied. 
The screening procedure applied to a SLAM that covers the two-parameter SLAM as a submodel has the sure screening property. {Specifically, there exists a constant $\beta_{\min}>0$ such that $\mathbb P(\widehat \mca_{\text{screen}}\supseteq \mca_0)\geq 1- |\mca_0|\exp(-N \beta_{\min})\to 1$ as $N\to\infty$.}
\end{theorem}

{Theorem \ref{prop-ra} shows that the probability of the screening procedure failing to include all true patterns has an exponential decay with the sample size $N$.}
We point out that despite having the nice property of sure screening, the screening procedure does not guarantee consistency in selecting exactly the set $\mca_0$ of true patterns, if the number of observed variables per subject $J$ is not large enough.   
Generally speaking, as $N$ goes large but $J$ does not, the set $\widehat\mca_{\text{screen}}$ will include many false attribute patterns, although it will contain the true set $\mca_0$ with probability tending to one.
Therefore the shrinkage estimation approach in Section \ref{sec-shrink} is still essential to performing pattern selection.

In Algorithm 3, we present the proposed screening algorithm with stochastic approximations based on a number of $M_{\eff}$ Gibbs samples of $\mathbf A$ in the E step.
Alternatively, we can also use  an   even faster screening procedure by just updating the conditional probability of each subject possessing each attribute (i.e., each $A_{i,k}$) in each E step, conditioning on everything else; we term this alternative procedure  the variational screening procedure.
As stated before, the screening algorithm is derived based on the log-likelihood of the two-parameter SLAM, but can be applied to a multi-parameter SLAM that covers the two-parameter counterpart as a submodel.
After the screening stage, the set of attribute patterns as input to the shrinkage Algorithms 1 or 2 is taken as $\mca_{\text{input}} = \widehat\mca_{\text{screen}}$. 
Screening drastically lowers down the computational cost of the subsequent shrinkage estimation, and the number of candidate patterns fed to the shrinkage stage is kept at the order of $N$.

\begin{algorithm}[h!]
\caption{Stochastic Approximation Gibbs Screening}\label{algo-screen}

\SetKwInOut{Input}{input}
\SetKwInOut{Output}{Output}

\KwData{$Q$, $\RR$}
\KwResult{Candidate attribute patterns $\widehat\mca_{\text{screen}}$.}
 Initialize attribute patterns $\ma=(A_{i,k})_{N\times K}\in\{0,1\}^{N\times K}$, and $\ttt^+$ and $\ttt^-$. 
 
 Set $t=1$, $\ma^{\text{ave}}=\mz$,\quad $\II^{\text{ave}}=\mz$.\\
 \While{not converged}{
     $\ma^{\text{s}}\leftarrow\mz$,
    \quad $\II^{\text{s}}\leftarrow\mz$,
    \quad $M_{\text{eff}}\leftarrow0$.\\
  \For{$r\in [M_{\max}]$}{
  
    \For{$(i,k)\in[N]\times[K]$}{
    Draw $A_{i,k}\sim\text{Bernoulli}\Big(\text{logit}^{-1}\Big(\sum_{j} q_{j,k} \prod_{m\neq k} A_{i,m}^{q_{j,m}} \Big[R_{i,j}\log\frac{\theta^+_j}{\theta^-_j} + (1-R_{i,j})\log\frac{1-\theta^+_j}{1-\theta^-_j}\Big] \Big)\Big).$
    }

    \If{$r\geq M_{\max}-M_{\eff}$}{
     $\ma^{\text{s}} \leftarrow \ma^{\text{s}}+\ma$,\qquad
     $\II^{\text{s}} \leftarrow \II^{\text{s}}+\Big(\prod_{k} A_{i,k}^{q_{j,k}}\Big)_{N\times J}$.\\
    }
  }
  $\ma^{\text{ave}} \leftarrow \frac{1}{t}\ma^{\text{s}}/M_{\text{eff}}
  + \Big(1-\frac{1}{t}\Big)\ma^{\text{ave}},\quad
    \II^{\text{ave}} \leftarrow \frac{1}{t}\II^{\text{s}}/M_{\text{eff}}+ \Big(1-\frac{1}{t}\Big)\II^{\text{ave}},\quad t=t+1.$ \\
   \For{$j\in [J]$}{
     $\theta_j^+ \leftarrow (\sum_{i}R_{i,j} I^{\text{ave}}_{i,j})/(\sum_{i}I_{i,j}^{\text{ave}}),
     \qquad
     \theta_j^- \leftarrow (\sum_{i}R_{i,j} (1-I^{\text{ave}}_{i,j}))(\sum_{i}(1-I_{i,j}^{\text{ave}})).$
   }
 }
   \For{$(i,k)\in[N]\times[K]$}{
     $w_{i,k}\leftarrow I(A^{\text{ave}}_{i,k}>\frac{1}{2})$.
   }
\Output{include all the unique row vectors of $W$ in the set $\widehat\mca_{\text{screen}}$.}
\end{algorithm}

\begin{remark}\label{rmk-enhance}
The screening algorithm can   be modified  to be more conservative in order to reduce the risk of excluding true patterns. In particular, after each stochastic E step in the screening algorithm, based on the current iterate of $\mathbf A^{\ave}$ we can obtain a $N\times K$ binary matrix with the $(i,k)$th entry being $I(A^{\ave}_{i,k})>1/2$. 
The unique row vectors of this binary matrix can be viewed as the current candidate latent patterns.
To make the screening procedure more conservative, we recommend saving this set of candidate patterns after every $M$ stochastic EM iterations ($M$ is a positive integer), and take the union of these saved sets in the end of the algorithm to form $\widehat\mca_{\text{screen}}$ as the output. We call this strategy  ``\textit{screening enhanced by Gibbs exploration}'', since it takes advantage of the latent patterns that the Gibbs sampling explores along the stochastic EM iterations.
\end{remark}
\color{black}

\section{Simulation Studies}\label{sec-simu}
We next present simulation results with the two-parameter SLAM  and the multi-parameter all-effect SLAM, respectively.

\paragraph{Two-parameter SLAM.}
 Consider the two-parameter SLAM with  a  $3K\times K$ $Q$-matrix $Q=(Q_1^\top,Q_2^\top,Q_3^\top)^\top$, where the three  submatrices $Q_1$, $Q_2$ and $Q_3$ are specified in \eqref{eq-3q}. 
 We consider three dimensions of possible attribute patterns with $2^K=2^{10}, 2^{15}, $ and $2^{20}$,  three sample sizes with $N=150, 500$ and $1000$, and two different signal levels with true item parameters:  $\{\theta_j^+=0.8, \,\theta_j^-=0.2;\, j\in [J]\}$, the relatively weak signals;  and $ \{\theta_j^+=0.9,\, \theta_j^-=0.1;\, j\in [J]\}$, the relatively strong signals.
We randomly generate the set of true attribute patterns $\mca_0\subseteq\{0,1\}^K$ with cardinality  $|\mca_0|=10$ and $p_{\aaa}=0.1$ for all $\aaa\in \mca_0$.
In the simulations, for $K=10$ the $\mca_{\text{input}}$ is   taken to be $\{0,1\}^K$; while for $K=15$ and $20$, the $\mca_{\text{input}}$ is   taken to be $\widehat\mca_{\text{screen}}$, i.e., the set of candidate patterns output by the screening procedure.
\color{black}

In each scenario we perform 200 independent simulations. 
For shrinkage estimation, we apply the proposed  Algorithm 1 ``Penalized EM (PEM)" and Algorithm 2 ``Fractional Power Variational EM (FP-VEM)", and also apply the plain EM algorithm with thresholding for comparison. 
For $2^K=2^{10}$, we directly apply the shrinkage algorithms; and for large $2^K=2^{15}$ and $2^K=2^{20}$, we run Algorithm \ref{algo-screen} for screening first, and then run the shrinkage algorithms with $\widehat\mca_{\text{screen}}$ as input patterns.
When running PEM we compute a solution path by varying $\lambda$ in the range of $\lambda\in\{-0.2,\, -0.4, \,\ldots,\, -3.8,\,-4.0\}$, and select the $\lambda$ that gives the smallest EBIC.
When running  FP-VEM we fix $\beta=\lambda+1=0.01$ and compute a solution path by varying $\Upsilon$ in $\{1.0,0.9,\ldots,0.4,0.3\}$ and also select  $\Upsilon$ using EBIC.
We use the threshold value $\rho_N=1/(2N)$ for the estimated proportions in the last step for all three shrinkage algorithms to select patterns (other smaller $\rho_N$ values give  similar results). 
 
\begin{table}[h!]
\caption{Pattern selection accuracies for two-parameter SLAM. Tuning parameter $\lambda\in\{-0.2,\, -0.4, \,\ldots,\, -3.8,\,-4.0\}$ in PEM (Algorithm \ref{algo-pem}) and $\Upsilon\in\{1.0,\, 0.9,\,\ldots,\,0.4,\,0.3\}$ in FP-VEM (Algorithm \ref{algo-fpvem}) are selected based on EBIC.}
\label{tab-dina}
\centering
\begin{tabular}{ccccccccc}
\toprule
\multirow{2}{*}{signal strength} & \multirow{2}{*}{$2^K$} & \multirow{2}{*}{$N$} 
& \multicolumn{3}{c}{$1-$FDR} 
& \multicolumn{3}{c}{TPR} \\
\cmidrule(rl){4-6}\cmidrule(lr){7-9}
&  &  & EM & Algo.\,\ref{algo-pem} & Algo.\,\ref{algo-fpvem} & EM & Algo.\,\ref{algo-pem} & Algo.\,\ref{algo-fpvem}  \\
\midrule
\multirow{9}{3cm}{\centering $\theta^+_j=0.8$, $\theta^-_j=0.2$.} & \multirow{3}{*}{$2^{10}$} 
  & 150  &  0.139  & 0.883 &  0.896   &   0.930  &  0.885  &  0.895   \\
& & 500  &  0.115 & 0.995 &  0.992    &   1.000 &  1.000   &  0.999   \\
& & 1000 &  0.100  &  1.000 &  0.996   &   1.000  &  1.000  &  1.000 \\
\cmidrule(lr){2-9}
& \multirow{3}{*}{$2^{15}$} 
  & 150   &  0.049  &  0.523  &  0.544 &  0.539 &  0.530   &  0.543   \\
& & 500   &  0.089  &  0.924 & 0.928   &  0.934  &  0.930 &  0.932   \\
& & 1000  &  0.078 &  0.984  &  0.988   &   0.991 &  0.991  &   0.991   \\ 
\cmidrule(lr){2-9}
& \multirow{3}{*}{$2^{20}$}
  & 150   & 0.019  &   0.213  & 0.264  &   0.270 &  0.255  &  0.271  \\
& & 500  &  0.019   &  0.609  & 0.633 &  0.636  & 0.641   &  0.642   \\ 
& & 1000  &  0.038 &   0.816  & 0.848  &   0.864 & 0.864   &  0.863   \\ 
\midrule
\multirow{9}{3cm}{\centering $\theta^+_j=0.9$, $\theta^-_j=0.1$.} & \multirow{3}{*}{$2^{10}$} 

  & 150 & 0.323 &  0.909 &  1.000    &  1.000  &  1.000 &   1.000 \\
  & & 500 & 0.208  &   1.000   &  1.000   &  1.000     &1.000    & 1.000 \\
  & & 1000 & 0.167  &  1.000 &   1.000  &  1.000   & 1.000  & 1.000 \\
\cmidrule(lr){2-9}
& \multirow{3}{*}{$2^{15}$} 
  & 150  &  0.317 &  0.989 &  0.974   &  0.993  &  0.991  &  0.992 \\
& & 500  &  0.220 &  1.000  &  0.995   &  1.000  &  1.000  &  1.000  \\
& & 1000 &  0.205 &  1.000 &  0.994  &  1.000  &  1.000  & 1.000  \\ \cmidrule(lr){2-9}
& \multirow{3}{*}{$2^{20}$}
  & 150  &  0.232   &  0.968  & 0.941 &   0.972 &   0.971 &  0.970  \\
& & 500 & 0.159  & 1.000   & 0.999  &  1.000  &  1.000  &  1.000   \\
& & 1000   & 0.146 &  1.000 &  0.997   &  1.000 &   1.000  &  1.000  \\
\bottomrule
\end{tabular}

\end{table}

The simulation results on selection accuracies are presented in Table \ref{tab-dina}.
The ``TPR'' stands for True Positive Rate, which denotes the proportion of true patterns that are selected. The ``1-FDR'' stands for ``$1-$False Discovery Rate (FDR)", which denotes the proportion of selected patterns that are true patterns.
Table \ref{tab-dina} shows the proposed PEM and FP-VEM yield good selection results in various scenarios, while the  EM algorithm with direct thresholding at $\rho_N$  suffers from high FDR, i.e., selecting too many non-existing attribute patterns.
We would like to point out that the plain VEM as presented in Example \ref{exp-fpvem} is a special case of the   proposed FP-VEM, by just taking the fractional power $\Upsilon$ to be $\Upsilon=1$. So in each simulation run, the result given by VEM is included in the solution path given by FP-VEM with $\Upsilon\in\{1.0,0.9,\ldots,0.4,0.3\}$, and in the final step EBIC selects the best $\Upsilon$ from the entire solution path. Indeed, in all our simulations about FP-VEM, the result given by $\Upsilon=1$ is never selected by EBIC, which means the selection result given by plain VEM is never favored over the proposed FP-VEM.
We also remark here that the proposed methods are computationally efficient.
All the algorithms are implemented in Matlab.
 In particular, in the case of relatively strong signal with $1-\theta_j^+=\theta_j^-=0.10$, screening and computing an entire solution path for $(2^K,N)=(2^{20},1000)$ takes $<2$ minutes on average on a laptop with a 2.8 GHz processor,
and yields almost perfect pattern selection results, as shown in the last row of Table \ref{tab-dina}.

We give some discussions on the comparison of the PEM and the FP-VEM algorithms. The estimation accuracies presented in Table \ref{tab-dina} generally show the two algorithms have comparable performance on pattern selection. In terms of selecting the tuning parameter, the FP-VEM can be easier to tune because the fractional power $\Upsilon$ is always between 0 and 1, while the PEM algorithm has a negative tuning parameter $\lambda\in(-\infty,0)$ that can have an arbitrarily large magnitude. Specifically, the scenario of an increasing sparsity corresponds to $\Upsilon\to 0$ and $\lambda\to-\infty$, and   when extremal sparsity exists, the FP-VEM needs to choose $\Upsilon$ close to zero with a small magnitude and the PEM needs to choose $\lambda$ with a large magnitude. Therefore, in such cases the tuning of PEM may take more time, since $\lambda<0$ needs to be searched over a relatively large interval;
an exponential grid search may be of help in this case while a further exploration would be needed.
Meanwhile, we find in simulation studies that choosing a small $\Upsilon$ in FP-VEM too close to zero may result in the algorithm to be less stable in some cases. In practice, if the computation time is not a primary concern, we recommend first considering the PEM algorithm for the better stability.
\color{black}

We further conduct a simulation study to investigate how the threshold value $\rho_N$ for the estimated proportions impact the pattern selection results of different methods. In the setting with $1-\theta_j^+=\theta_j^-=0.2$ and $N=150$ (the same setting as the first line in Table \ref{tab-dina}), we simulate 200 independent datasets, and apply the proposed PEM (Algorithm 1), FP-VEM (Algorithm 2) and the usual EM algorithm with various thresholds $\rho_N\in\{1/(50N)\}\cup\{i/(2N),~ i=1,3,5,\ldots,15\}$. Figure \ref{fig-thres} plots the average ``TPR'' and average ``$1-$FDR" versus the threshold values. It can be seen that directly thresholding the MLE of the proportions (corresponding to the thresholding after EM) does not yield good selection results. For a small threshold $\rho_N=1/(2N)$, the FDR of thresholded EM is quite high.
When further decreasing the threshold $\rho_N$ from $1/(2N)$ to $1/(50N)$, the FDR of thresholded EM becomes worse while the proposed methods have stable performance.
On the other hand, as the threshold $\rho_N$ increases from $1/(2N)$ to larger values, the TPR of EM quickly decreases. In contrast, the proposed methods PEM and FP-VEM give reasonably good selection results across all the threshold values, and have slightly better performance for smaller thresholds. Even the best selection result given by thresholding EM corresponding to the threshold $\rho_N=7/(2N)$ is not comparable to those given by the proposed methods.

\begin{figure}[h!]
\centering
\includegraphics[width=0.7\linewidth]{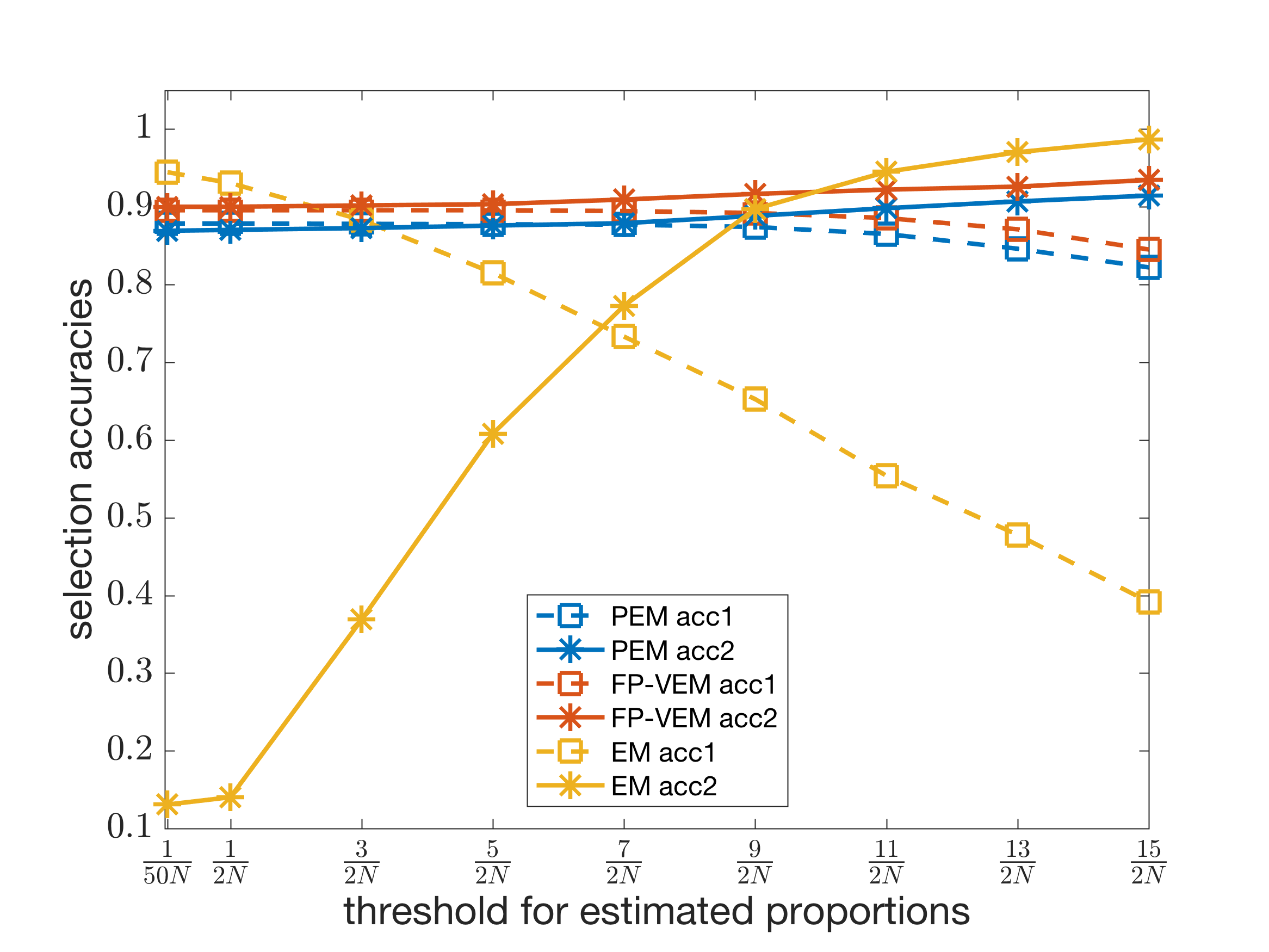}
\caption[]{Selection accuracies versus thresholds for the two-parameter SLAM with $1-\theta_j^+=\theta_j^-=0.2$ and $N=150$. For each method, ``acc1" denotes the True Positive Rate (TPR), the proportion of true patterns that are selected; and ``acc2" denotes ``$1-$False Discovery Rate (FDR)", the proportion of selected patterns that are true.} 
\label{fig-thres}
\end{figure}

We next evaluate the performance of the screening procedure. 
We find that the screening procedure drastically reduces the computational cost in the subsequent shrinkage estimation stage. For instance, in the setting $(N,K)=(150,15)$ when noise rate is $1-\theta_j^+=\theta_j^-=20\%$, based on 200 runs, the variational screening procedure takes $1.55$ seconds on average, and the subsequent PEM algorithm takes $6.42$ seconds on average; while if no screening is performed, the PEM algorithm takes $7.96\times 10^3$ seconds on average.

\begin{figure}[h!]
\centering

    \begin{minipage}[b]{0.24\textwidth}
	\centering
	\includegraphics[width=0.98\textwidth]{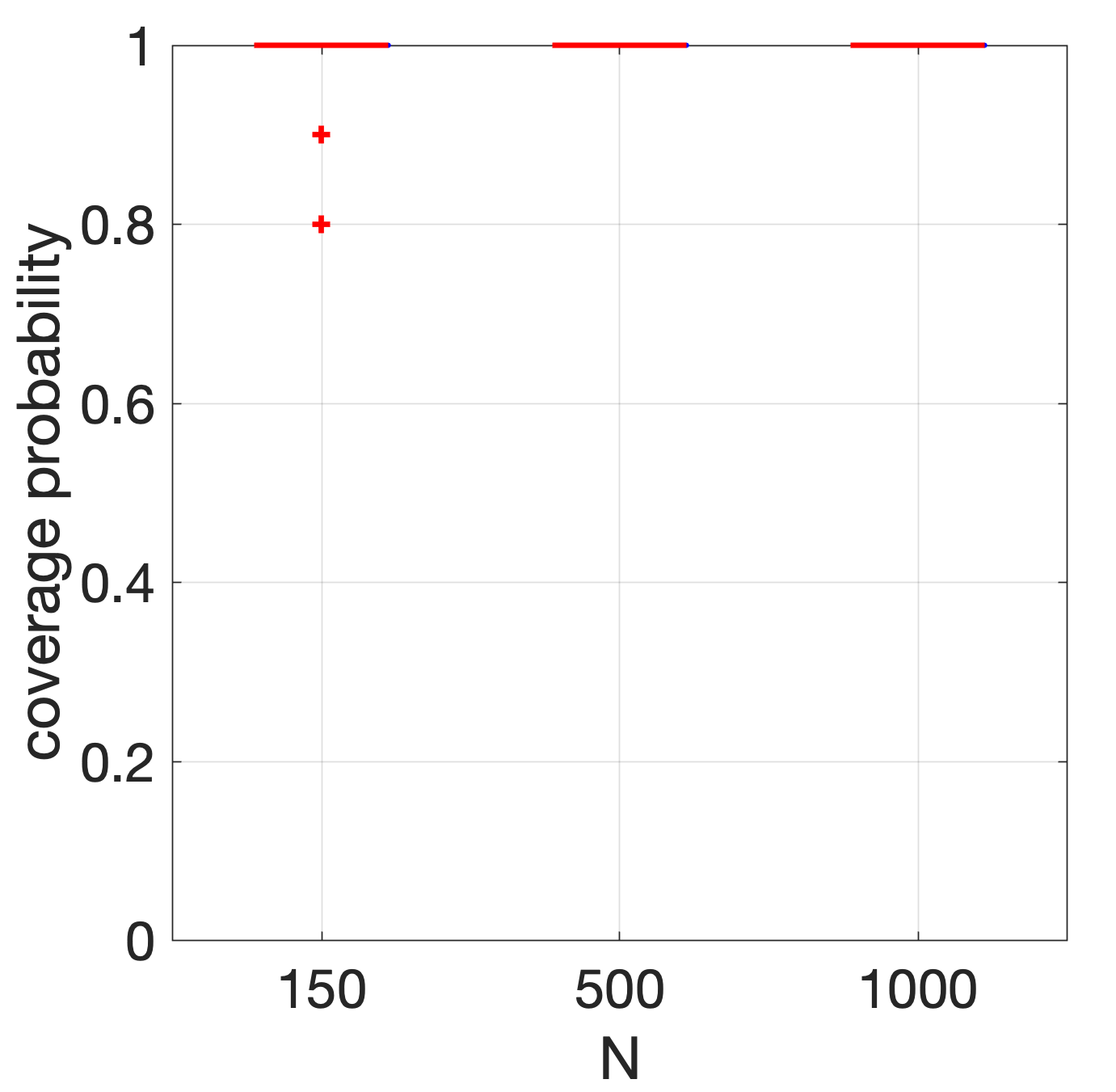}
	\end{minipage}
	\hfill
    \begin{minipage}[b]{0.24\textwidth}
    \centering
    \includegraphics[width=\textwidth]{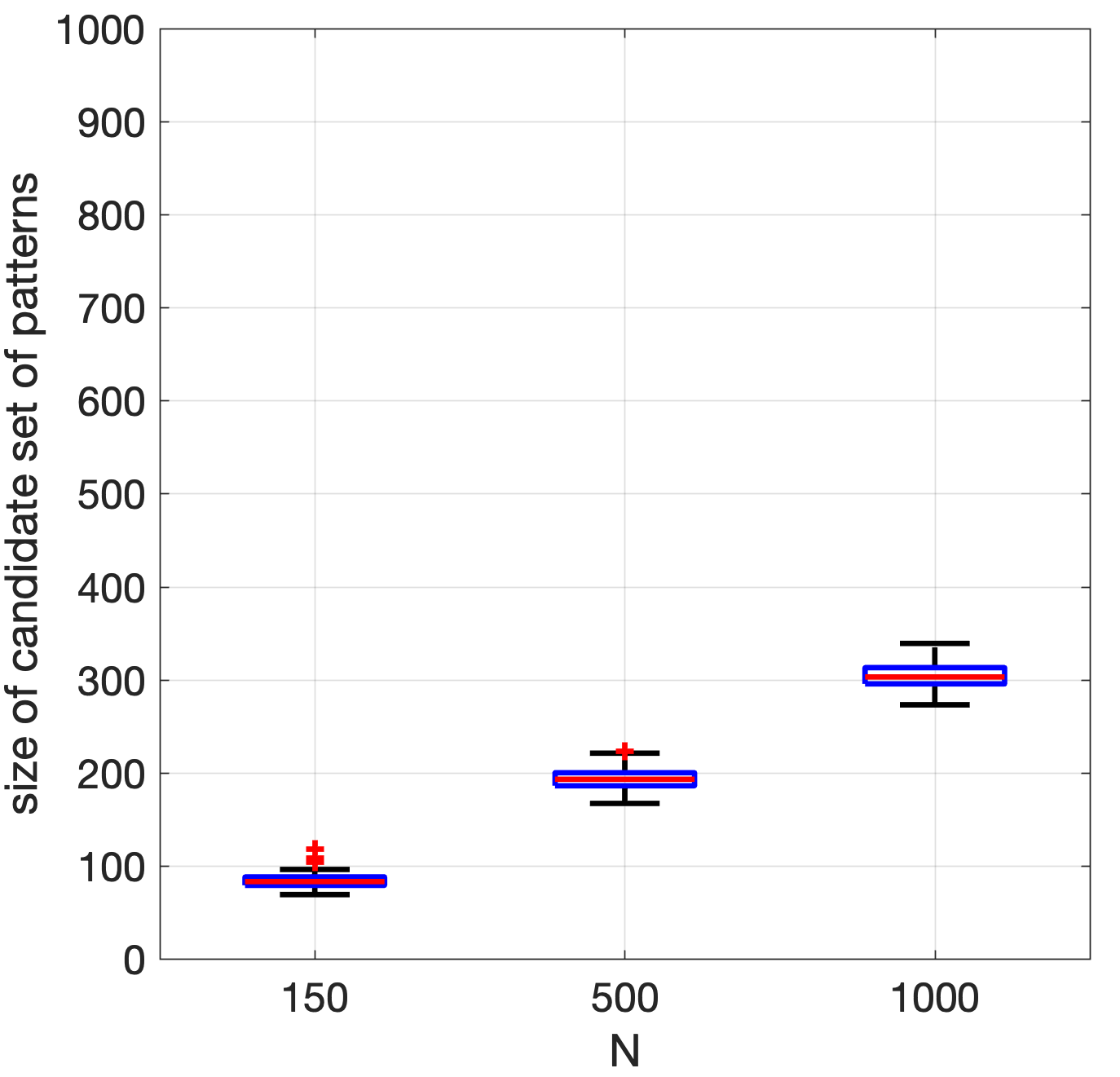}
	\end{minipage}
	\hfill
	\begin{minipage}[b]{0.24\textwidth}
	\centering
	  \includegraphics[width=0.98\textwidth]{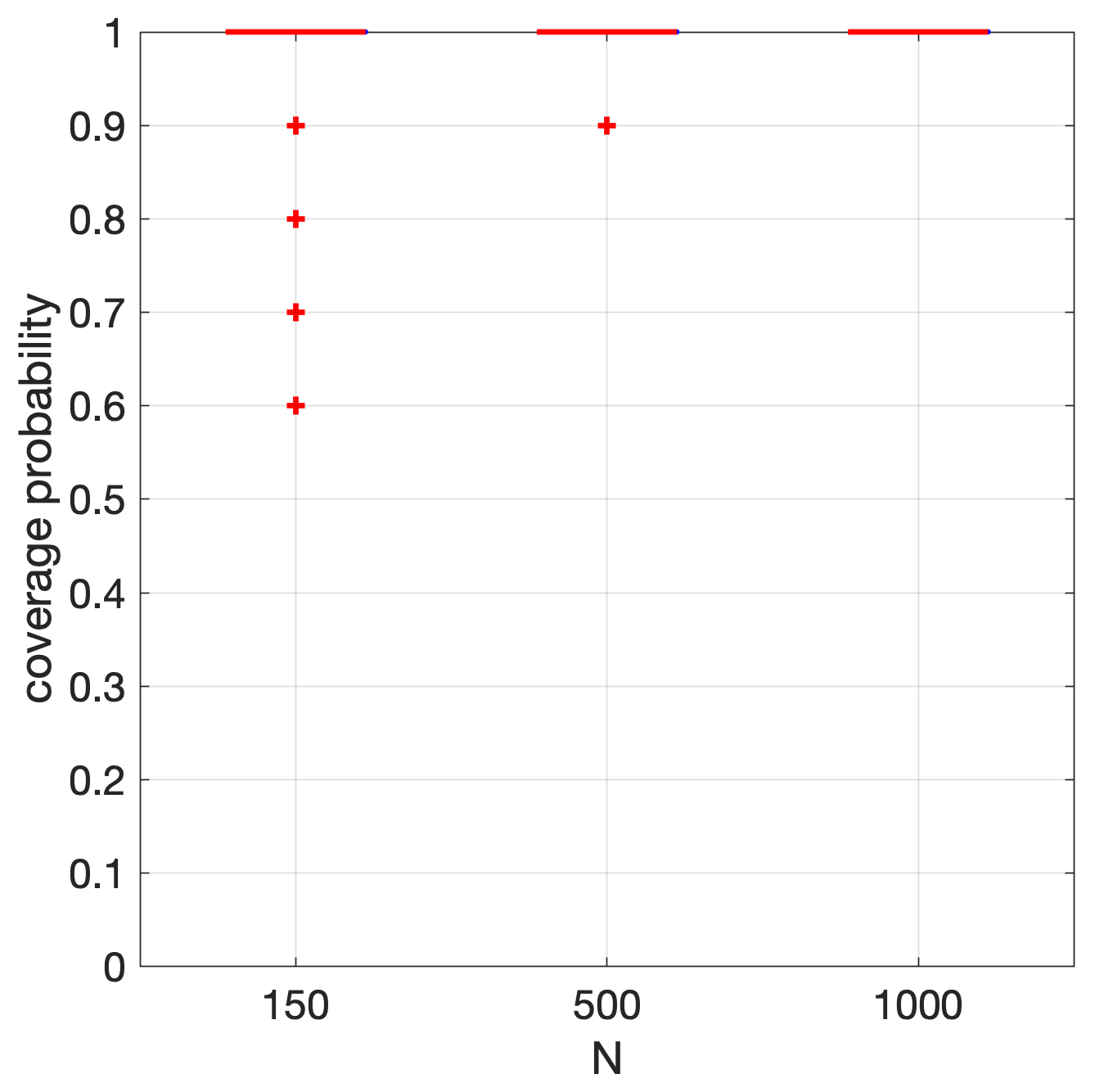}
	\end{minipage}
	\hfill
    \begin{minipage}[b]{0.24\textwidth}
    \centering
    \includegraphics[width=\textwidth]{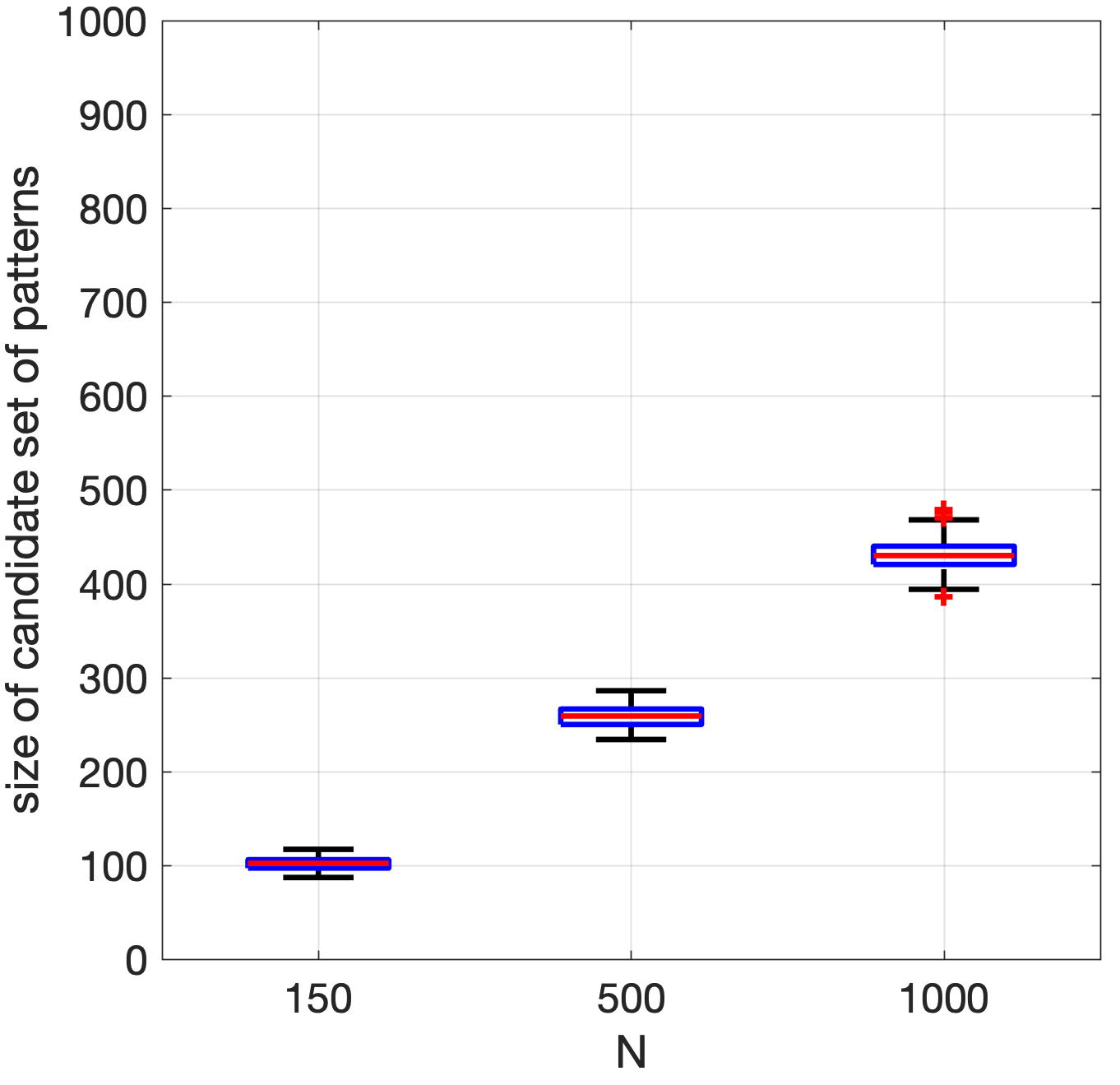}
	\end{minipage}
	
	\begin{minipage}[b]{0.24\textwidth}
	\centering
	  \small{(a) $K=15$, noise $10\%$;\\ screening accuracy}
	\end{minipage}
	\hfill
    \begin{minipage}[b]{0.24\textwidth}
\centering
    \small{(b) $K=15$, noise $10\%$;\\ size of $\widehat\mca_{\screen}$}
	\end{minipage}
	\hfill
	\begin{minipage}[b]{0.24\textwidth}
	\centering
	\small{(c) $K=20$, noise $10\%$;\\ screening accuracy}
	\end{minipage}
	\hfill
    \begin{minipage}[b]{0.24\textwidth}
\centering
    \small{(d) $K=20$, noise $10\%$;\\ size of $\widehat\mca_{\screen}$}
	\end{minipage}

	\begin{minipage}[b]{0.24\textwidth}
	\centering
	  \includegraphics[width=0.98\textwidth]{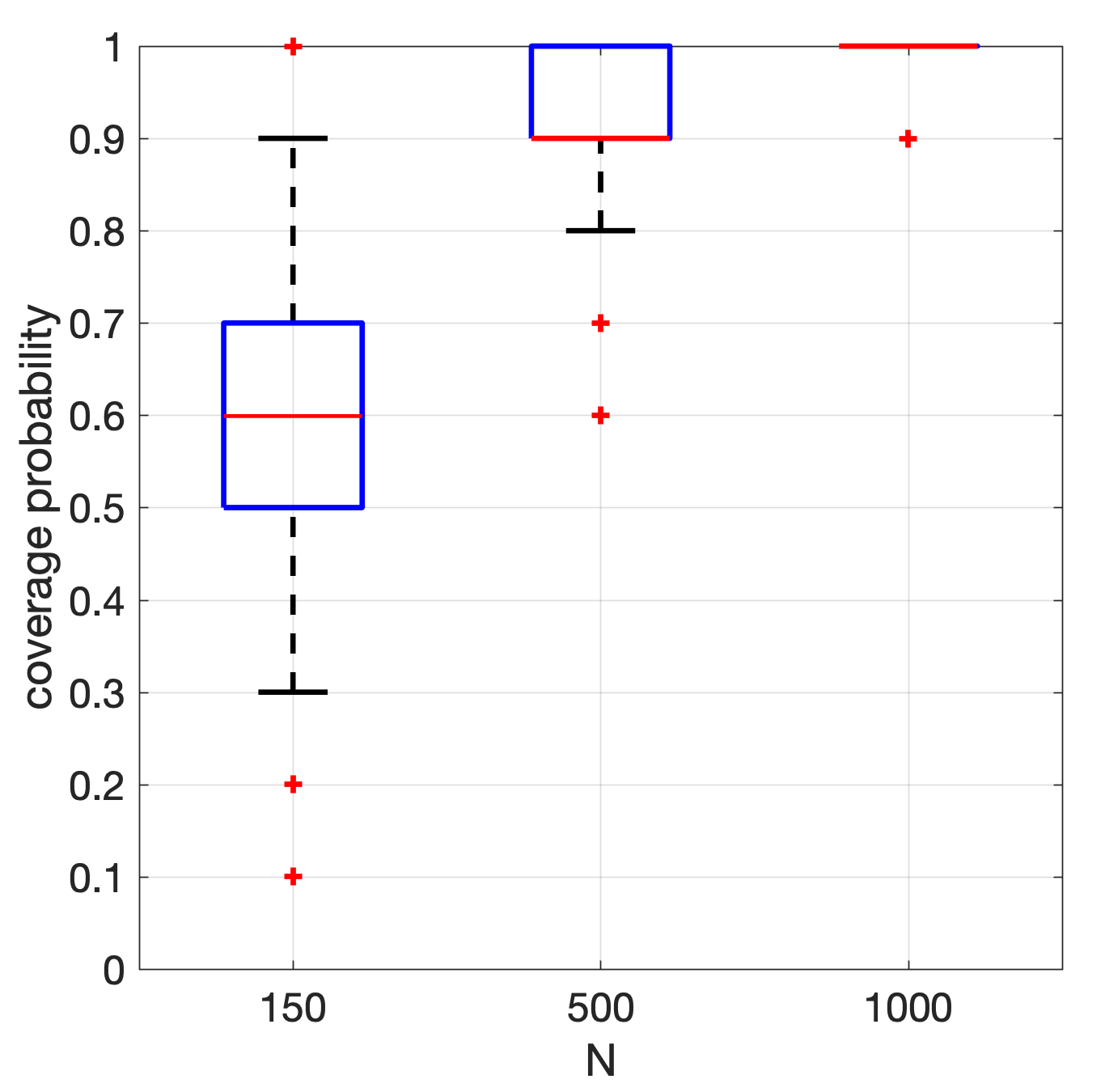}
	\end{minipage}
	\hfill
    \begin{minipage}[b]{0.24\textwidth}
\centering
\includegraphics[width=\textwidth]{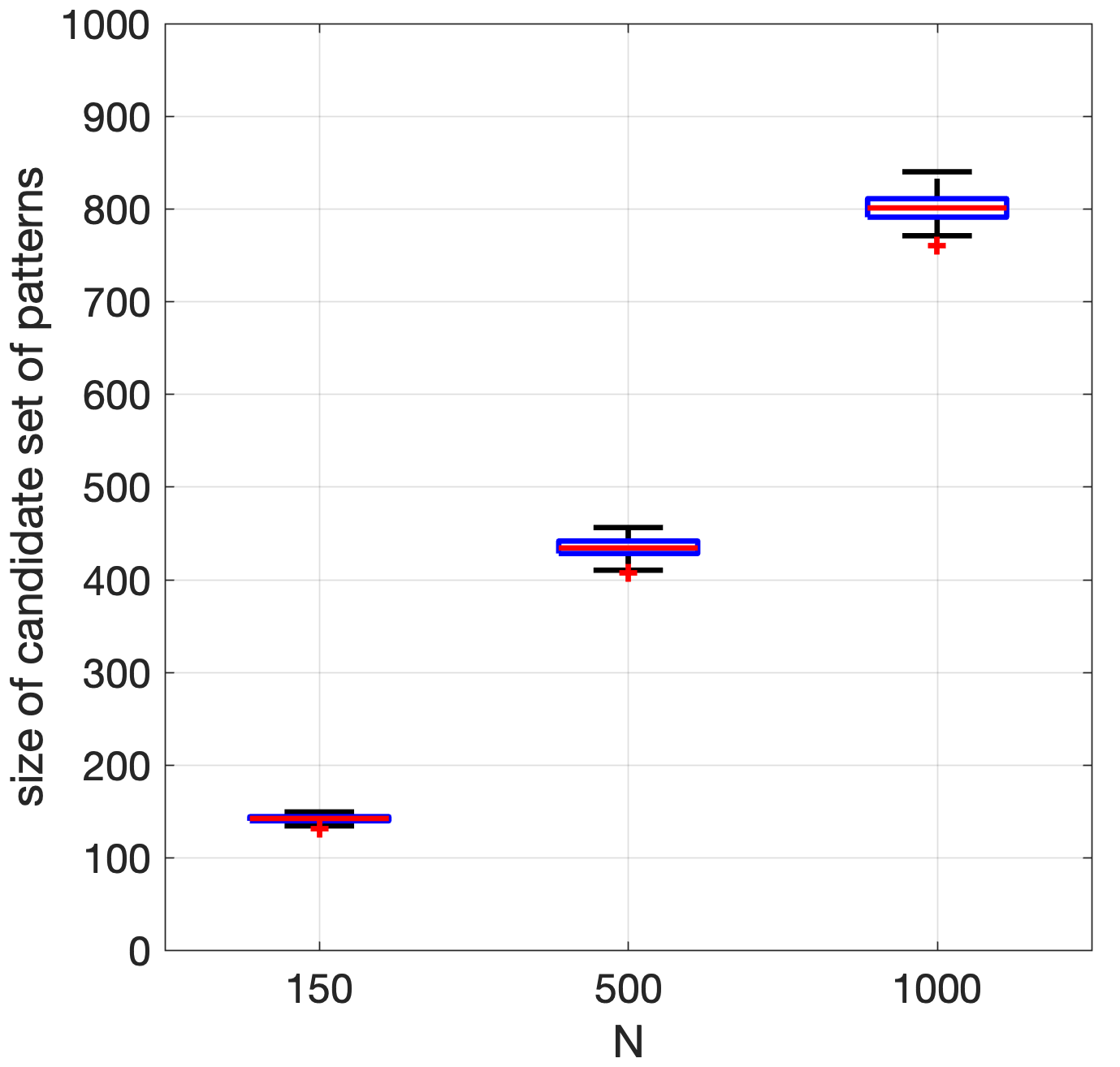}
	\end{minipage}
	\hfill
	\begin{minipage}[b]{0.24\textwidth}
	\centering
	  \includegraphics[width=0.98\textwidth]{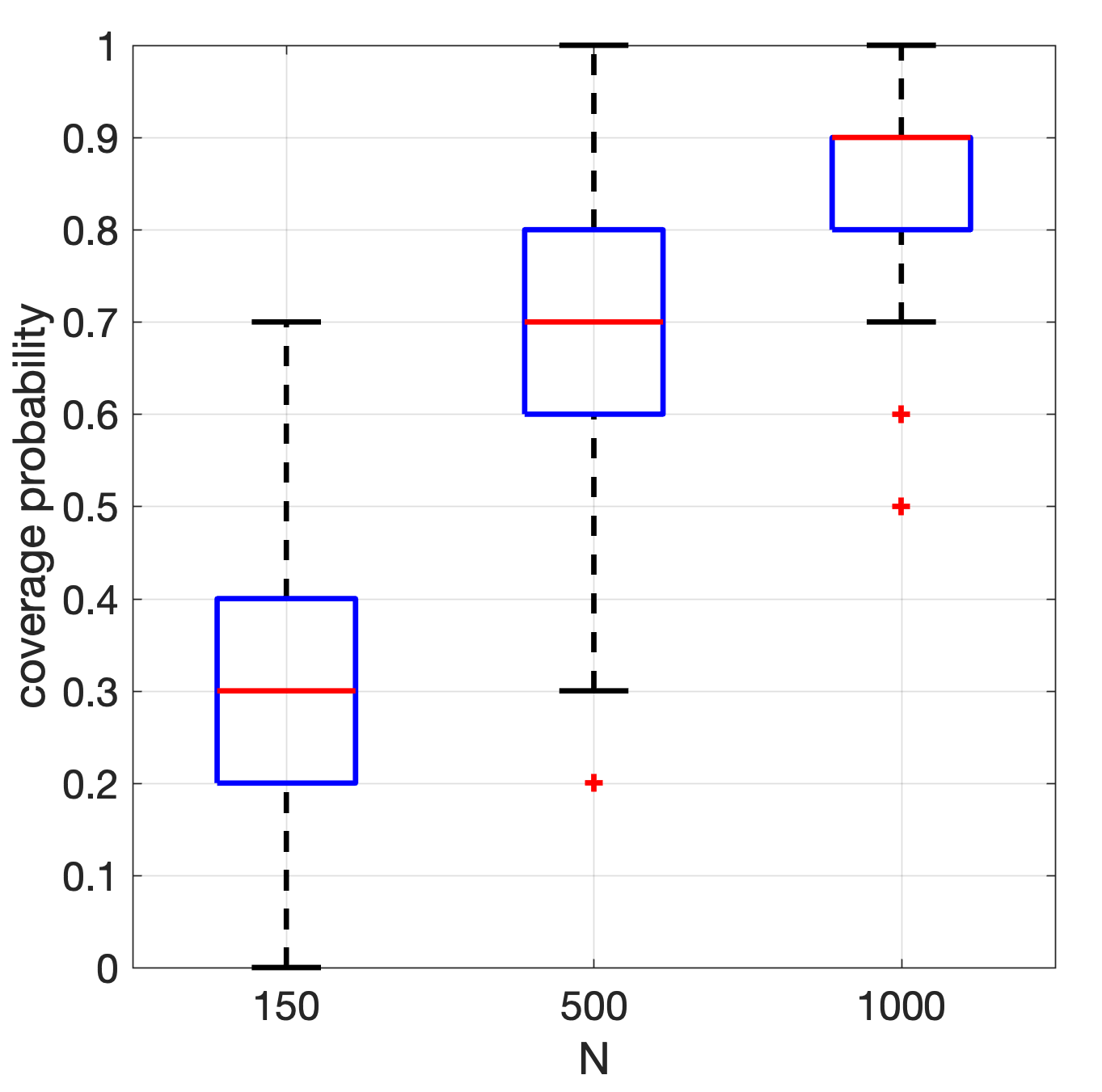}
	\end{minipage}
	\hfill
    \begin{minipage}[b]{0.24\textwidth}
\centering
\includegraphics[width=\textwidth]{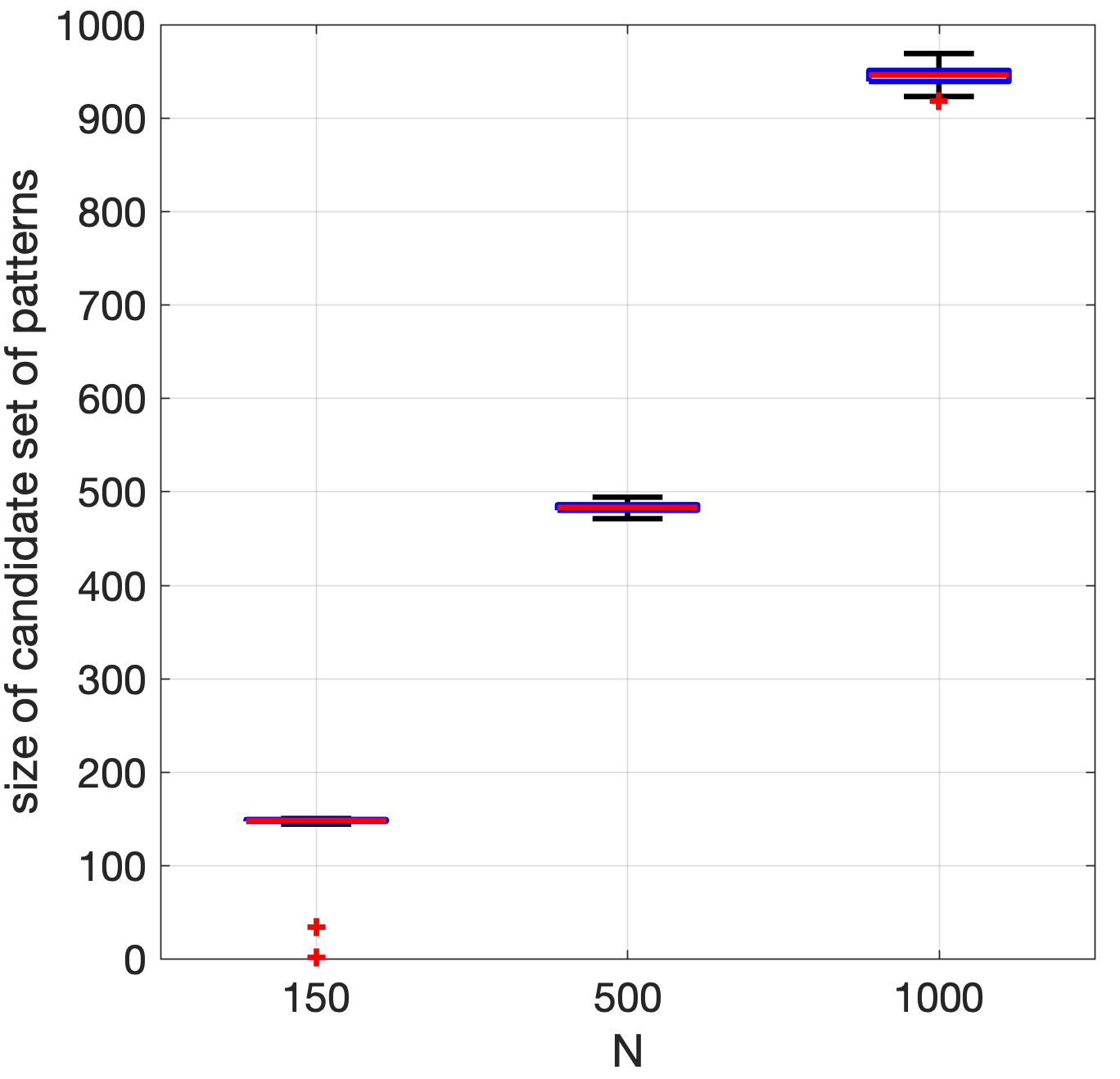}
	\end{minipage}
	
	\begin{minipage}[b]{0.24\textwidth}
	\centering
	  \small{(e) $K=15$, noise $20\%$; \\screening accuracy}
	\end{minipage}
	\hfill
    \begin{minipage}[b]{0.24\textwidth}
\centering
    \small{(f) $K=15$, noise $20\%$; \\size of $\widehat\mca_{\screen}$}
	\end{minipage}
	\hfill
	\begin{minipage}[b]{0.24\textwidth}
	\centering
	\small{(g)  $K=20$, noise $20\%$; \\screening accuracy}
	\end{minipage}
	\hfill
    \begin{minipage}[b]{0.24\textwidth}
\centering
    \small{(h)  $K=20$, noise $20\%$; \\size of $\widehat\mca_{\screen}$}
	\end{minipage}

\caption{
{Screening: plots (a), (c), (e) and (g) are coverage probabilities of the true patterns, from the screening procedure under the two-parameter SLAM; 
plots (b), (d), (f) and (h) are sizes of $\widehat\mca_{\screen}$.
The ``noise'' refers to the value of $1-\theta_j^+=\theta_j^-$.}
}
\label{fig-screen0}
\end{figure}

\begin{figure}[h!]
\centering
\begin{minipage}[b]{0.24\textwidth}
	\centering
	  \includegraphics[width=0.97\textwidth]{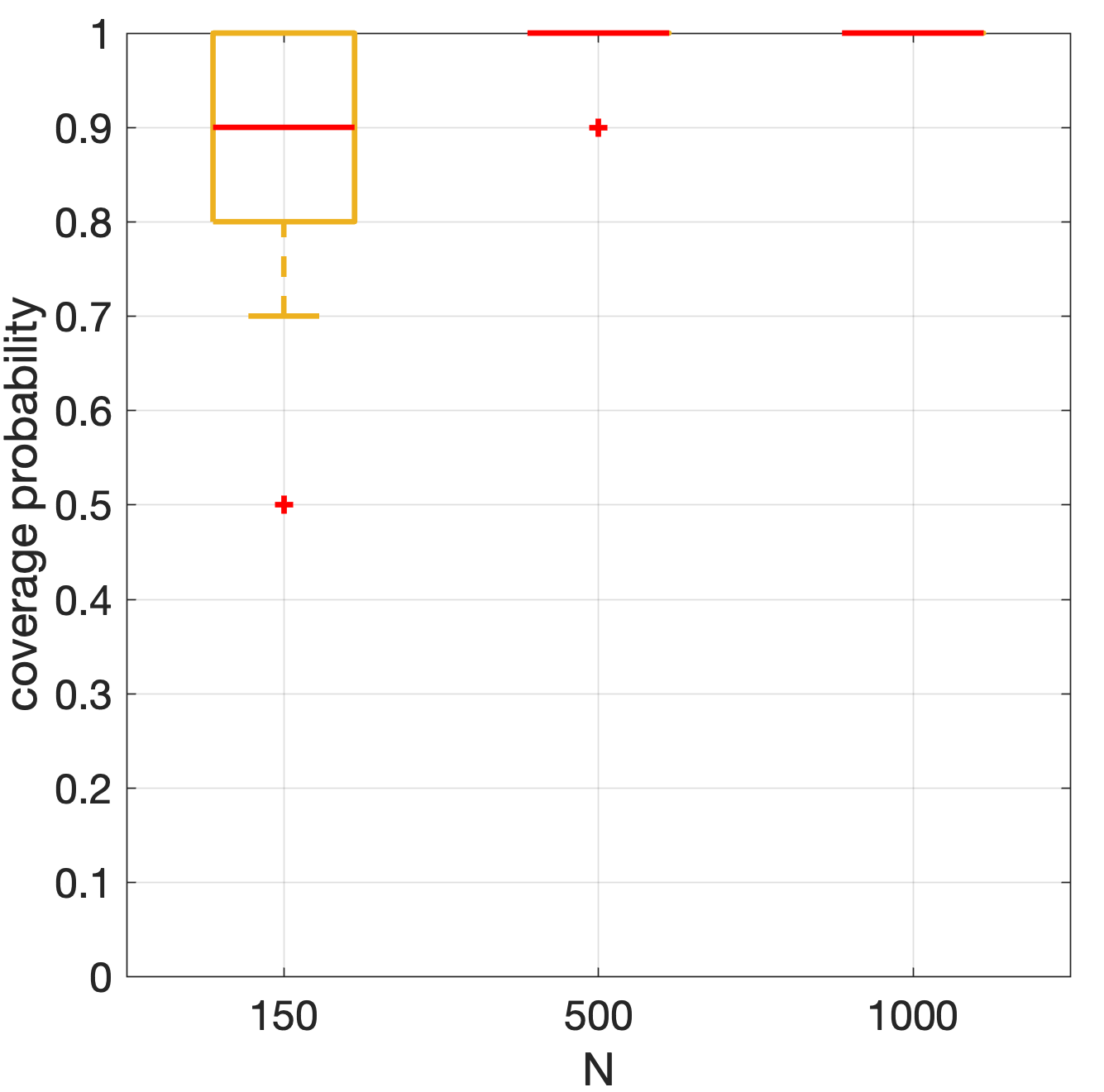}
	\end{minipage}
	\hfill
    \begin{minipage}[b]{0.24\textwidth}
\centering
\includegraphics[width=\textwidth]{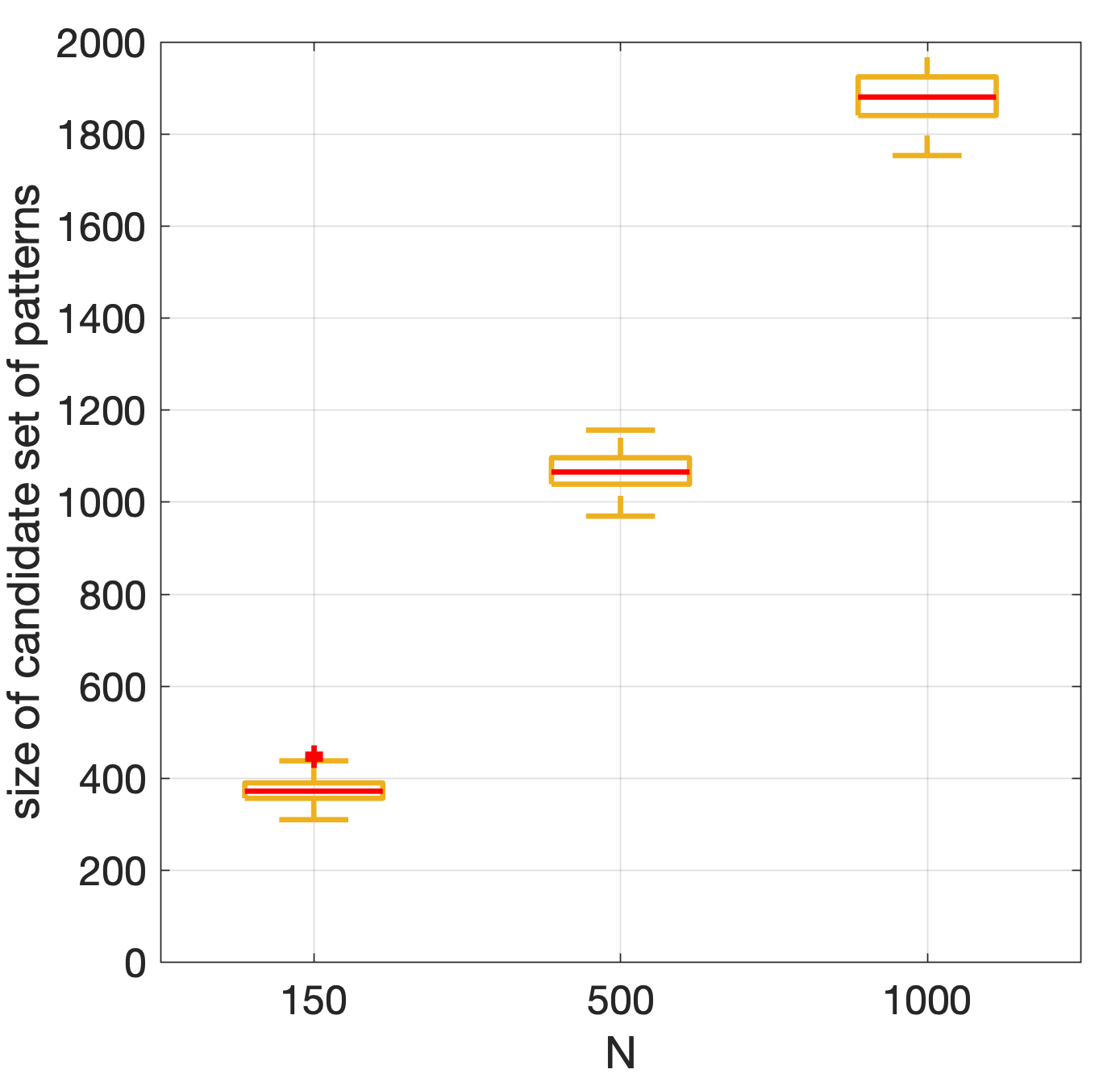}
	\end{minipage}
	\hfill
	\begin{minipage}[b]{0.24\textwidth}
	\centering
	  \includegraphics[width=0.97\textwidth]{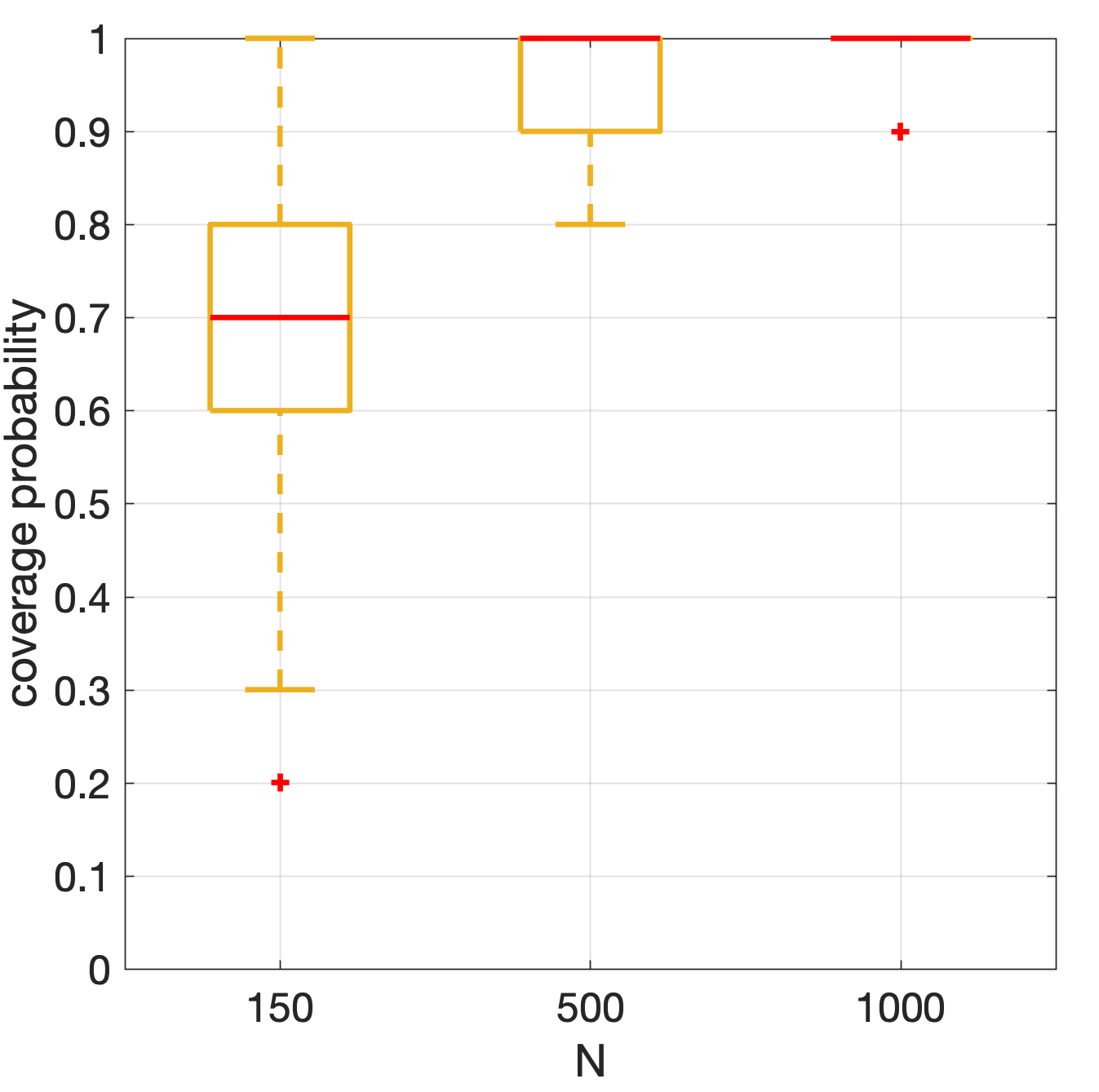}
	\end{minipage}
	\hfill
    \begin{minipage}[b]{0.24\textwidth}
\centering
\includegraphics[width=\textwidth]{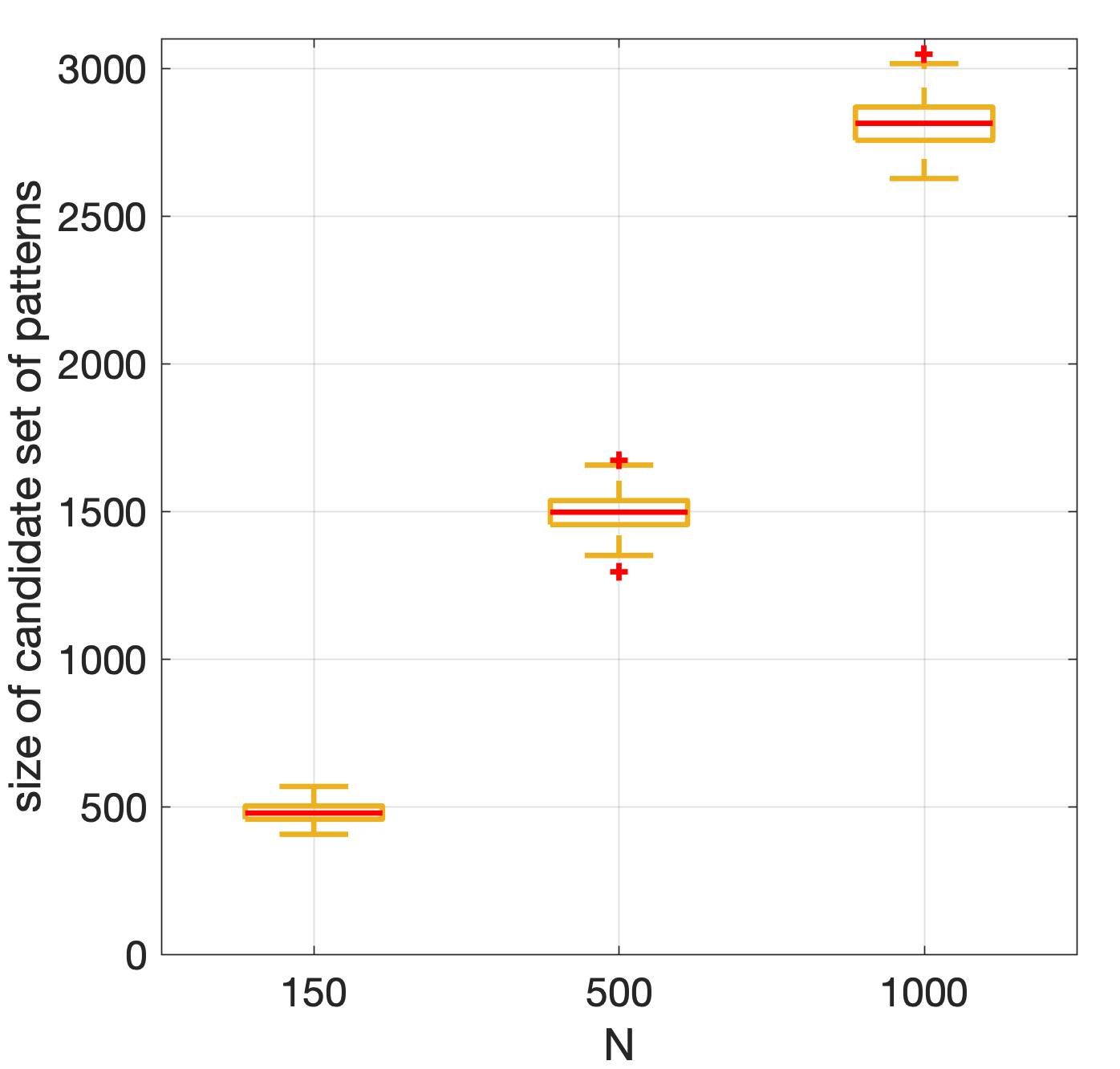}
	\end{minipage}
	
	\begin{minipage}[b]{0.24\textwidth}
	\centering
	  \small{(a) $K=15$, noise $20\%$;  screening accuracy}
	\end{minipage}
	\hfill
    \begin{minipage}[b]{0.24\textwidth}
\centering
    \small{(b) $K=15$, noise $20\%$;  size of $\widehat\mca_{\screen}$}
	\end{minipage}
	\hfill
	\begin{minipage}[b]{0.24\textwidth}
	\centering
	\small{(c) $K=20$, noise $20\%$;   screening accuracy}
	\end{minipage}
	\hfill
    \begin{minipage}[b]{0.24\textwidth}
\centering
    \small{(d) $K=20$, noise $20\%$;  size of $\widehat\mca_{\screen}$}
	\end{minipage}

\caption{
{Screening enhanced by Gibbs exploration: screening accuracy and size of $\widehat\mca_{\screen}$. Noise rate is $1-\theta_j^+=\theta_j^-=20\%$.
}
}
\label{fig-enhance}
\end{figure}

As described earlier, the screening is considered successful if all true patterns are included in the candidate set $\widehat\mca_{\screen}$. Under each simulation scenario in Table \ref{tab-dina} corresponding to $K=15$ or $K=20$, we record the coverage probabilities of the true patterns for each of 200 runs, where in each run $\sum_{\aaa\in\mca_0}I(\aaa\in\widehat\mca_{\screen})/|\mca_0|$ is recorded as the coverage probability. The boxplots of coverage probabilities under these scenarios are presented in Figure \ref{fig-screen0}(a), (c), (e) and (g). We also record the size of $\widehat\mca_{\screen}$, i.e., the number of candidate patterns given by the screening procedure in each run, and present their boxplots in Figure \ref{fig-screen0}(b), (d), (f) and (h).
The screening procedure generally works well. On the other hand, Figure \ref{fig-screen0}(e) and (g) show that for the relatively large noise rate and small sample size, the screening accuracy is not  high. 

To improve the performance of screening, we apply the strategy of \textit{screening enhanced by Gibbs exploration} described in Remark \ref{rmk-enhance} with  $M=3$. That is, along the stochastic EM iterations of the screening algorithm, after every three iterations we add the current set of latent patterns to the candidate set $\widehat\mca_{\screen}$. The resulting screening accuracies and sizes of $\widehat\mca_{\screen}$ are presented in Figure \ref{fig-enhance}. Compared to the second row of plots in Figure \ref{fig-screen0}, one can clearly see that   the enhancing procedure improves the screening accuracy significantly, while the size of $\mathcal A_{\text{screen}}$ also increases but still remains quite manageable. Under the noise rate $1-\theta_j^+=\theta_j^-=20\%$, the size of $\mathcal A_{\text{screen}}$ is always below $N$ for screening without enhancing, while for screening with enhancing, the size of $\mathcal A_{\text{screen}}$ is around $2N$ for $K=15$ and around $3N$ for $K=20$. 
The enhancing by Gibbs exploration would not sacrifice the efficiency of the screening procedure itself, though it results in a larger set of $\widehat\mca_{\screen}$ which incurs higher computational cost in the shrinkage stage.
In practice, one should leverage this tradeoff according to the sample size. Specifically,  when sample size $N$ is small, choosing a more conservative screening procedure (with a smaller integer $M$) is recommended, because this would increase the screening accuracy without causing much computational burden for the shrinkage algorithm.
With the enhanced screening procedure, in the relatively weak signal case $1-\theta_j^+=\theta_j^-=0.2$ and under  $(K,N)=(15,150)$,
the two accuracy measures $1-$FDR and TPR for the PEM algorithm, become $(0.850, 0.860)$ (previously it was $(0.523,0.530)$ in Table \ref{tab-dina}), and those under the FP-VEM algorithm become $(0.839, 0.853)$ (previously $(0.544,0.543)$ in Table \ref{tab-dina}). Under $(K,N)=(20,150)$, the two accuracy measures for the PEM become $(0.608, 0.648)$ (previously $(0.213,0.255)$ in Table \ref{tab-dina}) and those for the FP-VEM become $(0.620, 0.634)$ (previously $(0.264,0.271)$ in Table \ref{tab-dina}).

\color{black}

\paragraph{Multi-parameter all-effect SLAM.} Consider the multi-parameter all-effect SLAM  introduced in \eqref{alleff} in Example \ref{exp-gdina} with an identity link function $f(\cdot)$. Let the $Q$-matrix be in the form $Q=(Q_1^\top, Q_2^\top, Q_2^\top)$ with $Q_1$ and $Q_2$ specified in \eqref{eq-3q}. 
Similar to the two-parameter simulation study, we consider three dimensions of possible attribute patterns with $2^K=2^{10}, 2^{15}, $ and $2^{20}$, and  three sample sizes with $N=150, 500$ and $1000$.
For each item, set the baseline probability, the positive response probability of the all-zero attribute pattern $\aaa=\mathbf 0_K$, to  $0.2$ (i.e., $\theta_{j,\zero_K}=0.2$), and the positive response probability of $\aaa=\mathbf 1_K$ to  $0.8$ (i.e., $\theta_{j,\one_K}=0.8$). And we set all the main effects and interaction effects parameters of the item to be equal (i.e., $\beta_{j,S_1}=\beta_{j,S_2}$ for any $\varnothing\neq S_1,S_2\subset\mathcal K_j$ for the $\beta$-coefficients in \eqref{alleff}). 
We randomly generate the set of true attribute patterns, $\mca_0\subseteq\{0,1\}^K$ with cardinality  $|\mca_0|=10$ and set $p_{\aaa}=0.1$ for all $\aaa\in \mca_0$.

\begin{table}[h!]
\caption{Pattern selection accuracies for multi-parameter all-effect SLAM. Tuning parameter $\lambda\in\{-0.2, -0.4, \ldots,  -4.0\}$ in PEM (Algorithm \ref{algo-pem}) and $\Upsilon\in\{1.0, 0.9,\ldots,\allowbreak  0.3\}$ in FP-VEM (Algorithm \ref{algo-fpvem}) are selected using EBIC. Signal strengths are $\theta_{j,\zero_K}=0.1,~\theta_{j,\one_K}=0.9$.}
\label{tab-gdina}

\centering
\begin{tabular}{cccccccc}
\toprule
 \multirow{2}{*}{$2^K$} & \multirow{2}{*}{$N$} & \multicolumn{3}{c}{$1-$FDR} & \multicolumn{3}{c}{TPR} \\
 \cmidrule(lr){3-5}\cmidrule(lr){6-8}
 &  & EM & Algo.\,\ref{algo-pem} & Algo.\,\ref{algo-fpvem} & EM & Algo.\,\ref{algo-pem} & Algo.\,\ref{algo-fpvem}  \\
\midrule
\multirow{3}{*}{$2^{10}$} 
& 150 &  0.277  &  0.983  & 0.953  & 0.996 &   0.980  &   0.974   \\
& 500 &  0.214  &  0.988  &   0.976  & 1.000  &  1.000   & 1.000  \\
& 1000 &  0.193  &  0.992  &   0.986 & 1.000  &  1.000   &  1.000   \\
\midrule
 \multirow{3}{*}{$2^{15}$} 
  & 150  &  0.198 &  0.900  &  0.893  & 0.904 &  0.902 &   0.902 \\
 & 500 &   0.166 &   0.999 &   0.997  & 1.000 &  1.000 &   1.000  \\
 & 1000  &   0.134   &  1.000  &  0.996 & 1.000  &  1.000  &  1.000\\
\midrule
 \multirow{3}{*}{$2^{20}$} 
  & 150  &  0.109 &  0.723  &   0.741  &   0.739 &  0.734  &   0.743   \\
 & 500 &  0.129 &  0.980  &   0.981   &   0.980  &  0.982  &   0.983 \\
 & 1000 &   0.104 &   1.000 &   0.998  &  1.000 &   1.000 &  1.000\\
\bottomrule
\end{tabular}

\end{table}

Similar to the observations in Table \ref{tab-dina}, Table \ref{tab-gdina} shows that the proposed methods also have   good pattern selection performance for the more complicated multi-parameter all-effect model. The approximate screening algorithm based on the two-parameter submodel is quite effective here for screening out candidate patterns under the multi-parameter model. And similarly to the two-parameter case, the EM algorithm tends to severely overselects the attribute patterns. 
{Please see Appendix A.2 for additional results on  the performance of the   screening procedure.}

\section{Data Analysis}
In this section, we apply the proposed methodology to two real world datasets in educational assessment to uncover the knowledge structure of the population. 
 
 \paragraph{Analysis of Fraction Subtraction Data.}
The fraction subtraction dataset is     widely analyzed   in the psychometrics literature \citep{dela,decarlo2011analysis,HensonTemplin09,dela2011}. 
The dataset contains $N=536$ middle school students' binary responses to $20$ questions that were designed for diagnosis assessment of 8 attributes related to fraction and subtraction. 
Table \ref{tab-q8} presents the $Q$-matrix specified in \cite{dela}.
The eight attributes   are  ($\alpha_1$)	Convert a whole number to a fraction;
 ($\alpha_2$)	Separate a whole number from a fraction;
 ($\alpha_3$)	Simplify before subtracting;
  ($\alpha_4$)	Find a common denominator;
  ($\alpha_5$)	Borrow from whole number part;
 ($\alpha_6$)	Column borrow to subtract the second numerator from the first;
  ($\alpha_7$)	Subtract numerators;
  ($\alpha_8$)	Reduce answers to simplest form.

{\tiny  \begin{table}[h!]
\centering
\caption{$Q$-matrix, Fraction Subtraction Data}
\begin{tabular}{lcccccccc}
\toprule
Item ID &  $\alpha_1$ & $\alpha_2$ & $\alpha_3$ & $\alpha_4$ & $\alpha_5$ & $\alpha_6$ & $\alpha_7$ & $\alpha_8$\\
\midrule
1&  0 &	0&	0&	1&	0&	1&	1&	0\\
2& 0 &   0&	0&	1&	0&	0&	1&	0\\
3&  0&	0&	0&	1&	0&	0&	1&	0\\
\vdots  & \vdots &	\vdots &	\vdots &	\vdots &	\vdots &	\vdots &	\vdots &	\vdots \\
18&  0 &	1 &	0&	0&	1&	1&	1&	0\\
19& 1&	1&	1&	0&	1&	0&	1&	0\\
20&  0&	1&	1&	0&	1&	0&	1&	0\\
\bottomrule
\end{tabular}
\label{tab-q8}
\end{table}
} 

 \begin{figure}[h!]
\begin{subfigure}{0.5\textwidth}\centering
\includegraphics[width=0.8\linewidth]{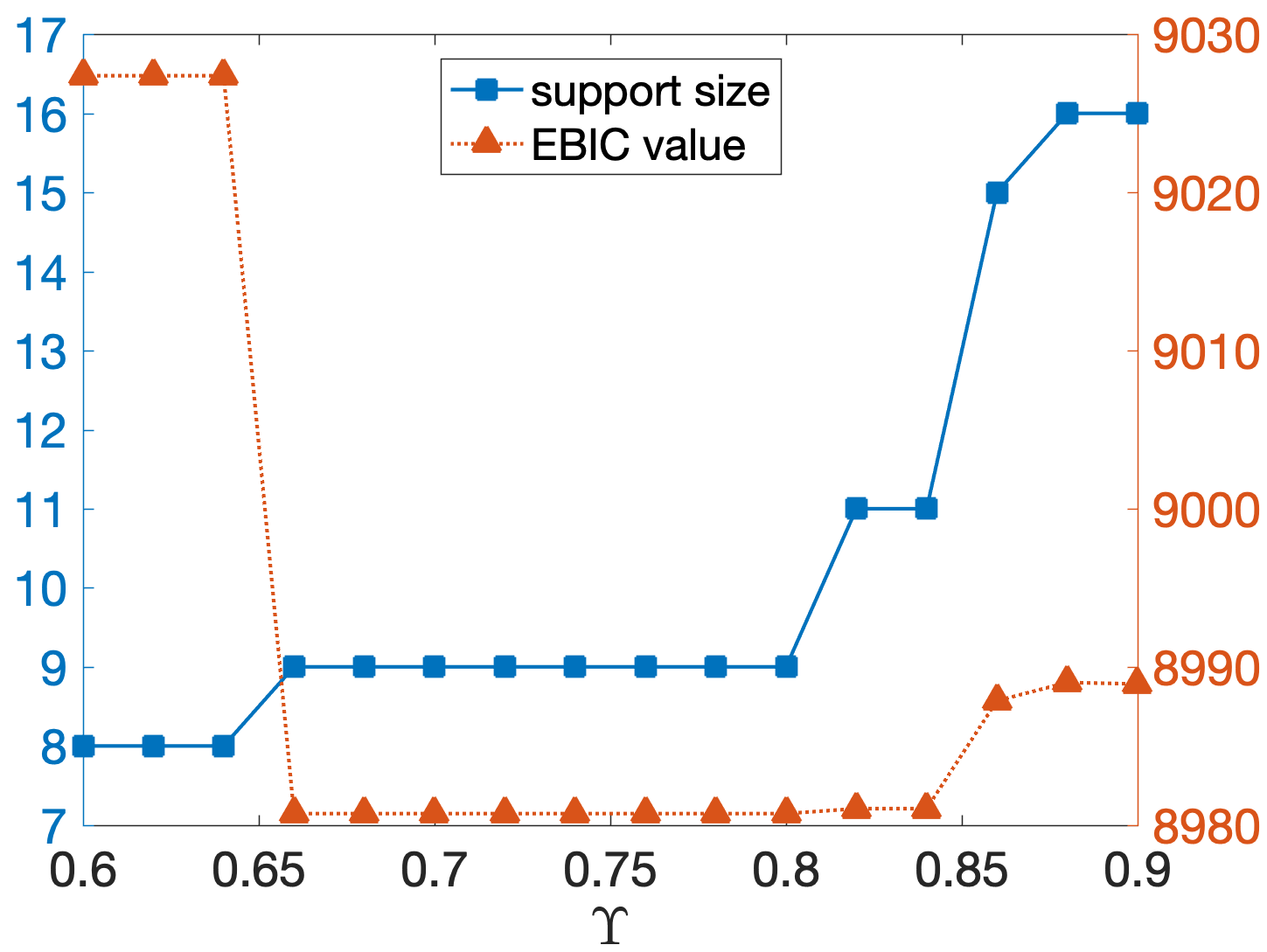}
\end{subfigure}\hspace*{\fill}
\begin{subfigure}{0.5\textwidth}
\centering
\begin{tikzpicture}[scale=0.8]
\draw [group1] (-0.5,3.5) rectangle (0.5,4.5);

\draw [group2] (-3.5,1.5) rectangle (-2.5,2.5);

\draw [group3] (-1,1.5) rectangle (1,2.5);

\draw [group5] (-3.5,-0.5) rectangle (-2.5,0.5);

\draw [group4] (-1.5,-0.5) rectangle (1.5,0.5);

\path
node at ( 0,4) [place] (7) {7}

node at (-0.5,2) [place] (2) {2}
node at ( 0.5,2) [place] (8) {8}

node at ( -1,0) [place] (1) {1}
node at ( 0,0) [place] (3) {3}
node at ( 1,0) [place] (5) {5}

node at ( -3,2) [place] (6) {6}
node at ( -3,0) [place] (4) {4};

\draw [->, dotted, thick] (-0.1,3.5) to (-3,2.5); 
\draw [->, dotted, thick] (0,3.5) to (0,2.5);

\draw [->, dotted, thick] (-3,1.5) to (-3,0.5);
\draw [->, dotted, thick] (-2.9,1.5) to (-0.1,0.5);

\draw [->, dotted, thick] (0,1.5) to (0,0.5);
\end{tikzpicture}

\end{subfigure}
\begin{minipage}[t]{.5\linewidth}
        \caption*{(a) EBIC values and support sizes versus $\Upsilon$}
\end{minipage}%
\hspace*{\fill}
\begin{minipage}[t]{.5\linewidth}
        \caption*{(b) attribute structure selected by EBIC}
\end{minipage}%
\caption[]{Results of Fraction Subtraction Data analyzed using two-parameter SLAM.}
\label{fig-frac}
\end{figure}

\begin{figure}[h!]
\centering

\begin{minipage}[t]{.24\linewidth}
\centering
\includegraphics[width=0.9\textwidth]{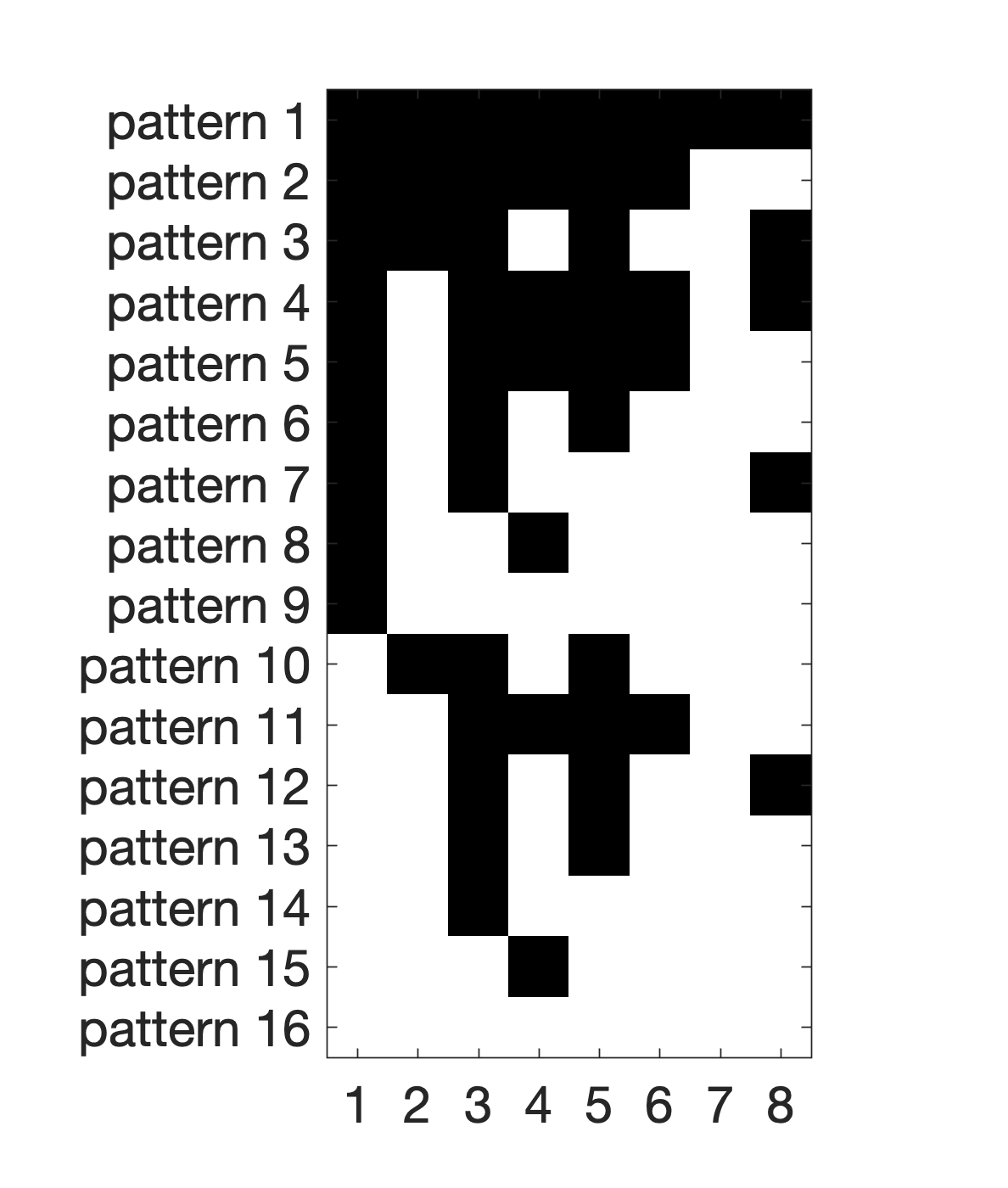}
\end{minipage}%
\hfill
\begin{minipage}[t]{.24\linewidth}
\includegraphics[width=0.9\textwidth]{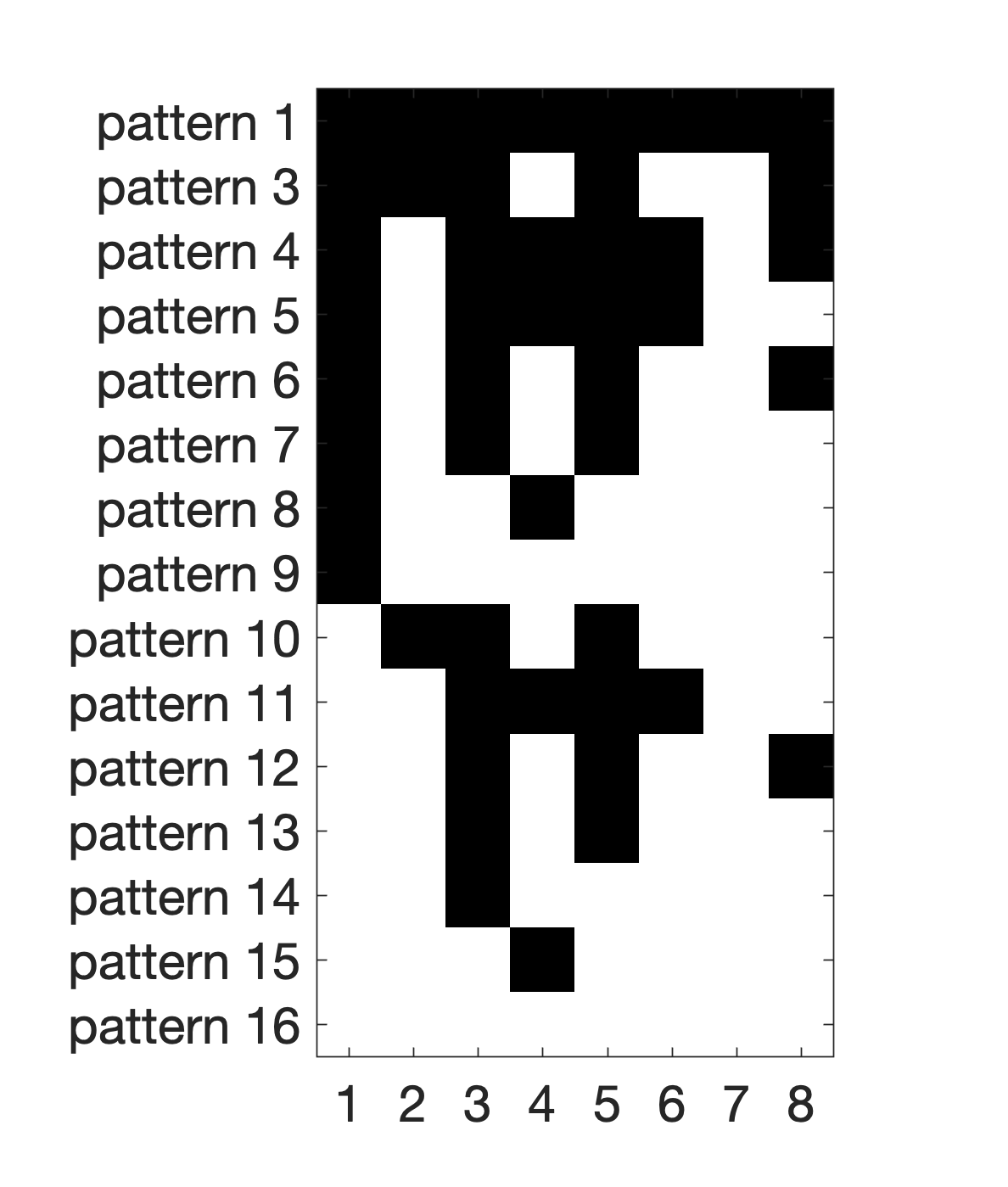}
\end{minipage}%
\hfill
\begin{minipage}[t]{.24\linewidth}
\includegraphics[width=\textwidth]{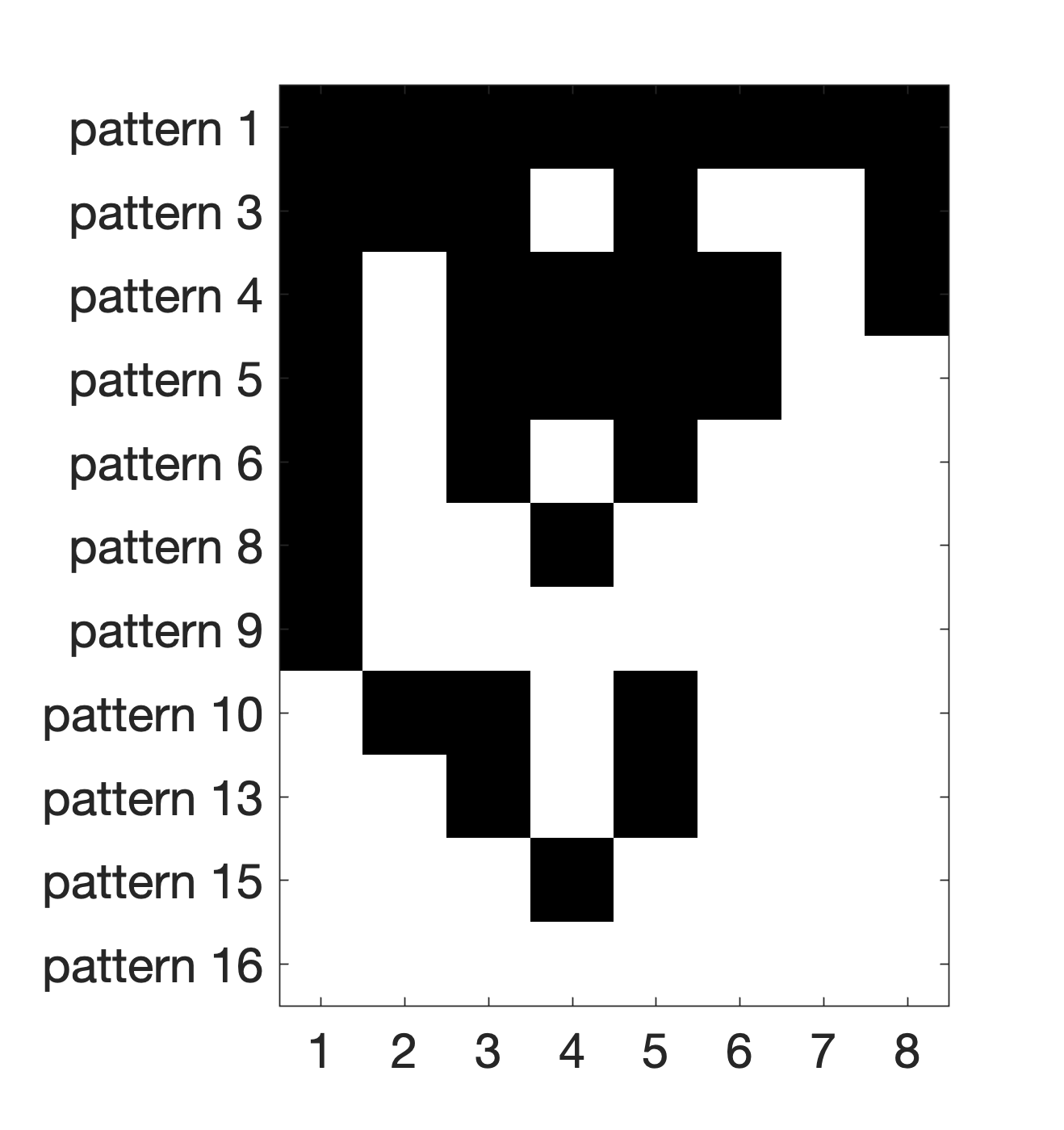}
\end{minipage}%
\hfill
\begin{minipage}[t]{.24\linewidth}
\includegraphics[width=\textwidth]{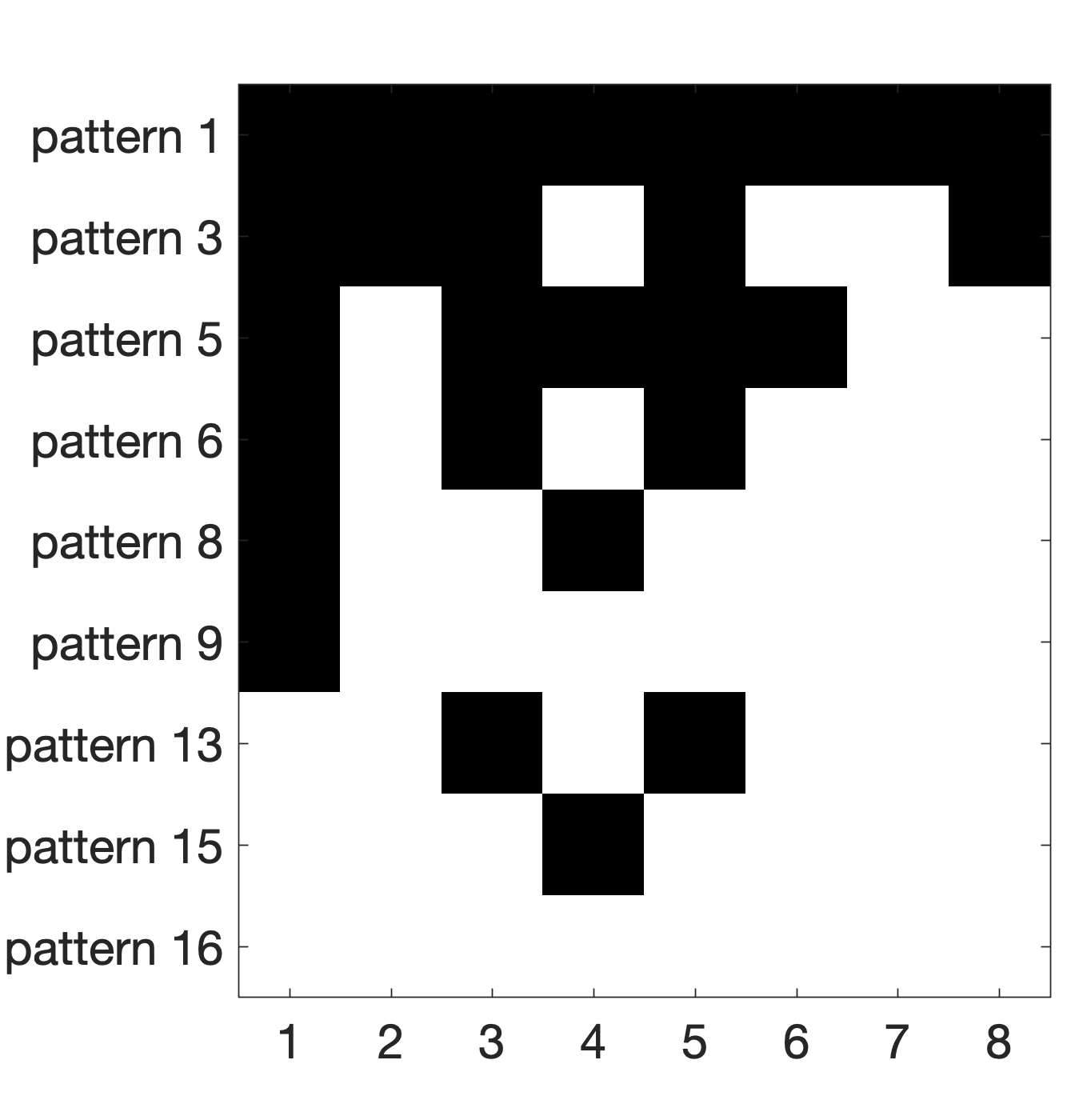}
\end{minipage}%

\begin{minipage}[t]{.24\linewidth}
\centering
\small{(a) $\Upsilon=0.90$,\\ 16 patterns}
\end{minipage}%
\hfill
\begin{minipage}[t]{.24\linewidth}\centering
\small{(b) $\Upsilon=0.86$,\\ 15 patterns}
\end{minipage}%
\hfill
\begin{minipage}[t]{.24\linewidth}\centering
\small{(c) $\Upsilon\in[0.82,0.84]$,\\ 11 patterns}
\end{minipage}%
\hfill
\begin{minipage}[t]{.24\linewidth}\centering
\small{(d) $\Upsilon\in[0.66,0.80]$,\\ 9 patterns}
\end{minipage}%

\begin{minipage}[t]{.24\linewidth}
\centering
\includegraphics[width=0.9\textwidth]{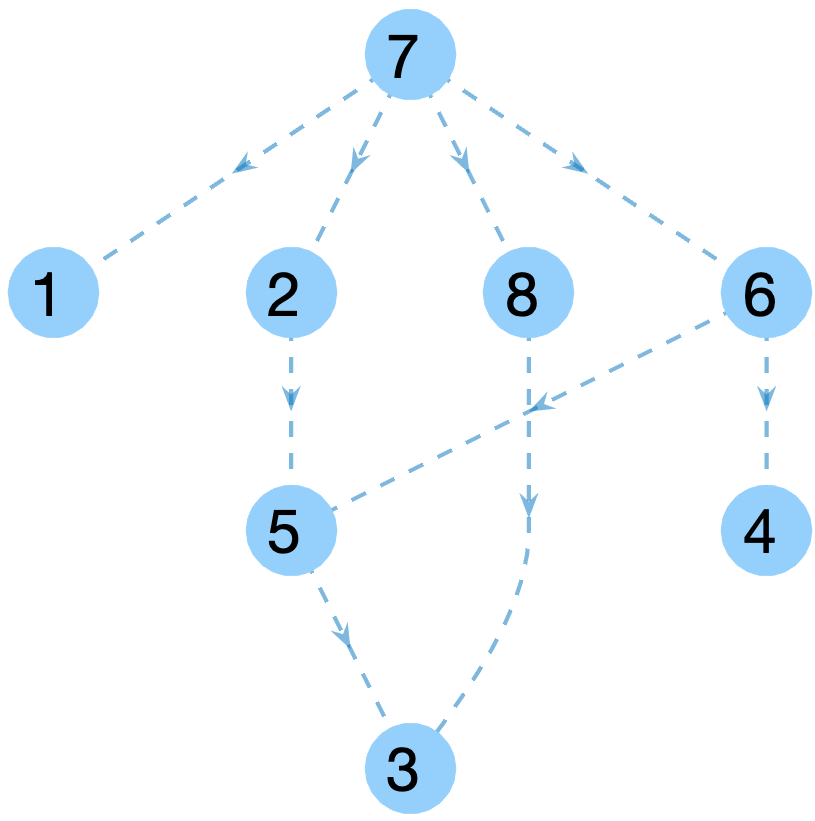}
\end{minipage}%
\hfill
\begin{minipage}[t]{.24\linewidth}
\includegraphics[width=0.9\textwidth]{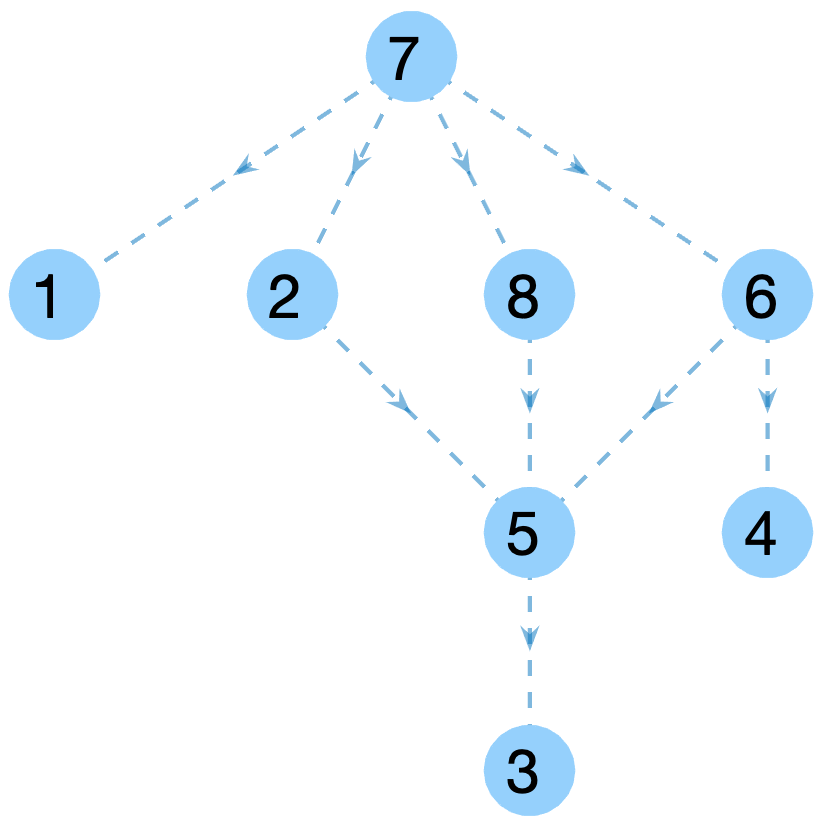}
\end{minipage}%
\hfill
\begin{minipage}[t]{.24\linewidth}
\includegraphics[width=0.95\textwidth]{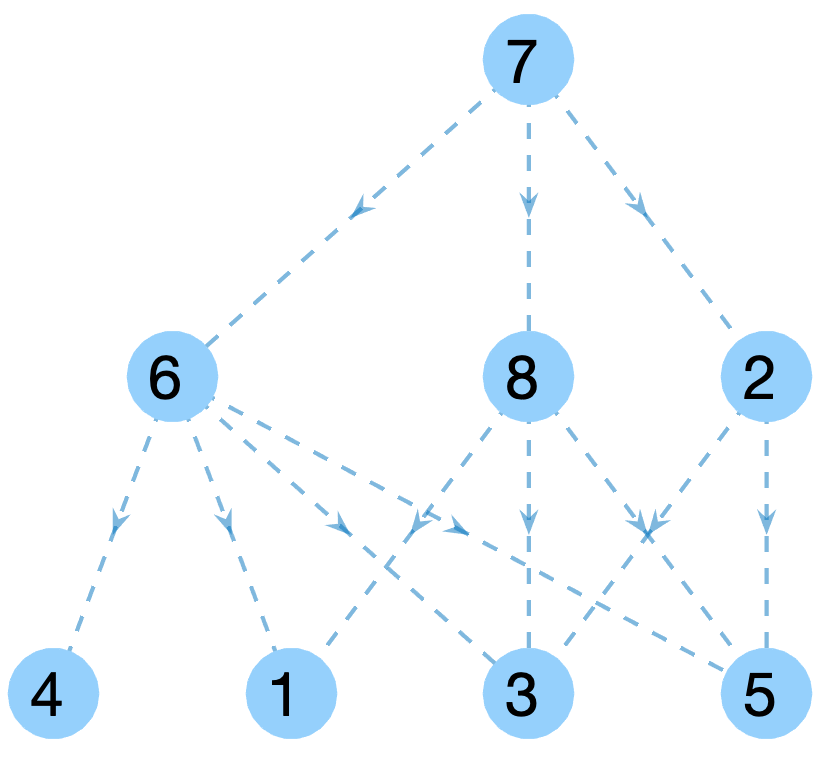}
\end{minipage}%
\hfill
\begin{minipage}[t]{.24\linewidth}
\includegraphics[width=\textwidth]{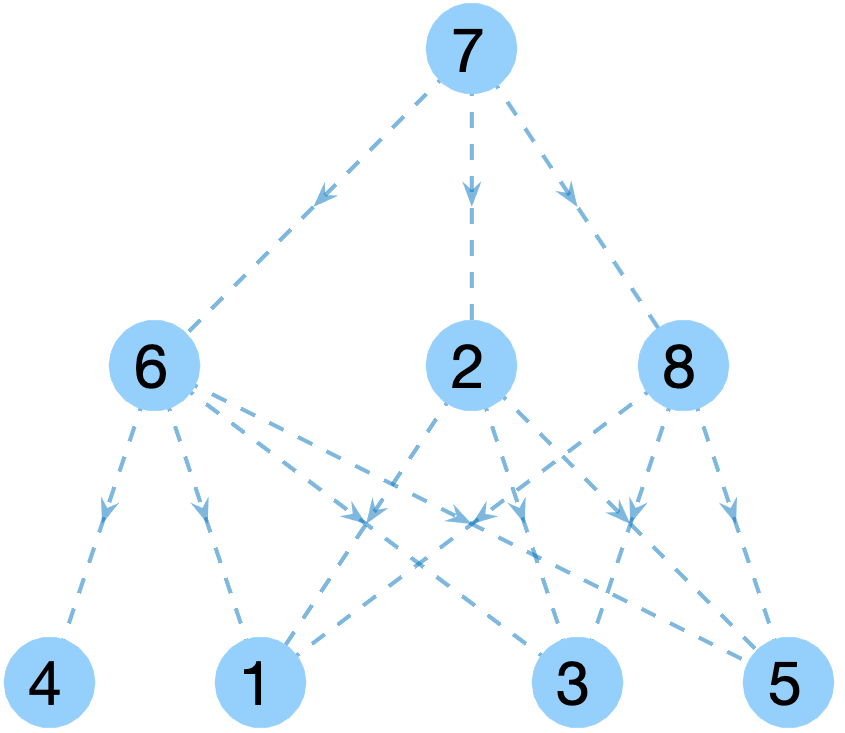}
\end{minipage}%

\begin{minipage}[t]{.24\linewidth}
\centering
\small{(e) $\Upsilon=0.90$,\\ 8 groups of attributes}
\end{minipage}%
\hfill
\begin{minipage}[t]{.24\linewidth}\centering
\small{(f) $\Upsilon=0.86$,\\ 7 groups of attributes}
\end{minipage}%
\hfill
\begin{minipage}[t]{.24\linewidth}\centering
\small{(g) $\Upsilon\in[0.82,0.84]$,\\ 7 groups of attributes}
\end{minipage}%
\hfill
\begin{minipage}[t]{.24\linewidth}\centering
\small{(h) $\Upsilon\in[0.66,0.80]$,\\ 5 groups of attributes}
\end{minipage}%

\caption{
{
Fraction Subtraction Data: different sets of estimated patterns (a)--(d) (black for ``0'' and white for ``1'') and the corresponding attribute structures (e)--(h) under various $\Upsilon$'s in Algorithm \ref{algo-fpvem}. Plot (h) here is equivalent to Figure \ref{fig-frac}(b).
}
}
\label{fig-frac-evolve}
\end{figure}

Many studies in the literature use the two-parameter SLAM to fit the dataset, mostly due to that it is reasonable to assume the required attributes of each item act together to form a ``capable" knowledge state and an ``incapable" knowledge state. This results in two levels of item parameters for each item. We   first use the two-parameter model to analyze the data.  Given this $20\times 8$ $Q$-matrix, the number of equivalence classes induced by the $Q$-matrix $Q_{20\times 8}$ under the two-parameter model is $|\{\vee_{j\in S}\qq_j:\,S\subseteq\{1,\ldots,J\}\}|=58$. 
 We apply Algorithm 2, the FP-VEM algorithm with a sequence of fractional power values $\Upsilon\in\{0.90,0.89,\cdots,0.60\}$ and use  EBIC to select the tuning parameter $\Upsilon$ while keeping the Dirichlet hyper-parameter $\beta=0.01$.  
Figure \ref{fig-frac}(a) plots the EBIC values and the support sizes of $\pp$, both against the $\Upsilon$ values. It can be seen that $\Upsilon = 0.8$ yields the smallest EBIC value $8.98\times 10^{3}$, and it is the largest $\Upsilon$ value in the flat window of $[0.66,0.8]$ that gives 9 equivalence classes of attribute patterns. 
We also use the multi-parameter all-effect model introduced in Example \ref{exp-gdina} to fit the dataset. For a range of values of the tuning parameters $\Upsilon$, the smallest EBIC value is above $1.02\times 10^{4}$, which is much higher than the smallest EBIC $8.98\times 10^{3}$ given by the two-parameter model. This also aligns with the results in the literature that the two-parameter model fits the fraction subtraction dataset better than other models \citep{decarlo2011analysis, dela}.
Therefore next we only present and discuss the results given by the two-parameter model.

Figure \ref{fig-frac}(b) plots the attribute structure corresponding to the 9 equivalence classes of attribute patterns selected by EBIC.
We obtain this attribute structure using the following procedure.
First,  we obtain the representatives of these 9 equivalence classes and construct a $9\times 8$ matrix of selected attribute patterns. We denote this $9\times 8$ matrix by $\widehat \ma$, with each row of $\widehat\ma$   a 8-dimensional binary vector denoting one selected knowledge state.
If $\widehat\ma(\bcdot,k_1)\succeq \widehat\ma(\bcdot,k_2)$, then attribute $k_1$ is considered as a prerequisite for attribute $k_2$.
 Examining these 9 selected knowledge states, 
we find that the total number of 8 attributes are separated into 5 groups $G_1=\{7\}$, $G_2=\{2,8\}$, $G_3=\{6\}$ and $G_4=\{4\}$ and $G_5=\{1,3,5\}$, such that the attributes in the same group play the same role in clustering the students population into the 9 knowledge states. 
In particular, based on the observed data, attributes 2 and 8 are equivalent in distinguishing the students population's knowledge states; and so are attributes 1, 3, 5.
The estimated prerequisite relationship among these 5 groups is depicted in Figure \ref{fig-frac}(b). 
 Figure \ref{fig-frac}(b) implies that attribute ($\alpha_7$) \textit{Subtract numerators}, is a quite basic skill attribute and serves as prerequisite for all the remaining attributes. This suits the common sense that in the problems about fraction and subtraction, the ability of subtracting integers should be the most basic.
 Figure \ref{fig-frac} also shows that attributes ($\alpha_2$), ($\alpha_6$), ($\alpha_8$) are middle level skills that only has one prerequisite attribute ($\alpha_7$), and serve as prerequisites for multiple other skills. Finally, the remaining attributes ($\alpha_4$), ($\alpha_1$), ($\alpha_3$) and ($\alpha_5$) are high level skills in the hierarchical structure.
We would like to point out that the directed edges in the  attribute hierarchy in Figure \ref{fig-frac}(b) (and also in the later Figure \ref{fig-tim} for the TIMSS dataset) do not necessarily correspond to causal relations between the skill attributes. Instead, the attribute hierarchy results from the learned subset of attribute patterns, and it just reflects the estimated cognitive structure of the students  being measured.

For the Fraction Subtraction data, in addition to the attribute structure chosen by EBIC shown in Figure \ref{fig-frac}(b),
we also present those sets of attribute patterns selected by different $\Upsilon$'s in the solution path. The four sets of patterns and their corresponding attribute structures are presented in Figure \ref{fig-frac-evolve}. As shown in Figure \ref{fig-frac-evolve}(a)--(d), the latent patterns selected by a smaller $\Upsilon$ always form a subset of those patterns selected by a larger $\Upsilon$. Also, the attribute structures selected by different $\Upsilon$'s share some commonalities. Among the second row of Figure \ref{fig-frac-evolve}, plot (h) is equivalent to the attribute structure in Figure \ref{fig-frac}(b).
\color{black}

\paragraph{Analysis of TIMSS Data.} We also apply the proposed method to the TIMSS 2003 8th grade data.  The dataset contains $N=757$ students' responses to $J=23$ test items, and the $Q$-matrix is of size $23\times 13$.
Under the two-parameter SLAM, the $Q$-matrix gives  $|\{\vee_{j\in S}\qq_j:\,S\subseteq\{1,\ldots,J\}\}|=1625$ equivalence classes. 
 Figure \ref{fig-tim} shows the results of fitting the two-parameter SLAM with $\beta=0.01$.
 The power parameter $\Upsilon$ selected by EBIC is 0.84 and the corresponding number of equivalence classes is 5. The smallest EBIC value in Figure \ref{fig-tim}(a) is $1.96\times 10^4$.
 We remark here that we also fit the general multi-parameter all-effect SLAM to the dataset, while the smallest EBIC given by the multi-parameter model is $7.38\times 10^4$, which is much larger than the best EBIC given by the two-parameter SLAM. So we next focus on the results given by the two-parameter SLAM.
 
 \begin{figure}[h!]
\begin{subfigure}{0.48\textwidth}
\includegraphics[width=\linewidth]{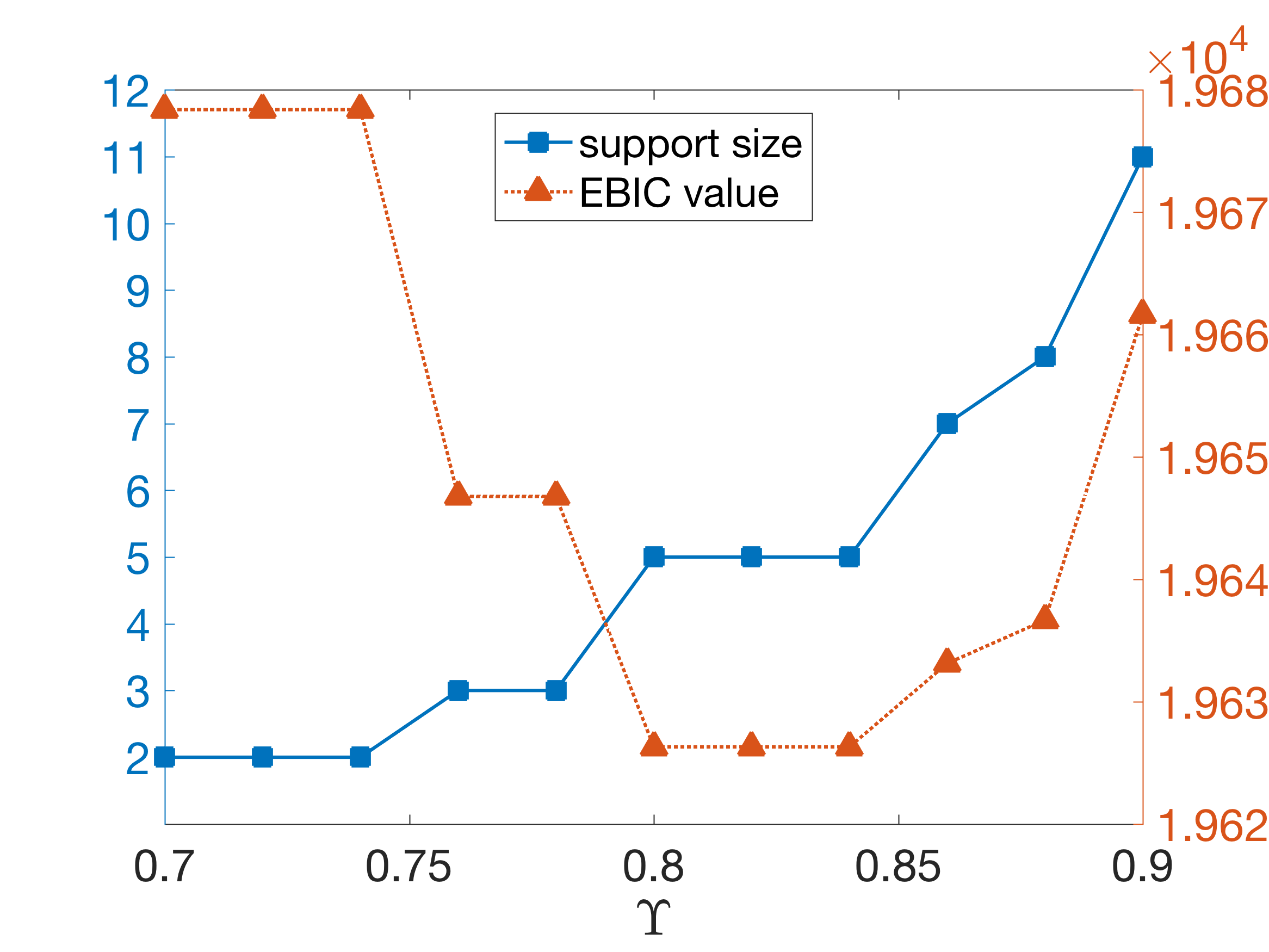}
\end{subfigure}\hspace*{\fill}
\begin{subfigure}{0.47\textwidth}
\centering
\begin{tikzpicture}[scale=0.75]
\draw [group1] (-1.5,5.5) rectangle (1.5,6.5); 

\draw [group2] (-3,3.5) rectangle (-1,4.5); 

\draw [group3] (-0.5,3.5) rectangle (3.5,4.5); 

\draw [group5] (0,1.5) rectangle (3,2.5); 

\draw [group4] (-0.5,-0.5) rectangle (0.5,0.5); 

\path
node at ( 0,0) [place] (a4) {4}

node at ( 0.5,2) [place] (a1) {1}
node at ( 1.5,2) [place] (a2) {2}
node at ( 2.5,2) [place] (a8) {8}

node at ( 0,4) [place] (a6) {6}
node at ( 1,4) [place] (a7) {7}
node at ( 2,4) [place] (a10) {10}
node at ( 3,4) [place] (a12) {12}

node at ( -2.5,4) [place] (a5) {5}
node at ( -1.5,4) [place] (a9) {9}

node at ( -1,6) [place] (a3) {3}
node at ( 0,6) [place] (a11) {11}
node at ( 1,6) [place] (a13) {13};

\draw [->, dotted, thick] (-0.1,5.5) to (-1.5,4.5); 
\draw [->, dotted, thick] (0,5.5) to (1.5,4.5); 

\draw [->, dotted, thick] (1.5,3.5) to (1.5,2.5); 

\draw [->, dotted, thick] (1.5,1.5) to (0,0.5); 

\draw [->, dotted, thick] (-1.5,3.5) to (0,0.5); 

\end{tikzpicture}
\end{subfigure}
\begin{minipage}[t]{.48\linewidth}
        \caption*{(a) EBIC values and support sizes versus $\Upsilon$}
\end{minipage}%
\hspace*{\fill}
\begin{minipage}[t]{.47\linewidth}
        \caption*{(b) attribute structure selected by EBIC}
\end{minipage}%
\caption[]{Results of TIMSS 2003 8th Grade Data analyzed using two-parameter SLAM.}
\label{fig-tim}
\end{figure}

Figure \ref{fig-tim}(b) plots the attribute structure given by the selected 5 knowledge states.
The 13 attributes are separated into five groups $G_1=\{3,11,13\}$, $G_2=\{5,9\}$, $G_3=\{6,7,10,12\}$ and $G_4=\{1,2,8\}$ and $G_5=\{4\}$, such that the attributes in the same group play the same role in clustering the students population into the five knowledge states. The prerequisite relationships among groups of attributes is also shown in  Figure \ref{fig-tim}(b).
Attribute $(\alpha_3)$ \textit{compute fluently with multi-digit numbers and find common factors and multiples}, attribute $(\alpha_{11})$ \textit{compare two fractions with different numerators and different denominators}, attribute ($\alpha_{13}$) \textit{use equivalent fraction as a strategy to add and subtract fractions}, are the most basic skills in the attribute hierarchy and serve as the prerequisites for all the remaining attributes. Indeed, these three are basic algorithmic operations   needed to solve the mathematical problems in the TIMSS test. {In addition to the structure selected by EBIC presented in Figure \ref{fig-tim}(b), other attribute structures corresponding to different $\Upsilon\in[0.7,0.9]$ are presented in Figure \ref{fig-timss-evolve} in Appendix A.2.}

\medskip
Existing works in the literature analyzing the fraction subtraction data and the TIMSS data either make the assumption that all possible configurations of latent attribute patterns exist in the population  or pre-specify the attribute structure  based on domain experts' judgements \citep{su2013hierarchical}. To our knowledge, there has not been a systematic approach to selecting a potentially small set of latent patterns from a high-dimensional space.   For the two real datasets, we also find that the EBIC values of the existing EM algorithm  are much larger than the proposed method, as indicated in Figures \ref{fig-frac} and \ref{fig-tim} when $\Upsilon$ close to 1; thus the proposed method provides a better fit of the two datasets. 
\color{black}

\section{Discussion}

In this paper we propose a penalized likelihood method to learn the attribute patterns in the structured latent attribute models, a special family of discrete latent variable models. 
We allow the number of latent patterns to go to infinity and perform pattern selection by penalizing the proportion parameters of the latent attribute patterns.
 The theory of pattern selection consistency is established for the proposed regularized MLE. 
The nice form of the penalty term facilitates the computation. Two algorithms are developed to solve the optimization problem, one being a modification of the EM algorithm, and the other being a variational EM algorithm that results from an alternative formulation of the objective function. The simulation study and real data analysis show the proposed methods have good pattern selection performance.

 This work assumes the design $Q$-matrix is prespecified and correct. 
In practice, if there is reason to suspect that the   $Q$-matrix could be misspecified, then one needs to simultaneously estimate the $Q$-matrix and learn the attribute patterns from data. 
Given fixed number of attribute patterns, previous works including \cite{xu2018} and \cite{chen2018} used the likelihood based methods and the Bayesian methods, respectively, to estimate $Q$. 
It is also desirable to develop  methods to jointly estimate $Q$ and learn attribute patterns with the existence of large number of attributes. We would like to point out that the identifiability results developed in this work (in Section \ref{sec-id}) directly apply to this case, and can guarantee both the design matrix $Q$ and the set of significant attribute patterns are learnable from data.

 The learnability theory developed in this paper guarantees one can reliably learn a SLAM with an arbitrary set of attribute patterns from data.
As mentioned earlier, SLAMs can be expressed as higher-order  probability tensors with special structures.
 Also, SLAMs share similarities with the restricted Boltzmann machines and the deep Boltzmann machines in the bipartite graph structure among the latent and observed multivariate binary variables.
 Current techniques for proving identifiability of SLAMs could be adapted to develop theory for uniqueness of structured tensor decompositions and learnability of some more complicated latent variable models.
We leave these directions for future study.

\section*{Appendix}

\subsection*{A.1 Technical Proofs}

We introduce a useful notation, the $T$-matrix, before proving the identifiability theory.
We consider a marginal probability matrix $T(\Gamma^{\mca},\, \TT^{\mca})$ of size $2^J\times |\mca|$ as follows. When it causes no confusion, we also write $T(\Gamma^{\mca},\, \TT^{\mca})$ simply as $T(\Gamma, \TT)$. Rows of $T(\Gamma,\TT)$ are indexed by the $2^J$ possible response patterns $\rr = (r_1,\ldots,r_J)^\top\in\{0,1\}^J$ and columns of $T(\Gamma, \TT)$ are indexed by latent attribute patterns $\aaa\in\mca$, while the $(\rr, \aaa)$th entry of $T(\Gamma, \TT)$, denoted by $T_{\rr, \aaa}(\Gamma, \TT)$, represents the marginal probability that subjects in latent class $\aaa$ provide positive responses to the set of items $\{j: r_j=1\}$, namely 
$T_{\rr,\aaa}(\Gamma, \TT) = P(\boldsymbol R \succeq \rr\mid\TT, \aaa) = \prod_{j=1}^J \theta_{j,\aaa}^{r_j}.$
Denote the $\aaa$th column vector and the $\rr$th row vector of the $T$-matrix by $T_{\Cdot,\aaa}(\Gamma, \TT)$ and $T_{\rr,\Cdot}(\Gamma, \TT)$ respectively. Let $\ee_j$ denote the $J$-dimensional unit vector with the $j$th element being one and all the other elements being zero, then any response pattern $\rr$ can be written as a sum of some $\ee$-vectors, namely $\rr=\sum_{j:r_j=1}\ee_j$.
The $\rr$th element of the $2^J$-dimensional vector $T(\Gamma, \TT)\pp$ is
$
\{T(\Gamma, \TT)\pp\}_{\rr}= T_{\rr,\Cdot}(\Gamma, \TT)\pp = \sum_{\aaa\in\mca} T_{\rr,\aaa}(\Gamma, \TT) p_{\aaa} = P(\boldsymbol R \succeq \rr \mid \Gamma,\TT).
$
The $T$-matrix have some nice algebraic properties that will be useful in later proofs. We state them in the following lemma, the proof of which is similar to that of Proposition 3 in \cite{xu2017} and hence is omitted.

\begin{lemma}\label{lem}
Under a SLAM with constraint matrix $\Gamma$, $(\Gamma,\TT, \pp)$ are jointly identifiable if and only if for any $(\Gamma,\TT, \pp)$ and  $(\bar \Gamma,\bar\TT, \bar\pp)$, 
\begin{equation}\label{eq1}
T(\Gamma,\TT)\pp = T(\bar \Gamma, \bar\TT)\bar\pp
\end{equation}
implies $(\Gamma,\TT,\pp) = (\bar \Gamma,\bar\TT,\bar\pp)$.
For any $\ttt^*=(\theta_1,\ldots,\theta_J)^\top\in\mathbb R^J$, there exists an invertible matrix $D(\ttt^*)$  only depending on $\ttt^*$, such that
\begin{equation}\label{eq-trans}
T(\Gamma,\TT-\ttt^*\mathbf1^\top) = D(\ttt^*)T(\Gamma,\TT),
\end{equation}
where $\one^\top$ denotes an all-one vector and $\ttt^*\mathbf1^\top$ is a matrix of same size as $\TT$.
\end{lemma}

\paragraph{Proof of Theorem \ref{thm-id} and Corollary \ref{cor1}.}
We aim to prove that if $\Gamma:=\Gamma^{\mca_0}$ of size $J\times L_0$ ($L_0=|\mca_0|$) satisfies Conditions $A$ and $B$, then for any binary matrix $\bar\Gamma$ also of size $J\times L_0$, which can be viewed as a constraint matrix imposing restrictions on the parameter space of the $J\times L_0$ item parameter matrix $\bar \TT$, and for any $L_0$-dimensional vector $\bar\pp:=(\bar p_{1},\ldots,\bar p_{L_0})$ with $\bar p_{l}\geq 0$ and $\sum_{l=1}^{L_0} \bar p_l=1$, which can be viewed as a population proportion vector giving proportions of the $L_0$ latent classes, if
\begin{equation}\label{eq-tp}
T(\Gamma,\TT)\pp = T(\bar\Gamma,\bar\TT)\bar\pp
\end{equation}
holds, then $(\Gamma,\TT,\pp) = (\bar\Gamma,\bar\TT,\bar\pp)$ up to a label swapping of the latent classes.
If this is proved, then combining Condition $C$ that any column vector of $\Gamma^{\mca_0}$ is different from any column vector of $\Gamma^{\mca_0^c}$, we would have the conclusion that the identified $\Gamma^{\mca_0}$ uniquely maps to the true set of attribute patterns $\mca_0$.

We add a remark here that  given \eqref{eq-tp}, the columns of the $\bar \Gamma$ do not necessarily have the interpretation of representing some $K$-dimensional binary attribute patterns; instead, these columns  just correspond to $L_0$ latent classes. And after we obtain $(\Gamma,\TT,\pp) = (\bar\Gamma,\bar\TT,\bar\pp)$ up to a label swapping, we would have the conclusion that $\bar\Gamma$ equals $\Gamma$ up to column permutation; Then with Condition $C$, the $\bar\Gamma$ would have the interpretation of being the constraint matrix for the attribute patterns in $\mca_0$. Because of this, in the following proof, we sometimes will also ignore the interpretation of the columns of the true $\Gamma^{\mca_0}$, and simply denote the columns of it  by the column index integer $l$, i.e., $\Gamma^{\mca_0}$ has columns $\Gamma^{\mca_0}_{\Cdot,l}$ for $l=1,\ldots,L_0$.

For notational simplicity, we denote $\Gamma^{({S_i},\mca_0)}$ by $\Gamma^i$ for $i=1,2$ and $\Gamma^{({(S_1\cup S_2)^c},\mca_0)}$ by $\Gamma^3$. 
We also denote item parameter matrix $\TT^{({S_1},\mca_0)}$, $\TT^{({S_2},\mca_0)}$ and $\TT^{({(S_1\cup S_2)^c},\mca_0)}$ by $\TT^1$, $\TT^2$ and $\TT^3$, respectively. So each $\TT^i$ has the same size as $\Gamma^i$ and respects the constraints specified by $\Gamma^i$.
Without loss of generality, suppose $\Gamma$ takes the form
 $
\Gamma^\top = [
(\Gamma^1)^\top,
(\Gamma^2)^\top, 
(\Gamma^3)^\top
],
 $ 
where each $\Gamma^i$ is of size $J_i\times L_0$ and $J_1+J_2+J_3 = J$.
For any item $j$, by the definition of SLAM we have all those $\aaa$ with $\Gamma^{\mca_0}_{j,\aaa}=1$ have the same highest value of item parameter. For simplicity, we denote this value of the item parameter by $\theta_{j,H}$, where ``$H$" stands for ``highest" level item parameter for item $j$.

\bigskip

We first show $T(\bar\Gamma^1,\bar\TT^1)$ and $T(\bar\Gamma^2,\bar\TT^2)$ both have full column rank $L_0$, and that $\bar p_l>0$ for all $l\in\{1,\ldots,L_0\}$. 
By Proposition 3 in \cite{partial}, Condition $A$ ensures that $T(\Gamma^1,\TT^1)$ of size $2^{J_1}$ and $T(\Gamma^2,\TT^2)$ of size $2^{J_2}$ both have full column rank $L_0$, since $\Gamma^1$ and $\Gamma^2$ are both separable. Moreover, in the proof of that conclusion,  an invertible square matrix $W_1$ of size $2^{J_1}\times 2^{J_1}$ as well as $L_0$ response patterns $\rr_1,~\ldots,~\rr_{L_0}\in\{0,1\}^L$ were constructed such that the row vectors in the transformed $W_1\boldsymbol\cdot T(\Gamma^1,\TT^1)$, which are indexed by the chosen $\rr_1,~\ldots,~\rr_{L_0}$, form a $L_0\times L_0$ lower triangular matrix with nonzero diagonal elements. In other words, in the $2^{J_1}\times L_0$ rectangular matrix $W_1 T(\Gamma^1,\TT^1)$, there is a $L_0\times L_0$ submatrix that is lower triangular and full-rank. For notational simplicity, we denote this submatrix by $\{W_1 T(\Gamma^1,\TT^1)\}_{\rr_{1:L_0}}$. Similarly, there exists $W_2$ and $\rr'_1,~\ldots,~\rr'_{L_0}\in\{0,1\}^{L_0}$ such that there is a $L\times L$ full-rank submatrix of $W_2 T(\Gamma^2,\TT^2)$ with rows indexed by $\rr'_1,~\ldots,~\rr'_{L_0}$, which we denote by $\{W_2 T(\Gamma^2,\TT^2)\}_{\rr'_{1:L_0}}$.

Based on the above constructions, there exist two invertible square matrices $U_1$ and $U_2$ such that
$U_1\boldsymbol\cdot \{W_1 T(\Gamma^1,\TT^1)\}_{\rr_{1:L_0}} = I_{L_0}$ and  $U_2\boldsymbol\cdot \{W_2 T(\Gamma^2,\TT^2)\}_{\rr'_{1:L_0}} = I_{L_0}$.
Denote the $C$ row vectors of $U_1$ by $\{\uu_{l}^\top,~l\in[L_0]\}$, then we have that for any $l\in[L_0]$,
\begin{equation}\label{eq-ua}
\uu_{l}^\top\boldsymbol\cdot \{W_1 T(\Gamma^1,\TT^1)\}_{\rr_{1:L_0}} = (\mz, \underbrace{1}_{\text{column}~ l},\mz).
\end{equation}
Next we prove by contradiction that $\{W_1 T(\bar\Gamma^1,\bar\TT^1)\}_{\rr_{1:{L_0}}}$ and $\{W_2 T(\bar\Gamma^2,\bar\TT^2)\}_{\rr'_{1:{L_0}}}$ must also be invertible. We focus on $\{W_2 T(\bar\Gamma^2,\bar\TT^2)\}_{\rr'_{1:{L_0}}}$ and conclusion for the other is the same.
If $\{W_2 T(\bar\Gamma^2,\bar\TT^2)\}_{\rr'_{1:L_0}}$ does not have full rank, then $U_2\bcdot\{W_2 T(\bar\Gamma^2,\bar\TT^2)\}_{\rr'_{1:L_0}}$ also does not have full rank, so there exists a nonzero vector $\boldsymbol x = (x_1,\ldots,x_{L_0})$ such that
$$
\boldsymbol x^\top \bcdot U_2 \bcdot \{W_2 T(\bar\Gamma^2,\bar\TT^2)\}_{\rr'_{1:L_0}} = \zero.
$$
Note that 
$\boldsymbol x^\top\bcdot U_2\bcdot \{W_2 T(\bar\Gamma^2,\bar\TT^2)\}_{\rr'_{1:L_0}} = \boldsymbol x$
 from the previous construction of $W_2$. Since $\boldsymbol x\neq \zero$, suppose without loss of generality that $x_l\neq 0$ for some $l$, then we have
\begin{align*}
&[\uu_{\aaa}^\top\boldsymbol\cdot \{W_1 T(\Gamma^1,\TT^1)\}_{\rr_{1:L_0}} ] \odot [\boldsymbol x^\top \bcdot U_2 \bcdot \{W_2 T(\Gamma^2,\TT^2)\}_{\rr'_{1:L_0}}]\boldsymbol\cdot \pp = x_{l}p_{l}\neq0,\\
&[\uu_{\aaa}^\top\boldsymbol\cdot \{W_1 T(\bar\Gamma^1,\bar\TT^1)\}_{\rr_{1:L_0}} ] \odot [\boldsymbol x^\top \bcdot U_2 \bcdot \{W_2 T(\bar\Gamma^2,\bar\TT^2)\}_{\rr'_{1:L_0}}]\boldsymbol\cdot \bar \pp=0,
\end{align*}
which contradicts \eqref{eq-tp}. Here $\boldsymbol a\odot \boldsymbol b$ denotes the elementwise product of two vectors $\boldsymbol a$ and $\boldsymbol b$ of the same length. Therefore $\{W_2 T(\bar\Gamma^2,\bar\TT^2)\}_{\rr'_{1:L_0}}$ must have full rank $C$, and so as $\{W_1 T(\bar\Gamma^1,\bar\TT^1)\}_{\rr_{1:L_0}}$.

{Based on the above conclusion, we next show that $\bar p_{l}>0$ for any $l\in[L_0]$.} Suppose this is not true and $\bar p_{l}=0$ for some $l$, then there exists a nonzero vector $\boldsymbol y  = (y_{1},\ldots,y_{L_0})^\top$ such that
$$
\boldsymbol y^\top \bcdot \{W_2 T(\bar\Gamma^2,\bar\TT^2)\}_{\rr'_{1:L_0}} =  (\zero, \underbrace{1}_{\text{column}~ l},\zero).
$$
Since $\{W_2 T(\Gamma^2,\TT^2)\}_{\rr'_{1:L_0}}$ has full rank and $\boldsymbol y\neq\zero$, we have $\boldsymbol y^\top \bcdot \{W_2 T(\Gamma^2,\TT^2)\}_{\rr'_{1:L_0}}\neq\zero$. Without loss of generality, suppose the $l^\star$-th column of this product vector is nonzero and denote the nonzero value by $b_{l^\star}$, then using the  $\uu$-vectors constructed previously in \eqref{eq-ua}, we have
\begin{align*}
&[\uu_{\aaa^\star}^\top\boldsymbol\cdot \{W_1 T(\Gamma^1,\TT^1)\}_{\rr_{1:L_0}} ] \odot [\boldsymbol y^\top \bcdot \{W_2 T(\Gamma^2,\TT^2)\}_{\rr'_{1:L_0}}]\boldsymbol\cdot \pp 
= b_{l^\star} p_{l^\star}\neq0,\\
&[\uu_{\aaa^\star}^\top\boldsymbol\cdot \{W_1 T(\bar\Gamma^1,\bar\TT^1)\}_{\rr_{1:L_0}} ] \odot [\boldsymbol y^\top \bcdot \{W_2 T(\bar\Gamma^2,\bar\TT^2)\}_{\rr'_{1:L_0}}]\boldsymbol\cdot \bar \pp=0,
\end{align*}
which contradicts \eqref{eq-tp}. This shows that $\bar p_{l}>0$ must hold for all $l\in[L_0]$.

\bigskip
We next show that for any $j\in(S_1\cup S_2)^c$ and any $l\in\{1,\ldots,L_0\}$,   $\theta_{j,l} = \theta_{j,\sigma(l)}$, where $\sigma(\cdot)$ is a permutation map from $\{1,\ldots,L_0\}$ to $\{1,\ldots,L_0\}$.
There must exist a permutation map $\sigma:\{1,\ldots,L\}\to \{1,\ldots,L\}$ such that for each $l\in[L_0]$, 
\begin{equation}\label{eq-sigma}
\bar f_{\sigma(l)} := [\uu_{l}^\top\boldsymbol\cdot \{W_1 T(\bar\Gamma^1,\bar\TT^1)\}_{\rr_{1:L_0}} ]_{\sigma(l)}\neq 0.
\end{equation}
This is because otherwise there would exist  $l\in[L_0]$ such that $\{U_1\boldsymbol\cdot T(\bar\Gamma^1,\bar\TT^1)\}_{\bcdot,l}$ equals the zero vector, which contradicts the fact that both $U_1$ and $\{W_1 T(\bar\Gamma^1,\bar\TT^1)\}_{\rr_{1:L_0}}$ are invertible matrices.
Given the permutation $\sigma$, there exists a $L_0\times L_0$ invertible matrix $V$ with row vectors denoted by $\{\vv_{l},~l\in[L_0]\}$ such that for each $\aaa\in\mca$, 
\begin{align}\label{eq-va}
\vv_{l}^\top \boldsymbol\cdot \{W_2 T(\bar\Gamma^2,\bar\TT^2)\}_{\rr'_{1:L_0}} = (\mz, \underbrace{1}_{\text{column}~ \sigma(l)},\mz).
\end{align}
Then we have 
\begin{align}
&[\uu_{l}^\top \boldsymbol\cdot \{W_1 T(\Gamma^1,\TT^1)\}_{\rr_{1:L_0}}]\odot
[\vv_{l}^\top \boldsymbol\cdot \{W_2 T(\Gamma^2,\TT^2)\}_{\rr'_{1:L_0}}]\bcdot \pp = f_{l}p_{l},\label{eq-r1}\\
&[\uu_{l}^\top \boldsymbol\cdot \{W_1 T(\bar\Gamma^1,\bar\TT^1)\}_{\rr_{1:L_0}}]\odot
[\vv_{l}^\top \boldsymbol\cdot \{W_2 T(\bar\Gamma^2,\bar\TT^2)\}_{\rr'_{1:L_0}}]\bcdot \bar\pp = \bar f_{\sigma(l)}\bar p_{\sigma(l)}\neq0, \label{eq-r2}
\end{align}
where $f_{l}=[\vv_{l}^\top \boldsymbol\cdot \{W_2 T(\Gamma^2,\TT^2)\}_{\rr'_{1:L_0}}]_{l}$. Now we have $f_{l}p_{l} = \bar f_{\sigma(l)}\bar p_{\sigma(l)}\neq0$. Next further consider an arbitrary item $j\in (S_1\cup S_2)^c$. Equation \eqref{eq-tp} indicates that
$$
\theta_{j,l}=\frac{T_{\ee_j,\bcdot}(\Gamma,\TT)\odot\eqref{eq-r1} }{\eqref{eq-r1}} = 
\frac{T_{\ee_j,\bcdot}(\bar\Gamma,\bar\TT)\odot\eqref{eq-r2} }{\eqref{eq-r2}}
=\bar\theta_{j,\sigma(l)}.
$$

We next show that for any $j\in S_1\cup S_2$ and any $l\in\{1,\ldots,L_0\}$ such that $\Gamma_{j,l}=1$,   $\theta_{j,l} = \theta_{j,H} = \bar \theta_{j,\sigma(l)} = \bar\theta_{j,H}$.
We introduce a lemma before proceeding with the proof.
\begin{lemma}\label{lem-order-a} 
Under the assumptions of Theorem \ref{thm-id}, 
the vectors $\{\vv_{l},~l\in\mca_0\}$ constructed in \eqref{eq-va} satisfy that
\begin{equation}\label{eq-order}
\begin{aligned}
\{\vv_{l}^\top \cdot \{W_2 T(\Gamma^2,\TT^2)\}_{\rr'_{1:L_0}}\}_{l'} &= 0,\quad \forall\aaa_{l'}\precneqq_{S_1}\aaa_l~\text{under}~\Gamma^{\mca_0}.\\\end{aligned}
\end{equation}
\end{lemma}
\paragraph{Proof of Lemma \ref{lem-order-a}}
If $\{\vv_{l}^\top \cdot \{W_2 T(\Gamma^2,\TT^2)\}_{\rr'_{1:L_0}}\}_{l'}=z_{l'}\neq 0$, then similar to \eqref{eq-r1} and \eqref{eq-r2} we have
\begin{align*}
&[\uu_{l'}^\top \boldsymbol\cdot \{W_1 T(\Gamma^1,\TT^1)\}_{\rr_{1:L_0}}]\odot
[\vv_{l}^\top \boldsymbol\cdot \{W_2 T(\Gamma^2,\TT^2)\}_{\rr'_{1:L_0}}]\bcdot \pp = z_{l'}p_{l'}\neq 0,\\
&[\uu_{l'}^\top \boldsymbol\cdot \{W_1 T(\bar\Gamma^1,\bar\TT^1)\}_{\rr_{1:L_0}}]\odot
[\vv_{\aaa}^\top \boldsymbol\cdot \{W_2 T(\bar\Gamma^2,\bar\TT^2)\}_{\rr'_{1:L_0}}]\bcdot \bar\pp = \bar f_{\sigma(l)}\bar p_{\sigma(l)},
\end{align*}
and further we have $\theta_{j,l'} = \bar\theta_{j,\sigma(l)} = \theta_{j,l}$ for $j\in(S_1\cup S_2)^c$, which contradicts condition (C2). This completes the proof of the lemma.
\qed

\medskip
We proceed with the proof.
For any $l\in[L_0]$, define $\ttt^* = \sum_{h\in S_1:\, \Gamma_{h,l}=1} \theta_{h,1}\ee_h$.
With $\ttt^*$, the row vector corresponding to $\rr^*=\sum_{h\in S_1:\Gamma_{h,l}=0}\ee_h$ in the transformed $T$-matrix  
satisfies that 
\begin{align}\label{eq-ba}
&b_{l}:=T_{\rr^*,l}(\Gamma^1,\TT^1-\ttt^*\one^\top)\neq 0;\\ \notag
&T_{\rr^*,l'}(\Gamma^1,\TT^1-\ttt^*\one^\top)= 0,\quad\forall \aaa_{l'}\npreceq_{S_1}\aaa_l
~\text{under}~\Gamma^{\mca_0}.
\end{align}
The proof of Step 2 as well as Lemma \ref{lem-order-a} ensures
\begin{align}\label{eq-fa}
 f_{l}=
&[\vv^\top_{l}\cdot \{W_2 T(\Gamma^2,\TT^{2})\}_{\rr'_{1:L_0}}]_{l} \neq 0;\\ \notag
&  [\vv^\top_{l}\cdot \{W_2T(\Gamma^2,\TT^{2})\}_{\rr'_{1:L_0}}]_{l'} = 0,\quad \forall\aaa_{l'}\preceq_{S_1}\aaa_l~\text{under}~\Gamma^{\mca_0}.
\end{align}
Consider any $j\in S_1\cup S_2$ such that $\Gamma_{j,l}=1$, then obviously $\ee_j$ is not included in the sum in the previously defined response pattern $\rr^*$, because $\rr^*$ only contains those items that $\aaa_l$ is not capable of, i.e., those $j$ s.t. $\Gamma^{\mca_0}_{j,l}=0$. The above two equations \eqref{eq-ba} and \eqref{eq-fa} indicate
\begin{equation}\label{eq-s3-1}
\begin{aligned}
& T_{\rr^*,\cdot}(\Gamma^1,\TT^1-\ttt^*\one^\top)
\odot
[\vv^\top_{l}\cdot \{W_2T(\Gamma^2,\TT^{2})\}] =\Big( \mz^\top, 
\underbrace{b_{l}\cdot f_{l}}_\text{column $l$},
 \mz^\top \Big),
\end{aligned}\end{equation}
\begin{equation}\label{eq-s3-2}
\begin{aligned}
& T_{\rr^*+\ee_j,\cdot}(\Gamma^1,\TT^1-\ttt^*\one^\top)
\odot
[\vv^\top_{l}\cdot \{W_2T(\Gamma^2,\TT^2)\}] =\Big( \mz^\top, 
\underbrace{\theta_{j,H}\cdot b_{l}\cdot f_{l}}_\text{column $l$},
 \mz^\top \Big).
\end{aligned}\end{equation}
Similarly for $(\bar\TT,\bar\pp)$ we have
\begin{align}\label{eq-s3-3}
& T_{\rr^*,\cdot}(\bar\TT-\ttt^*\one^\top)
\odot
\{\vv^\top_{l}\cdot T(\bar\TT^2)\} =\Big( \mz^\top, 
\underbrace{\prod_{h\in S_1:\Gamma_{h,l}=0}(\bar\theta_{h,\sigma(l)} - \theta_{h,H})}_\text{column $\sigma(l)$},
 \mz^\top \Big),
\end{align}
\begin{align}\label{eq-s3-4}
& T_{\rr^*+\ee_j,\cdot}(\bar\TT-\ttt^*\one^\top)
\odot
\{\vv^\top_{l}\cdot T(\bar\TT^2)\} =\Big( \mz^\top, \underbrace{\bar\theta_{j,H}\cdot\prod_{h\in S_1:\Gamma_{h,l}=0}
(\bar\theta_{h,\sigma(l)} - \theta_{h,H})}_\text{column $\sigma(l)$}, \mz^\top \Big).
\end{align}
Equation \eqref{eq-tp} implies $(\ref{eq-s3-1})\cdot\pp = (\ref{eq-s3-3})\cdot\bar\pp$. 
By (\ref{eq-tp}), the above four equations give that
\[
\theta_{j,H} = \theta_{j,l} =
\frac{(\ref{eq-s3-2})\cdot\pp}{(\ref{eq-s3-1})\cdot\pp} = \frac{(\ref{eq-s3-4})\cdot\bar\pp}{(\ref{eq-s3-3})\cdot\bar\pp}
= \bar\theta_{j,\sigma(l)} = \bar\theta_{j,H},\quad \forall j\in S_2.
\]
Note that the above equality $\theta_{{j},H} = \bar\theta_{{j},H}$ holds for any $l$ and any item $j$ such that $\Gamma_{j,l}=1$. 
Therefore we have shown $\theta_{{j},H} = \bar\theta_{{j},H}$ holds for any $j\in S_1\cup S_2$.

\bigskip
We next show that for any $j\in S_1\cup S_2$ and any $l\in\{1,\ldots,L_0\}$ such that $\Gamma_{j,l}=0$,   $\theta_{j,l} = \bar \theta_{j,\sigma(l)}$, and show $p_{l} = \bar p_{\sigma(l)}$ for any $l\in\{1,\ldots,L_0\}$.
We use an induction method to show for any $l\in[L_0]$,
\begin{equation}\label{eq-theta-p}
\forall j\in S_1\cup S_2,\quad
\theta_{j,l} = \bar\theta_{j,\sigma(l)},\quad p_{l} = \bar p_{\sigma(l)}.
\end{equation}
 We first introduce the \textit{lexicographic order} between two binary vectors of the same length. For two vectors $\bo a=(a_1,\ldots,a_L)$ and $\bo b=(b_1,\ldots,b_L)$, we say $\bo a$ has smaller lexicographic order than $\bo b$ and denote by $\bo a\prec_{\text{lex}}\bo b$, if either $a_1<b_1$, or $a_l<b_l$ for some integer $l\leq L$ and $a_m=b_m$ for all $m=1,\ldots,l-1$.
 By Condition $A$, $\Gamma^{(S_i,\mca_0)}$ has distinct column vectors for $i=1,2$, so without loss of generality, we can assume the columns of it are sorted in an increasing lexicographic order, i.e.,
\begin{equation}\label{eq-glex}
\Gamma^{(S_1,\mca_0)}_{\Cdot,1}\prec_{\text{lex}}\cdots
\prec_{\text{lex}}
\Gamma^{(S_1,\mca_0)}_{\Cdot,L_0}.
\end{equation}
Firstly, we prove \eqref{eq-theta-p} hold for $l=1$, where from \eqref{eq-glex} we have $\Gamma^{(S_1,\mca_0)}_{\Cdot,1}$ has the smallest lexicographical order among the column vectors of $\Gamma^{(S_1,\mca_0)}$.
We claim that $\Gamma^{(S_2,\mca_0)}_{\Cdot,1}$ has the smallest lexicographical order among the column vectors of $\Gamma^{(S_2,\mca_0)}$, because otherwise ``~$\succeq_{S_1} = \succeq_{S_2}$'' under $\mca_0$ will not hold.
For $l=1$ we define
\[
\ttt^* = \sum_{h\in S_1:\Gamma_{h,1}=0}\theta_{h,H}\ee_h, 
\]
and consider the row vector of the transformed $T$-matrix $T(\TT-\ttt^*\mathbf1^\top)$ corresponding to $\rr=\sum_{h\in S_1:\Gamma_{h,1}=0}\ee_h$ has only one potentially nonzero element in the first column,
i.e.,
\[\begin{aligned}
T_{\rr,\Cdot}(\Gamma,\TT-\ttt^*\mathbf1^\top)  = \biggr(\prod_{h\in S_1:\Gamma_{h,1}=0} (\theta_{h,1}-\theta_{h,H}),~0,\ldots,0\biggr) 
\end{aligned}\]
Then similarly for parameters $(\bar\TT,\bar\pp)$ we have
\[\begin{aligned}
T_{\rr,\Cdot}(\bar\Gamma,\bar\TT-\ttt^*\mathbf1^\top)  = \biggr(0,\ldots,0,~\underbrace{\prod_{h\in S_1:\Gamma_{h,1}=0} (\bar\theta_{h,\sigma(1)}-\theta_{h,H})}_{\text{column}~\sigma(1)},~0,\ldots,0\biggr) 
\end{aligned}\]
and 
\[
\prod_{h\in S_1:\Gamma_{h,1}=0} (\theta_{h,1}-\bar\theta_{h,H})\neq 0,\quad
\prod_{h\in S_1:\Gamma_{h,1}=0} (\bar\theta_{h,1}-\theta_{h,H})\neq 0.
\]

Now consider $\theta_{j,1}$ for any $j\in S_2$ and $\Gamma_{j,1}=0$. The row vectors of $T(\Gamma,\TT-\ttt^*\mathbf1^\top)$ and $T(\bar\Gamma,\bar\TT-\ttt^*\mathbf1^\top)$ corresponding to the response pattern $\rr + \ee_j$ are
\begin{align}\label{eq-s1g1}
T_{\rr+\ee_j,\cdot}(\Gamma,\TT-\ttt^*\mathbf1^\top) 
= \Big(\prod_{h\in S_1:\Gamma_{h,1}=0} (\theta_{h,1}-\theta_{h,H})\cdot \theta_{j,1},~
0,\ldots,0\Big),
\end{align}
and
\begin{align}\label{eq-s1g2}
T_{\rr+\ee_j,\cdot}(\bar\Gamma,\bar\TT-\ttt^*\mathbf1^\top)
= \Big( 0,\ldots,0,~\underbrace{\prod_{h\in S_1:\Gamma_{h,1}=0} (\bar\theta_{h,\sigma(1)}-\theta_{h,H})\cdot \bar\theta_{j,\sigma(1)}}_{\text{column}~\sigma(1)},~
0,\ldots,0\Big),
\end{align}
respectively. The only potentially nonzero term in the first column of \eqref{eq-s1g1} is indeed nonzero, because we have $\theta_{h,1}<\theta_{h,H}$ for $h\in S_1$, $\Gamma_{h,1}=0$.
Now Equation (\ref{eq1}) implies that
\begin{eqnarray*}
	\theta_{j,1} =
\frac{T_{\rr+\ee_j,\Cdot}(\Gamma,\TT-\ttt^*\mathbf1^\top)\pp}{T_{\rr,\Cdot}(\Gamma,\TT-\ttt^*\mathbf1^\top)\pp}
= \frac{T_{\rr+\ee_j,\Cdot}(\bar\Gamma,\bar\TT-\ttt^*\mathbf1^\top)\bar\pp}{T_{\rr,\Cdot}(\bar\Gamma,\bar\TT-\ttt^*\mathbf1^\top)\bar\pp}
= \bar\theta_{j,\sigma(1)},
\end{eqnarray*}
for any $j\in S_2$ and $\Gamma_{j,1}=0$. Similarly we can obtain $\theta_{j,1}=\bar\theta_{j,\sigma(1)}$ for any $j\in S_1$ and $\Gamma_{j,\sigma(1)}=0$. 

After obtaining these $\bar \theta_{j,\sigma(1)}=\theta_{j,1}$ for $j\in(S_1\cup S_2)$ and $\Gamma_{j,1}=0$, the previous equations \eqref{eq-s1g1} and \eqref{eq-s1g2} just become the following,
\begin{align}\label{eq-s1g3}
T_{\rr+\ee_j,\cdot}(\Gamma,\TT-\ttt^*\mathbf1^\top) 
= &~\Big(\prod_{h\in S_1:\Gamma_{h,1}=0} (\theta_{h,1}-\theta_{h,H})\cdot \theta_{j,1},~0,\ldots,0\Big), \\ \label{eq-s1g4}
T_{\rr+\ee_j,\cdot}(\bar\Gamma,\bar\TT-\ttt^*\mathbf1^\top)
= &~\Big( 0,\ldots,0,~\underbrace{\prod_{h\in S_1:\Gamma_{h,1}=0} (\theta_{h,\sigma(1)}-\theta_{h,H})\cdot \theta_{j,\sigma(1)}}_{\text{column}~\sigma(1)},~
0,\ldots,0\Big).
\end{align}
Therefore $\eqref{eq-s1g3}\cdot\pp = \eqref{eq-s1g4}\cdot\bar\pp$ just gives $p_1=\bar p_{\sigma(1)}$.

Now as the inductive hypothesis, we assume for an $l\in[L_0]$,
\[
\forall \aaa_{l'} \text{ s.t. } \aaa_{l'}\preceq_{S_1}\aaa_l,
\quad \forall j\in S_1\cup S_2,\quad \theta_{j,{l'}} = \bar \theta_{j,\sigma({l'})}
, \quad p_{{l'}} = \bar p_{\sigma({l'})}.
\]
Recall that $\aaa_{l'}\preceq_{S_1}\aaa_l$ if and only if $\aaa_{l'}\preceq_{S_2}\aaa_l$ under $\mca_0$.
Define $\ttt^*$ as 
\[
\ttt^* = \sum_{h\in S_1:\Gamma_{h,l}=0}\theta_{h,H}\ee_h
+ \sum_{h\in S_1:\Gamma_{h,l}=1}\theta_{h,l}\ee_h,
\]
then for $\rr^* := \sum_{h\in S_1}\ee_h$ we have 
\begin{align}\label{eq-induc-nn1}
T_{\rr^*,\cdot}(\Gamma,\TT-\ttt^*\mo^\top) \pp
 = &\sum_{\aaa_{l'}\preceq_{S_1}\aaa_l}t_{\rr^*,{l'}}\cdot p_{{l'}} \\ \notag
 &+ \prod_{h\in S_1:\Gamma_{h,l}=0}(\theta_{h,l} - \theta_{h,H})
\prod_{h\in S_1:\Gamma_{h,l}=1}(\theta_{h,l} - \theta_{h,1})\cdot p_{l},\\
\label{eq-induc-nn2}
 T_{\rr^*,\cdot}(\bar\Gamma,\bar\TT-\ttt^*\mo^\top) \bar\pp
= &\sum_{\aaa_{l'}\preceq_{S_1}\aaa_l}\bar t_{\rr^*,\sigma(l')}\cdot \bar p_{\sigma(l')} \\ \notag
 &+  \prod_{h\in S_1:\Gamma_{h,l}=0}(\bar\theta_{h,\sigma(l)} - \theta_{h,H})
\prod_{h\in S_1:\Gamma_{h,l}=1}(\bar\theta_{h,\sigma(l)} - \theta_{h,1})\cdot \bar p_{\sigma(l)},
\end{align} 
where the notations $t_{\rr^*,l'}$ and $\bar t_{\rr^*,l'}$ are defined as
\[\begin{aligned}
t_{\rr^*,l'} &= \prod_{h\in S_1:\Gamma_{h,l}=0}(\theta_{h,l'} - \theta_{h,H})
\prod_{h\in S_1:\Gamma_{h,l}=1}(\theta_{h,l} - \theta_{h,1}), \\
\bar t_{\rr^*,l'} &= \prod_{h\in S_1:\Gamma_{h,l}=0}(\bar\theta_{h,\sigma(l')} - \theta_{h,H})
\prod_{h\in S_1:\Gamma_{h,l}=1}(\bar\theta_{h,\sigma(l)} - \theta_{h,1}).
\end{aligned}\]
Note that by induction assumption we have $\theta_{h,l'} = \bar\theta_{h,\sigma(l')}$ for any $l'$ such that $\aaa_{l'}\preceq_{S_1}\aaa_l$ under $\mca_0$. This implies $t_{\rr^*,l'} = \bar t_{\rr^*,\sigma(l')}$ and further implies
\[
\sum_{\aaa_{l'}\preceq_{S_1}\aaa_l}t_{\rr^*,l'}\cdot p_{l'}
=
\sum_{\aaa_{l'}\preceq_{S_1}\aaa_l}\bar t_{\rr^*,\sigma(l')}\cdot\bar p_{\sigma(l')}.
\]
So (\ref{eq-induc-nn1}) = (\ref{eq-induc-nn2}) gives
\begin{align}\label{eq-induc-nn3}
&\prod_{h\in S_1:\Gamma_{h,l}=0}(\theta_{h,l} - \theta_{h,H})
\prod_{h\in S_1:\Gamma_{h,l}=1}(\theta_{h,l} - \theta_{h,1})\cdot p_{l} \\ \notag
=&
\prod_{h\in S_1:\Gamma_{h,l}=0}(\bar\theta_{h,\sigma(l)} - \theta_{h,H})
\prod_{h\in S_1:\Gamma_{h,l}=1}(\bar\theta_{h,\sigma(l)} - \theta_{h,1})\cdot \bar p_{\sigma(l)},
\end{align}
and the two terms on both hand sides of the above equation are nonzero.
Now consider any $j\notin S_1$ and similarly $T_{\rr^*+\ee_j,\cdot}(\Gamma,\TT-\ttt^*\mo^\top) \pp = T_{\rr^*+\ee_j,\cdot}(\bar\Gamma,\bar\TT-\ttt^*\mo^\top)\bar \pp$ yields
\begin{align}\label{eq-induc-nn4}
&\theta_{j,l}\cdot\prod_{h\in S_1:\Gamma_{h,l}=0}(\theta_{h,l} - \theta_{h,H})
\prod_{h\in S_1:\Gamma_{h,l}=1}(\theta_{h,\aaa} - \theta_{h,1})\cdot p_{l} \\ \notag
=& ~\bar\theta_{j,\sigma(l)}\cdot
\prod_{h\in S_1:\Gamma_{h,l}=0}(\bar\theta_{h,\sigma(l)} - \theta_{h,H})
\prod_{h\in S_1:\Gamma_{h,l}=1}(\bar\theta_{h,\sigma(l)} - \theta_{h,1})\cdot \bar p_{\sigma(l)}.
\end{align}
Taking the ratio of the above two equations \eqref{eq-induc-nn4} and \eqref{eq-induc-nn3} gives
$\theta_{j,l}=\bar\theta_{j,\sigma(l)},\quad \forall j\notin S_1.$
Redefining $\rr^* := \sum_{h\in S_2}\ee_h$ similarly as above we have $\theta_{j,l}=\bar\theta_{j,\sigma(l)}$ for any $j\in S_1$.
Plug $\theta_{j,l}=\bar\theta_{j,\sigma(l)}$ for all $j\in S_1$ into (\ref{eq-induc-nn3}), then we have 
$p_{l} = \bar p_{\sigma(l)}$.
Now we have shown \eqref{eq-theta-p} hold for this particular $l$. 
Then the induction argument gives
\[
\forall l\in[L_0],\quad \forall j\in S_1\cup S_2,\quad
\theta_{j,l} = \bar\theta_{j,\sigma(l)},\quad p_{l} = \bar p_{\sigma(l)}.
\]

Now we have shown for any item $j$ and latent class index $l$, $\theta_{j,l} = \bar\theta_{j,\sigma(l)}$, which we denote by $\bar\TT = \sigma(\TT)$.
We claim that this result also indicates that the permutation $\sigma$ is unique. This is because $U_1\boldsymbol\cdot \{W_1 T(\Gamma^1,\TT^1)\}_{\rr_{1:L_0}} = I_L$ implies that
$$
U_1\bcdot \{W_1 T(\bar\Gamma^1,\bar\TT^1)\}_{\rr_{1:L_0}} = U_1 \bcdot \{W_1 T(\Gamma^1,\TT^1)\}_{\rr_{1:L_0}} \bcdot \sigma(I_L) = \sigma(I_L),
$$
which means given $U_1$ constructed from $(\Gamma,\TT)$, the form of $U_1\bcdot \{W_1 T(\bar\Gamma^1,\bar\TT^1)\}_{\rr_{1:L_0}}$ explicitly and uniquely determines $\sigma$.
Now we have shown $\bar\Gamma = \Gamma=\Gamma^{\mca_0}$ and $(\bar\TT,\bar\pp)=(\TT,\pp)$ must hold up to the column permutation $\sigma$.

As stated in the beginning of the proof, combining Condition $C$ that any column in $\Gamma^{\mca_0}$ is different from any column in $\Gamma^{\mca_0^c}$, the identification of $\Gamma^{\mca_0}$ uniquely identifies the set of true patterns $\mca_0$.
The proof of both Theorem \ref{thm-id} and Corollary \ref{cor1} is complete.
\qed

\paragraph{Proof of Theorem \ref{prop-genid} and Corollary \ref{cor-gid}.}
The following proofs of Theorem \ref{prop-genid} and Corollary \ref{cor-gid} use a similar proof idea as that of \cite{allman2009}; see also proofs of Theorems 4.2 and 4.3 in \cite{partial}.

\smallskip
\noindent{\it Proof of Theorem \ref{prop-genid}.}  We need to introduce the definition of \textit{algebraic variety}, a concept in algebraic geometry.
An algebraic variety $\mathcal V$ is defined as the simulateneous zero-set of a finite collection of multivariate polynomials $\{f_i\}_{i=1}^n\subseteq \mathbb R[x_1,x_2,\ldots,x_d]$,
$
\mathcal V =\mathcal V(f_1,\ldots,f_n) = \{\boldsymbol x\in\mathbb R^d\mid f_i(\boldsymbol x) = 0,~ 1\leq i\leq n.\}
$
 An algebraic variety ${\mathcal V}$ is all of $\mathbb R^d$ only when all the polynomials defining it are zero polynomials; otherwise, ${\mathcal V}$ is called a \textit{proper subvariety} and is of dimension less than $d$, hence necessarily of Lebesgue measure zero in $\mathbb R^d$. The same argument holds when $\mathbb R^d$ is replaced by the parameter space ${\Omega}\subseteq\mathbb R^d$ that has full dimension in $\mathbb R^d$.
For the structured latent attribute model, we consider the following parameter space,
\[
\Omega = 
\Big\{(\TT,\pp):\, \forall~j, \max_{\aaa:\Gamma_{j,\aaa}=1}\theta_{j,\aaa} = \min_{\aaa:\Gamma_{j,\aaa}=1} \theta_{j,\aaa}>\theta_{j,\aaa'},~\forall~ \Gamma_{j,\aaa'}=0 \Big\}.
\] 
On $\Omega$, altering some entries of zero to one in the $\Gamma$-matrix is equivalent to impose more affine constraints on the parameters and force them to be in a subset 
$\Omega^*$ of $\Omega$. 
Condition $A^\star$ guarantees that, there exists a $\Omega^*$ such that Condition $A$ holds for model parameters belonging to this $\Omega^*$, the proof of Theorem \ref{thm-id} gives that the matrix $T(\Gamma^{(S_i,\mca_0)},\TT^{(S_i,\mca_0)})$ has full column rank $C$ for $i=1,2$ for $(\TT^{(S_i,\mca_0)},\pp^{\mca_0})\in\Omega^*$.
Note that the statement that $2^{|S_i|}\times C$ matrix $T(\Gamma^{(S_i,\mca_0)},\TT^{(S_i,\mca_0)})$ has full column rank is equivalent to the statement that the map sending $T(\Gamma^{(S_i,\mca_0)},\TT^{(S_i,\mca_0)})$ to all its $\binom{2^{|S_i|}}{C}$ possible $C\times C$ minors $A_1^i,A_2^i,\ldots, A^i_{2^{|S_i|}}$ yields at least one nonzero minor, where $A^i_1,A^i_2,\ldots, A^i_{2^{|S_i|}}$ are all polynomials of the item parameters $\TT_{S_i}$. Define 
\[\mathcal V = \bigcup_{i=1,2} \Big\{\bigcap_{l=1}^{2^{|S_i|}} \{(\TT,\pp)\in \Omega: A^i_l(\TT^{(S_i,\mca_0)}) = 0\}\Big\},
\]
then $\mathcal V$ is a algebraic variety defined by polynomials of the model parameters. Moreover, $\mathcal V$ is a proper subvariety of $\Omega$, since the fact $T(\Gamma^{(S_i,\mca_0)},\TT^{(S_i,\mca_0)})$ has full column rank $C$ for $i=1,2$ for one particular set of  $(\TT,\pp)\in\Omega^*$ ensures that there exists one particular set of model parameters that give nonzero values when plugged into the polynomials defining $\mathcal V$. This indicates that the polynomials defining $\mathcal V$ are not all zero polynomials on $\Omega$. Then restricting parameters to $\Omega^*$ and proceeding in the same steps as the proof of Theorem \ref{thm-id} proves the conclusion of the proposition.
 
\smallskip
\noindent {\it Proof of Corollary \ref{cor-gid}.}
Consider a $Q$-matrix in the form of \eqref{eq-diag}. We denote $S_1=\{1,\ldots,K\}$, $S_2=\{K+1,\ldots,2K\}$ and $S_3 = \{2K+1,\ldots,J\}$, which are item sets corresponding to $Q_1$, $Q_2$ and $Q'$, respectively.
According to the proof of Theorem 4.3 in \cite{partial}, since the two submatrices $Q_1$ and $Q_2$ have all the diagonal elements equal to one, the  $2^K\times 2^K$ $T$-matrices $T(\Gamma^{(S_1,\text{all})},\TT^{(S_1,\text{all})})$ and $T(\Gamma^{(S_2,\text{all})},\TT^{(S_2,\text{all})})$ are generically full-rank. Furthermore, the matrix $T(\Gamma^{(S_3,\text{all})},\TT^{(S_3,\text{all})})\cdot \text{Diag}(\pp^{\text{all}})$ has Kruskal rank at least two. This means generically, any two columns of $T(\Gamma^{(S_3,\text{all})},\TT^{(S_3,\text{all})})\cdot \text{Diag}(\pp^{\text{all}})$ are linearly independent. 

Now consider an arbitrary set of attribute patterns $\mca_0\subseteq\{0,1\}$, we have the conclusion that $T(\Gamma^{(S_1,\mca_0)},\TT^{(S_1,\mca_0)})$ and $T(\Gamma^{(S_2,\mca_0)},\TT^{(S_2,\mca_0)})$ have  full column rank generically. This is because for $i=1,2$, the $T(\Gamma^{(S_i,\mca_0)},\TT^{(S_i,\mca_0)})$ is just a submatrix of $T(\Gamma^{(S_i,\text{all})},\TT^{(S_i,\text{all})})$ whose columns are a subset of different column vectors of the latter matrix. Therefore columns of $T(\Gamma^{(S_i,\mca_0)},\TT^{(S_i,\mca_0)})$ must be linearly independent, and hence the matrix must have full column rank generically.
 Also, the columns of $T(\Gamma^{(S_3,\mca_0)},\TT^{(S_3,\mca_0)})\cdot \text{Diag}(\pp^{\mca_0})$ can also be considered as a subset of different columns of $T(\Gamma^{(S_3,\text{all})},\TT^{(S_3,\text{all})})\cdot \text{Diag}(\pp^{\text{all}})$ up to a resealing of the columns. Therefore the former matrix must have any two different columns linearly independent generically and hence has Kruskal rank at least two. Now by Kruskal's conditions for unique tensor decomposition, a probability distribution of $\RR$ 
 with $T(\Gamma^{(S_i,\mca_0)},\TT^{(S_i,\mca_0)})$, $i=1,2,3$ having the above properties
uniquely determines $T(\Gamma^{(S_i,\mca_0)},\TT^{(S_i,\mca_0)})$ and also $\pp^{\mca_0}$ generically. Therefore $(\Gamma^{\mca_0},\TT^{\mca_0},\pp^{\mca_0})$ are generically identifiable. Then combined with Condition $C$, we have the conclusion that $\mca_0$ is generically identifiable.
 This completes the proof of the corollary.
\qed

\paragraph{Proof of Corollary \ref{cor-parid}.}
Under our definition of $\mca^{\text{rep}}$ and also Condition $C$, this matrix must have distinct column vectors, and each of its column corresponds to an equivalence class. We define $\TT^{\text{rep}}$ to be item parameters corresponding to the representative patterns in $\mca^{\text{rep}}$. We further define the proportion parameters of the equivalence classes $\nnu^{\text{rep}}=(\nu_{[\aaa_{\ell_1}]},\ldots, \nu_{[\aaa_{\ell_m}]})$, where $\nu_{[\aaa_{\ell_i}]}>0$ and $\sum_{i=1}^m \nu_{[\aaa_{\ell_i}]}=1$. Note that each $\nu_{[\aaa_{\ell_i}]}$ is a sum of population proportions of the attribute patterns that are in the same equivalence class of $\aaa_{\ell_i}$.
Since $\Gamma^{\mca^{\text{rep}}}$ also satisfies Conditions $A$ and $B$ by the assumption of the corollary.
So Theorem \ref{thm-id} gives that $\mca^{\text{rep}}$ is identifiable.
\qed

\paragraph{\textcolor{black}{Proof of Theorem \ref{thm-select} and Proposition \ref{prop-rate}.}}

We use $L=|\mca_{\text{input}}|$ to denote the number of attribute patterns as input given to the penalized likelihood method, then $L=2^K$ if there is no screening stage as preprocessing.
We denote the true proportion parameters by $\pp=(p_{\aaa}:\,\aaa\in\mca_{\inp})$, where $p_{\aaa}\geq 0$ for $\aaa\in\mca_{\inp}$ and $\sum_{\aaa\in\mca_{\inp}} p_{\aaa}=1$.
 Denote the number of true attribute patterns by $|\mca_0|$.
We now consider the following log likelihood with penalty parameter $\lambda_N$ for some $\gamma>0$,
\begin{align}\label{eq-mod}
\ell^{\lambda_N}(\pp,\TT) 
= & \frac{1}{N} \sum_{i=1}^N \log\Big\{ \sum_{\aaa\in\mca_{\inp}} p_{\aaa} \prod_{j=1}^J\theta_{j,\aaa}^{R_{i,j}}(1-\theta_{j,\aaa})^{1-R_{i,j}} \Big\} \\ \notag
& + \frac{\lambda_N}{N} 
\underbrace{\sum_{\aaa\in\mca_{\inp}}  \Big[\log p_{\aaa} \cdot I (p_{\aaa}>\rho_N) + \log \rho_N \cdot I (p_{\aaa}\leq \rho_N)\Big]}_{\log_{\rho_N}(\pp)}.
\end{align}

For a given $\lambda_N$, denote the estimated support of the proportion parameters $\widehat\pp$ by $\widehat \mca$, namely  $\widehat \mca=\{1\leq l\leq L: \widehat p_{\aaa_l}>\rho_N\}$. 
We denote the true and the estimated $|\mca_{\inp}|$-dimensional proportions by 
$\pp^{\mca_0}_{\text{full}}=(p_{\aaa},\aaa\in\mca_{\text{input}}:\,p_{\aaa}>0~\text{if and only if}~\aaa\in\mca_0)$ 
and 
$\widehat\pp^{\widehat\mca}_{\text{full}}=(\widehat p_{\aaa},\aaa\in\mca_{\text{input}}:\,\widehat p_{\aaa}>\rho_N~\text{if and only if}~\aaa\in\widehat\mca\,)$.
Denote the oracle MLE obtained assuming $\mca_0$ is known by $\widehat\TT^0:=\widehat\TT^{\mca_0}$ and $\widehat\pp^0:=\widehat\pp^{\mca_0}$, and denote $\widehat\eeta^0=(\widehat\TT^0,\widehat\pp^0)$.
Note that for $\widehat\mca\neq\mca_0$ the event $\{\ell^{\lambda_N}(\widehat\eeta^{\widehat\mca})>\ell^{\lambda_N}(\widehat\eeta^{0})\}$ implies the following event
\begin{align}\label{eq-lrt}
&~\frac{1}{N}\sum_{i=1}^N \log
\left[ \frac{\sum_{\aaa\in\mca_{\text{input}}} \widehat p_{\aaa}\prod_{j}\widehat\theta_{j,\aaa}^{R_{i,j}}(1-\widehat\theta_{j,\aaa})^{1-R_{i,j}}}
{\sum_{\aaa\in\mca_0}  \widehat p^0_{\aaa}\prod_{j}(\widehat\theta_{j,\aaa}^0)^{R_{i,j}}(1-\widehat\theta^0_{j,\aaa})^{1-R_{i,j}}} \right] \\ \notag
>&~
\frac{|\lambda_N|}{N}\Big\{ \log_{\rho_N}(\widehat\pp_{\text{full}}^{\widehat\mca}) - \log_{\rho_N}(\widehat\pp_{\text{full}}^{\mca_0}) \Big\}.
\end{align}
In the case of $|\widehat\mca\,|>|\mca_0|$ (which we call the overfitted case), the right hand side (RHS) of \eqref{eq-lrt} regarding the difference between the penalty terms has order 
$
O(N^{-1}|\lambda_N|\cdot|\mca_0|\cdot|\log \rho_N|).
$
In this overfitted case, we now consider the left hand side (LHS) of \eqref{eq-lrt},
\begin{align*}
\text{LHS of }\eqref{eq-lrt}=
&~\frac{1}{N}\sum_{i=1}^N \log
\left[ {\sum_{\aaa\in\mca_{\text{input}}} \widehat p_{\aaa}\prod_{j}\widehat\theta_{j,\aaa}^{R_{i,j}}(1-\widehat\theta_{j,\aaa})^{1-R_{i,j}}} \right] \\
&~
- \frac{1}{N}\sum_{i=1}^N \log \left[ {\sum_{\aaa\in\mca_0}  \widehat p^0_{\aaa}\prod_{j}(\widehat\theta_{j,\aaa}^0)^{R_{i,j}}(1-\widehat\theta^0_{j,\aaa})^{1-R_{i,j}}} \right] \equiv I_1 - I_0,
\end{align*}
where the $I_1$ part can be written as
\begin{align}
\notag
I_1
=&~ \frac{1}{N}\sum_{i=1}^N \log\Big[ {\sum_{\aaa\in\mca_{\text{input}},\atop \widehat p_{\aaa}>\rho_N} \widehat p_{\aaa}\prod_{j}\widehat\theta_{j,\aaa}^{R_{i,j}}(1-\widehat\theta_{j,\aaa})^{1-R_{i,j}}}\Big] +O(|\mca_{\text{input}}|\rho_N) \\
\label{eq-ndelta}
\allowbreak
=&~ \frac{1}{N}\sum_{i=1}^N \log
\Big[ {\sum_{\aaa\in\mca_{\text{input}},\atop \widehat p_{\aaa}>\rho_N} \widehat p_{\aaa}\prod_{j}\widehat\theta_{j,\aaa}^{R_{i,j}}(1-\widehat\theta_{j,\aaa})^{1-R_{i,j}}}\Big] +O(N^{-\delta}),
\end{align}
where the last equality follows from the assumption $|\mca_{\text{input}}|\cdot\rho_N = O(N^{-\delta})$ in the theorem.
So we further have the LHS of \eqref{eq-lrt} equal to
\begin{align*}
I_1-I_0 =&~ \frac{1}{N}\sum_{i=1}^N \log
\Big[ {\sum_{\aaa\in\mca_{\text{input}},\atop \widehat p_{\aaa}>\rho_N} \widehat p_{\aaa}\prod_{j}\widehat\theta_{j,\aaa}^{R_{i,j}}(1-\widehat\theta_{j,\aaa})^{1-R_{i,j}}}\Big] \\
&~ - \frac{1}{N}\sum_{i=1}^N \log \Big[ {\sum_{\aaa\in\mca_0}  \widehat p^0_{\aaa}\prod_{j}(\widehat\theta_{j,\aaa}^0)^{R_{i,j}}(1-\widehat\theta^0_{j,\aaa})^{1-R_{i,j}}} \Big] + O(N^{-\delta}).
\end{align*}
Note that  other than the last term $O(N^{-\delta})$ in the above display, the difference of the first two terms also has order $O_p(N^{-\delta})$ from assumption \eqref{eq-rate}, so LHS of \eqref{eq-lrt}$=I_1-I_0=O_p(N^{-\delta})$.
In order to have selection consistency in the overfitted case, we need the event described in \eqref{eq-lrt} to happen with probability tending to zero, so 
the $|\lambda_N|$ needs to be sufficiently large such that 
\begin{equation}\label{eq-delta}
N^{-\delta} \lesssim O\Big(N^{-1}|\lambda_N|\cdot|\mca_0|\cdot|\log \rho_N| \Big).
\end{equation} 
Note that by \eqref{eq-rho-N}, we have $\rho_N\asymp N^{-d}$ for some $d>0$.
So if $\delta<1$, i.e., if the convergence rate is slower than the $\sqrt N$ rate, then $\lambda_N$ must go to negative infinity as $N$ goes to infinity since $\delta<1$. 
Specifically, we obtain the following lower bound of the magnitude of the penalty parameter $\lambda_N$,
\begin{equation}\label{eq-lambda1}
|\lambda_N|\gtrsim N^{1-\delta}/|\log \rho_N|
\end{equation}
 would suffice for \eqref{eq-delta} to hold. 

\textit{We now prove the conclusion of Proposition \ref{prop-rate}.}
 A further implication of the above discussion is that, with $\rho_N\asymp N^{-d}$ as assumed in \eqref{eq-rho-N}, just imposing a proper Dirichlet prior with a positive hyperparameter  would fail to select the true model consistently. In particular, with a proper Dirichlet prior density with hyperparameter $\beta=\lambda_N+1\in(0,1)$, Equation \eqref{eq-delta} instead becomes $N^{1-\delta} = o(\log N)$. However, when $0<\delta<1$, $N^{1-\delta}/\log N\to\infty$. So \eqref{eq-delta} fails to hold, and one can not have consistent selection in the overfitted case. So if we denote  the set of attribute patterns estimated by maximizing \eqref{eq-orig} by $\widehat{\mathcal A}^{\lambda}$. Then for any $\{\lambda_N\}\subseteq[-1,0)$, $\mathbb P(\widehat{\mathcal A}^{\lambda}=\mathcal A_0)\not\to 1$ as $N\to\infty$.
{This proves Proposition \ref{prop-rate}.}


Now we consider the random set $\{\aaa\in\mca_{\text{input}}:\, \widehat p_{\aaa}>\rho_N\}=:\widehat\mca$ appearing in $I_1$ in \eqref{eq-ndelta}. With probability tending to one, the cardinality of this set is smaller than $|\mca_0|$. This is because if $|\widehat\mca|>|\mca_0|$, the log-penalty term corresponding to $\widehat\mca$ would be smaller than that corresponding to $\mca_0$ by $N^{-1}|\lambda_N|\cdot|\log\rho_N|$ which has order at least $N^{-\delta}$. Recall that the right hand side of \eqref{eq-lrt} has order $O_P(N^{-\delta})$, which means when $|\widehat\mca|>|\mca_0|$ the extent that the log-penalty part favors the a smaller model $\mca_0$ would dominate the extent that the likelihood part favors a larger model $\widehat\mca$ in the proposed penalized likelihood. Therefore any larger model $\widehat\mca$ with $|\widehat\mca|\geq|\mca_0|$ would be favored over $\mca_0$ with probability tending to zero. Therefore we have the conclusion that $\mathbb P(\widehat\mca\neq \mca_0)\not\to 0$ could only happen for $|\widehat\mca|\leq |\mca_0|$. So in the following discussion we will focus on the case where $|\widehat\mca|\leq |\mca_0|$ and prove consistency in this case. Namely, we aim to bound
\begin{align}\label{eq-noverfit}
\mathbb P\Big(\sup_{|\widehat\mca|\leq |\mca_0|,\,\widehat\mca\neq\mca_0}[\ell^{\lambda_N}(\eeta^{\widehat\mca})-\ell^{\lambda_N}(\eeta^{\mca_0})]>0\Big).
\end{align}

Next, we consider the upper bound of the magnitude of the penalty term. In order to have selection consistency in the case of $|\widehat\mca|\leq |\mca_0|$ and $\widehat\mca\neq \mca_0$, the log-penalty term can not be too large such that the extent that the penalty part favors a smaller model does not dominate the extent that the likelihood part favors the true model.
We follow a similar argument to \cite{shen2012}.  
Specifically, considering the term $-\epsilon_N^2\to 0$ in the large deviation inequality \eqref{eq-large} below; for a small constant $t>\epsilon_N$, we  need that the difference of the penalty part of the true and any alternative smaller model to be less than $t^2$, i.e., 
\begin{equation}\label{eq-eta-epsi}
|\lambda_N|\cdot|\mca_0|\cdot|\log \rho_N|/N\lesssim t^2,
\end{equation}
Equation \eqref{eq-eta-epsi} would hold if 
\begin{equation}\label{eq-lambda2}
|\lambda_N | = o({N}/|\log\rho_N|).
\end{equation}
We next show that such $\lambda_N$ can guarantee selection consistency. 
So we have a sample-size dependent $\lambda_N$ that penalizes the overfitted mixture and constrains the support size of the proportion parameters to be less than the true support size $|\mca_0|$.
As said, with such $\lambda_N$ it suffices to consider the case $|\widehat\mca|\leq|\mca_0|$.

In order to bound this mis-selection probability, we need to introduce the notion of bracketing Hellinger metric entropy $H(t,\mathcal B_{\mca})$. 
Let $h(\eeta^{\mca},\eeta^{\mca_0})$ denote the Hellinger distance between the probability mass functions of $\RR$ indexed by $\eeta^{\mca}$ and $\eeta^{\mca_0}$, i.e.,
$$
h(\eeta^{\mca},\eeta^{\mca_0})=\Big(\sum_{\rr\in\{0,1\}^J} \Big[\mathbb P(\RR=\rr\mid \TT^{\mca},\pp^{\mca})^{\frac{1}{2}} -  
\mathbb P(\RR=\rr\mid \TT^{\mca_0},\pp^{\mca_0})^{\frac{1}{2}}\Big]\Big)^{\frac{1}{2}}.
$$
Consider the local parameter space $\mathcal B_{\mca} = \{\eeta^{\mca}=(\TT^{\mca},\pp^{\mca}):\,|\mca|\leq |\mca_0|,\, h^2(\eeta^{\mca},\eeta^{\mca_0})\leq 2\epsilon_N^2\}$, the $H(t,\mathcal B_{\mca})$ is defined as the logarithm of the cardinality of the $t$-bracketing of $\mathcal B_{\mca}$ of the smallest size.
More specifically, following the definition in \cite{shen2012}, consider a bracket covering $S(t,m)=\{f_1^l,f_1^u,\ldots,f_m^l,f_m^u\}$ satisfying that $\max_{1\leq j\leq m}\norm{f_j^u-f_j^l}_2\leq t$ and for any $f\in\mathcal B_{\mca}$ there is some $j$ such that $f_j^l\leq f\leq f_j^u$ almost surely. Then $H(t,\mathcal B_{\mca})$ is $\log(\min\{m:S(t,m)\})$.
The $H(t,\mathcal B_{\mca})$ measures the complexity of the local parameter space.
The next lemma gives an upper bound for the bracketing Hellinger metric entropy $H(t,\mathcal B_{\mca})$ for $|\mca|\leq|\mca_0|$.

\begin{lemma}\label{lem-ent}
Denote $N_{[]}(t,\mathcal B_{\mca}) = \exp (H(t,\mathcal B_{\mca}) )$.
For the considered structured latent attribute model, 
denote the item parameter space of the $\ell$-th attribute pattern by $\mathcal F_\ell$. For $|\mca|\leq |\mca_0|$ and any $2^{-4}\epsilon<t<\epsilon$, there is
$H(t,\mathcal B_{\mca})
\lesssim |\mca_0|\log|\mca_{\inp}|\log(2\epsilon/t).$
\end{lemma}
By the assumption of the theorem there is $\log|\mca_{\inp}|/N\to 0$, so if we take 
$$\epsilon_{N}\allowbreak = \sqrt{1/N|\mca_0|\log |\mca_{\inp}|},$$ 
there is $\epsilon_N=o(1)$. 
We next verify the entropy integral condition in Theorem 1 of \cite{wong1995} is satisfied with this $\epsilon_N$, in order to obtain a large deviation inequality to bound the mis-selection probability. 
With Lemma \ref{lem-ent}, the integral of bracketing Hellinger metric entropy in the interval $[2^{-8}\epsilon_N^2,\, \sqrt{2}\epsilon_N]$  satisfies the following inequality
\begin{align}\notag
\int_{2^{-8}\epsilon_N^2}^{\sqrt{2}\epsilon_N}
H^{1/2}(t,\mathcal B_{\mathcal A}) dt
&\leq \int_{2^{-8}\epsilon_N^2}^{\sqrt{2}\epsilon_N} \sqrt{|\mca_0|\log|\mca_{\inp}|\log(2\epsilon_N/t)} dt \\ \notag
&=\sqrt{|\mca_0|\log|\mca_{\inp}|} \int_{\sqrt{\log\sqrt{2}}}^{\sqrt{\log\frac{2^9}{\epsilon_N}}}4\epsilon_N u^2 e^{-u^2}du \\ 
\label{eq-integ}
&=\sqrt{|\mca_0|\log|\mca_{\inp}|} \cdot2\epsilon_N \underbrace{\int_{\log\sqrt{2}}^{\log\frac{2^9}{\epsilon_N}}\sqrt{u} e^{-u} du}_{\text{bounded as }\epsilon_N\to 0} 
\lesssim \sqrt{N}\epsilon_N^2.
\end{align}
So the entropy integral condition in Theorem 1 in \cite{wong1995} is satisfied and the large deviation inequality there holds. In particular, we have
\begin{align}
\notag
&~\mathbb P\Big(
\sup_{
\atop 
h^2(\widehat\eeta^{\widehat\mca},\eeta^{\mca_0})\geq \epsilon_N^2} 
\Big[\frac{1}{N}\ell(\widehat\eeta^{\widehat\mca})-
\frac{1}{N}\ell(\widehat\eeta^{\mca_0})\Big]>-\epsilon_N^2\Big)\\
\label{eq-large}
\leq 
&~\mathbb P\Big(
\sup_{
h^2(\widehat\eeta^{\widehat\mca},\eeta^{\mca_0})\geq \epsilon_N^2} 
\Big[\frac{1}{N}\ell(\widehat\eeta^{\widehat\mca})-
\frac{1}{N}\ell(\eeta^{\mca_0})\Big]>-\epsilon_N^2\Big)
\leq 
\exp( -N\epsilon_N^2 ).
\end{align}
where $\eeta^{\mca_0}=(\TT^{\mca_0},\pp^{\mca_0})$ denote the true parameters.
Indeed, Theorem 1 in \cite{wong1995} guarantees the inequality \eqref{eq-large} holds with $\epsilon_N$ replaced by any $t>\epsilon_N=\sqrt{|\mca_0|\log|\mca_{\inp}|/N}$.
This large deviation inequality will be used later to bound the mis-selection probability in the case of $|\mca|\leq|\mca_0|$.

We next further look at 
the Hellinger distance between $\eeta^0:=\eeta^{\mca_0}$ and $\eeta^\mca$ for $|\mca|\leq|\mca_0|$, and investigate how the distance between a set of true patterns $\mca_0$ and an alternative set relate to identifiability of $\mca_0$.
\begin{align*}
  & ~~
  \frac{h^2( \eeta^{\mathcal A}, \eeta^{\mathcal A_0})}{\max(|\mathcal A_0\setminus\mathcal  A|,1)}\\
\asymp &~  [\max(|\mathcal A_0\setminus\mathcal  A|,1)]^{-1}
     \sum_{\rr\in\{0,1\}^J}\Big[\Big(\sum_{\aaa\in \mca}\mathbb P(\RR = \rr\mid \TT^{\mca}, \ma=\aaa)p^{\mca}_{\aaa}\Big)^{1/2} - \\
      &\qquad \qquad\qquad \qquad\qquad\qquad
      \Big(\sum_{\aaa\in \mca_0}\mathbb P(\RR = \rr\mid \TT^{\mca_0},\ma=\aaa)p^{\mca_0}_{\aaa}\Big)^{1/2}\Big]^2\\ \notag
   \asymp &~[\max(|\mathcal A_0\setminus\mathcal  A|,1)]^{-1}
       \sum_{\rr\in\{0,1\}^J} \Big(\sum_{\aaa\in \mca} T_{\rr,\aaa}(\TT^{\mca}) \, p^\mca_{\aaa} - \sum_{\aaa\in \mca_0} T_{\rr,\aaa} (\TT^{\mca_0})\,p_{\aaa}^{\mca_0} \Big)^2 \\
=&~ [\max(|\mathcal A_0\setminus\mathcal  A|,1)]^{-1}
 \norm{T(\Gamma^\mca, \TT^\mca)\pp^\mca - T(\Gamma^{\mca_0},\TT^{\mca_0})\pp^{\mca_0}}_2^2
\end{align*}
To proceed with the proof, we need to use Theorem \ref{thm-id} to establish an identifiability argument. Theorem \ref{thm-id} and Corollary \ref{cor1} state that if the true constraint matrix $\Gamma^{\mca_0}$ satisfies conditions $A$, $B$ and $C$, then $(\Gamma^{\mca_0},~\TT^{\mca_0},~\pp^{\mathcal A_0})$ are jointly identifiable.
This implies that given the set of true attribute patterns $\mathcal A_0$, for any other set $\mathcal A\neq \mathcal A_0$, $|\mathcal A|\leq |\mathcal A_0|$, and model parameters defined by $\mathcal A$ must lead to different $T(\TT^{\mathcal A})\pp^{\mathcal A}$ that is different from $T(\TT^{\mathcal A_0})\pp^{\mathcal A_0}$. 
Moreover, consider the parameter space
$\mathcal B = \{(\TT^{\mathcal A},\pp^{\mathcal A}):~|\mathcal A|\leq |\mathcal A_0|,~ p_{\aaa}>\rho_N~\forall \aaa\in\mca\}$.
Then $(\TT^{\mathcal A_0},\pp^{\mathcal A_0})\in \mathcal B$ and for any $(\TT^\mathcal A,\pp^{\mathcal A})\in\mathcal B$ with $\mathcal A\neq \mathcal A_0$, \textbf{either} some elements in $\TT^{\mathcal A}$ differs from those in $\TT^{\mathcal A_0}$ by a nonzero constant, \textbf{or} some elements in $\pp^{\mathcal A}$ differs from those in $\pp^{\mca_0}$ by a nonzero constant. Since $T^{\mathcal A}(\TT)\pp^{\mathcal A}$ is a continuous vector-valued function of the model parameters, we must have $[\max(|\mathcal A_0\setminus\mathcal  A|,1)]^{-1}\norm{T^{\mathcal A}(\TT)\pp^{\mathcal A} - T^{\mathcal A_0}(\TT)\pp^{\mathcal A_0}}_2^2 \geq C_0$ for some $C_0>0$. 
By the conditions of the theorem $\epsilon_N^2=o(1)$,  so we have obtained for some small constant $t>\epsilon_N$,
\begin{align}\label{eq-cmin}
C_{\min}(\eeta^0) \equiv
\inf_{\eeta^{\mca}:\, \mathcal A\neq \mathcal A_0, |\mathcal A|\leq |\mathcal A_0|}\,
\left\{ 
\frac{h^2( \eeta^{\mathcal A}, \eeta^{\mathcal A_0})}{\max(|\mathcal A_0\setminus\mathcal  A|,1)}
\right\}\geq C_0
\gtrsim t^2> \epsilon_N^2.
\end{align}

Finally, with the $\lambda_N$ of the previously specified order, we use the large deviation inequality \eqref{eq-large} and also the \eqref{eq-cmin} to   bound the false selection probability \eqref{eq-noverfit}.
The following argument uses a similar proof idea  as that of Theorem 1 in \cite{shen2012} which establishes finite sample mis-selection error bound of the $L_0$-constrained maximum likelihood estimation. Consider $|\widehat\mca\cap \mathcal A_0|=m\leq |\mca_0|-1$, by \eqref{eq-cmin} we have $h^2( \eeta^{\mathcal A}, \eeta^{\mathcal A_0})\geq(|\mca_0|-m)C_{\min}(\eeta^0)$. So
\begin{align*}
&~\mathbb P\Big(\sup_{|\widehat\mca|\leq |\mca_0|,\,\widehat\mca\neq\mca_0}
\Big[\frac{1}{N}\ell^{\lambda_N}(\eeta^{\widehat\mca})-\frac{1}{N}\ell^{\lambda_N}(\eeta^{0})\Big]>0\Big) \\
\leq &~ \sum_{m=0}^{|\mca_0|-1} \sum_{j=1}^{|\mca_0|-m} \mathbb P\Big(
\sup_{
 h^2( \eeta^{\mathcal A}, \eeta^{\mathcal A_0})\geq
 \atop
 (|\mca_0|-m)C_{\min}(\eeta^0)}
\frac{1}{N}\Big[\ell(\eeta^{\widehat\mca})-\ell(\eeta^{0})\Big]> 
-\frac{|\lambda_N|\cdot|\mca_0|\cdot|\log \rho_N|}{N} \Big) \\
\leq &~ \sum_{m=0}^{|\mca_0|-1} \sum_{j=1}^{|\mca_0|-m} \mathbb P\Big(\sup_{|\widehat\mca\cap \mathcal A_0|=m}
\frac{1}{N}\Big[\ell(\eeta^{\widehat\mca})-\ell(\eeta^{0})\Big]> 
-t^2 \Big) \quad(\text{by }\eqref{eq-lambda2}) \\
\leq &~ \sum_{m=0}^{|\mca_0|-1} \sum_{j=1}^{|\mca_0|-m} \mathbb P\Big(\sup_{|\widehat\mca\cap \mathcal A_0|=m}
\frac{1}{N}\Big[\ell(\eeta^{\widehat\mca})-\ell(\eeta^{0})\Big]> 
-(|\mca_0|-m)C_{\min}(\eeta^0) \Big)  \quad(\text{by }\eqref{eq-cmin}) \\
\leq &~ \sum_{m=0}^{|\mca_0|-1}{|\mca_0| \choose m}\exp\Big( -c_2 N (|\mca_0|-m)C_{\min}(\eeta^0) \Big)\sum_{j=1}^{|\mca_0|-m}{|\mca_{\inp}|-|\mca_0| \choose j}\quad (\text{by }\eqref{eq-large})\\
\leq &~ c_3\exp\Big( -c_2 N C_{\min}(\eeta^0) + 2\log(|\mca_{\inp}|+1) \Big),
\end{align*}
where the last but one line above uses the large deviation inequality in \eqref{eq-large}, and $c_2,c_3$ are some constants. And the last line follows from the calculations in the proof of Theorem 1 in \cite{shen2012} using some basic inequalities about binomial coefficients.
Since $C_{\min}(\eeta^0)\geq C_0$, and $\log|\mca_{\inp}|=o(N)$ by the assumption of the theorem, the right hand side of the above display goes to zero as $N\to\infty.$ Therefore $\mathbb P(\widehat\mca^{\lambda_N}\neq\mca_0,\,|\widehat\mca^{\lambda_N}|\leq|\mca_0|)\to 0$ as $N\to\infty$.
Combined with the previously shown result $\mathbb P(\widehat\mca^{\lambda_N}\neq \mca_0)\not\to 0$ could potentially happen only for $|\widehat\mca^{\lambda_N}|\leq |\mca_0|$, we have the conclusion $\mathbb P(\widehat\mca^{\lambda_N}\neq\mca_0)\to 0$ as $N\to\infty$.
The proof of the theorem is complete.
\qed

\paragraph{Proof of Theorem \ref{prop-ra}.}
Denote $\theta^+_j = \theta_{j,H}$ and $\theta^-_j = \max_{\aaa\nsucceq\qq_j}\theta_{j,\aaa}$ for each $j$.
Since the screening algorithm is developed for the two-parameter SLAM introduced in Example \ref{exp-dina}, for each item $j$ there are exactly two estimated item parameters, and we denote them by $\widehat\theta_j^+$ and $\widehat\theta_j^-$.
We claim that it suffices to prove that for any $\aaa\in \mca_0$, there exists a response pattern $\rr^{\aaa}\in\{0,1\}^J$ such that as $K\to \infty$,
\begin{equation}\label{eq-ir-claim}
\mathbb P(\RR = \rr^{\aaa},~ \ma=\aaa\mid \TT) >
\mathbb P(\RR = \rr^{\aaa},~ \ma=\wt\aaa\mid \TT),\quad \forall \wt\aaa\neq\aaa.
\end{equation}
For $\aaa\in \mca_0$, define $\rr^{\aaa} = (r^{\aaa}_1,\ldots,r^{\aaa}_J)$ to be 
$r^{\aaa}_j = I(\aaa\succeq \qq_j) =  \prod_{k}\alpha_{k}^{q_{j,k}}$.
For a general structured latent attribute model, consider 
the joint distribution of observed response vector $\RR$ and latent attribute pattern vector $\ma$ is
\begin{align}
&\mathbb P(\RR = \rr,~ \ma=\aaa\mid \TT) =
 \exp\Big\{\sum_{j=1}^J\Big[ r_{j}\Big(  \prod_{k}\alpha_{k}^{q_{j,k}} \log \theta^+_j + (1-\prod_{k}\alpha_{k}^{q_{j,k}}) \log \theta^-_{j,\aaa} \Big) + \\ \notag
& \quad\quad\quad (1-r_{j}) \Big(  \prod_{k}\alpha_{k}^{q_{j,k}} \log(1- \theta^+_j) + (1-\prod_{k}\alpha_{k}^{q_{j,k}}) \log(1- \theta^-_{j,\aaa}) \Big)\Big]\Big\}.
\end{align}
Therefore
\begin{align*}
&\mathbb P(\RR = \rr^{\aaa},~ \ma=\wt\aaa\mid \TT) =
 \exp\Big\{\sum_{j=1}^J\Big[ \prod_{k}\alpha_{k}^{q_{j,k}}\Big(  \prod_{k}\wt\alpha_{k}^{q_{j,k}} \log \theta^+_j + (1-\prod_{k}\wt\alpha_{k}^{q_{j,k}}) \log \theta^-_{j,\widetilde\aaa} \Big) + \\ \notag
& \quad\quad\quad (1-\prod_{k}\alpha_{k}^{q_{j,k}}) \Big(  \prod_{k}\wt\alpha_{k}^{q_{j,k}} \log(1- \theta^+_j) + (1-\prod_{k}\wt\alpha_{k}^{q_{j,k}}) \log(1- \theta^-_{j,\widetilde\aaa}) \Big)\Big]\Big\}.\\
&\mathbb P(\RR = \rr^{\aaa},~ \ma=\aaa\mid \TT) =
 \exp\Big\{\sum_{j=1}^J\Big[  \prod_{k}\alpha_{k}^{q_{j,k}} \log \theta^+_j  +  (1-\prod_{k}\alpha_{k}^{q_{j,k}}) \log(1- \theta^-_{j,\aaa})\Big] \Big\}.
\end{align*}
Then for any $\wt\aaa\neq \aaa$, 
\begin{align}\label{eq-domi}
&\log\mathbb P(\RR = \rr^{\aaa},~ \ma=\aaa\mid \TT) - \log \mathbb P(\RR = \rr^{\aaa},~ \ma=\widetilde\aaa\mid \TT) \\ \notag
\geq & \min_{j=1,\ldots,J}\{\log \theta^+_j - \log \theta^-_{j,\widetilde\aaa},~ \log (1-\theta^+_{j,\aaa}) - \log (1-\theta^+_j)\}\geq d >0.
\end{align}
That the above probability is bounded away from zero follows from the second part of  assumption \eqref{eq-min}. So the claim \eqref{eq-ir-claim} is proved.
We next bound the probability of failure of including all the true patterns in the screening stage. First, since $\ma_1,\ldots, \ma_N\stackrel{i.i.d.}{\sim} \text{Multinomial}(N,~(p_{\aaa},\aaa\in \mca_0))$, then $|\{i\in[N]: \ma_i=\aaa\}|$ denotes the number of subjects in the random sample whose attribute pattern is $\aaa$. By the concentration inequality of the multinomial distribution, for any $\aaa\in \mca_0$,
$$
\mathbb P\Big(\Big|\{i\in[N]: \ma_i=\aaa\}\Big| \geq Np_{\aaa}-2\sqrt{N}t\Big) \geq 1- 2^{|\mca_0|}\exp(-2t^2),\quad \forall t>0.
$$
Because of \eqref{eq-min}, we have $Np_{\aaa}\geq Nc_0\to\infty$ for all $\aaa\in\mca_0$.
Assume that $\widehat \theta^+_j-\widehat \theta^-_j>\delta>0$ for each $j\in[J]$. This constraint can be incorporated into the screening procedure or checked \textit{a posteiriori} after screening.
So with probability at least $1 - 2^{|\mca_0|}\exp(-2t^2)$ for a suitable $t$,
\begin{align}
&\mathbb P(\widehat\mca_{\text{screen}}\nsupseteq  \mca_0)
\leq  \sum_{\aaa\in \mca_0} \mathbb P( \widehat\ma_i\neq\aaa ~\forall i\in[N]~\text{s.t.}~\ma_i=\aaa) \\ \notag
\leq & \sum_{\aaa\in \mca_0}
\Big[\mathbb P\Big(\RR_{i}=\rr^{\aaa},~\exists\wt\aaa\neq\aaa,\\ \notag
&\quad \widehat{\mathbb P}(\RR=\rr^{\aaa}, \ma=\aaa) > \widehat{\mathbb P}(\RR=\rr^{\aaa}, \ma=\wt\aaa)\Big| \ma_{i}=\aaa\Big)\Big]^{N \big(p_{\aaa} - 2t/\sqrt{N}\big)}
\to 0,
\end{align}
as $N\to\infty$. Here $\widehat{\mathbb P}$ refers to the probability measure of $\RR$ and $\ma$ given the estimated item parameters $\widehat\ttt^+$ and $\widehat\ttt^-$.
This is because the probability inside the bracket in the above expression is strictly less than $1$ due to \eqref{eq-domi}; we denote this quantity by $C_{\delta}$ since it depends on $\delta$. Therefore there is 
\begin{align*}
\mathbb P(\widehat\mca_{\text{screen}}\nsupseteq  \mca_0) 
\leq 
\sum_{\aaa\in\mca_0} C_{\delta}^{N( p_{\aaa}+o(1))}
& =\sum_{\aaa\in\mca_0} \exp[-N (p_{\aaa} + o(1))\log(1/C_{\delta}) ]\\
& \leq |\mca_0|\exp(-N \beta_{\min}),
\end{align*}
where $\beta_{\min}$ is a positive constant which can be taken as $ c_0/2 \log(1/C_{\delta})$.
The last inequality above results from $p_{\aaa}\geq c_0$ for $\aaa\in\mca_0$ in \eqref{eq-min} and that $C_{\delta}<1$. Now we have obtained $\mathbb P(\widehat\mca_{\text{screen}}\supseteq  \mca_0) \geq 1- |\mca_0|\exp(-N\beta_{\min})$,
so the sure screening property holds and the proof is complete.
\qed

\paragraph{{\bf Proof of Lemma \ref{lem-ent}.}}
Following  the proof of Theorem 2 in \cite{genovese}, the overall bracketing entropy of the mixture distribution over $|\mca|$ mixture components (latent attribute patterns) can be bounded by the entropy of the $|\mca|-1$ dimensional simplex multiplied by the product of the entropy of the item parameter space for each mixture component. Since there are a total number of ${|\mca_{\inp}| \choose |\mca|}$ possibilities of choosing $|\mca|$ components from $|\mca_{\inp}|$ ones, we have
$$
N_{[]}(t,\mathcal B_{\mca}) \leq {|\mca_{\inp}| \choose |\mca|}  N_{[]}(t,\mathcal T^{|\mca| - 1})\prod_{l=1}^{|\mca|} N_{[]}(t/3,\mathcal F_l).
$$
Next,  Lemma 2 in \cite{genovese} gives the following bracketing entropy bound for the simplex,
$N_{[]}(t,\mathcal T^{|\mca| - 1}) \leq {|\mca| (2\pi e)^{|\mca|/2}}/{t^{|\mca| - 1}}.$
Since we consider the local parameter space around the true parameters (with squared Hellinger distance between the alternative model and the true model not greater than $2\epsilon^2$), the $1/t$ in the above display can be replaced by $\epsilon/t$.
Also, $N_{[]}(t/3,\mathcal F_l)\leq C_0\epsilon/t$ since the Hellinger distance is bounded by the $L_2$ distance and the $t$-bracketing number under the $L_2$ norm is bounded by $O(\epsilon/t)$. 
Therefore we have
\begin{align*}
H(t,\mathcal B_{\mca}) 
&\leq \log \Big\{ {|\mca_{\inp}| \choose |\mca|}  \frac{|\mca|(2\pi e)^{|\mca|/2}(\epsilon)^{|\mca|-1}}{t^{|\mca|-1}}
\Big( \frac{\epsilon}{t} \Big)^{|\mca|} \Big\} \\
&\lesssim |\mca|\log |\mca_{\inp}| + \log|\mca| + |\mca|\log(\epsilon/t) \\
&\lesssim |\mca_0|\log |\mca_{\inp}|\log(\epsilon/t).
\end{align*}
where $|\mca|\leq |\mca_0|$ and an elementary inequality ${a\choose b}\leq a^b$ are used.
\qed

\subsection*{A.2 Additional Experimental Results}

\paragraph{Impact of the value of the  pre-specified $c$ in Algorithm \ref{algo-pem}.}
In Algorithm \ref{algo-pem}, there is a pre-specified constant $c>0$ when updating the $\Delta_l$'s. This constant $c$ should be small, ideally close to zero. In all of our experiments in Section \ref{sec-simu}, we take $c=0.01$. Next we examine how the value of $c$ impacts the selection result of Algorithm \ref{algo-pem}. Since the performance of Algorithm \ref{algo-pem} is the focus here, we choose the simulation setting with $K=10$ such that  screening  can be omitted.
Under sample sizes $N=150$ and $N=500$, the plots of the two accuracy measures versus $c$ are presented in Figure \ref{fig-thres-c}. We observe that the results of Algorithm \ref{algo-pem} are generally not that sensitive to the choice of $c$, though smaller $c$ gives slightly better results for both accuracy measures under a small sample size $N=150$. For $N$ as large as $500$, for all the values of $c\in\{0.001,0.005\}\cup\{0.01\times i:\,i=1,2,\ldots,10\}$, the two accuracy measures are very close to one and do not have much variation.
In practice, we recommend fixing $c$ to a value no greater than $0.01$.

\begin{figure}[h!]
\centering
\begin{minipage}[b]{0.24\textwidth}
	\centering
	  $\vcenter{\includegraphics[width=\textwidth]{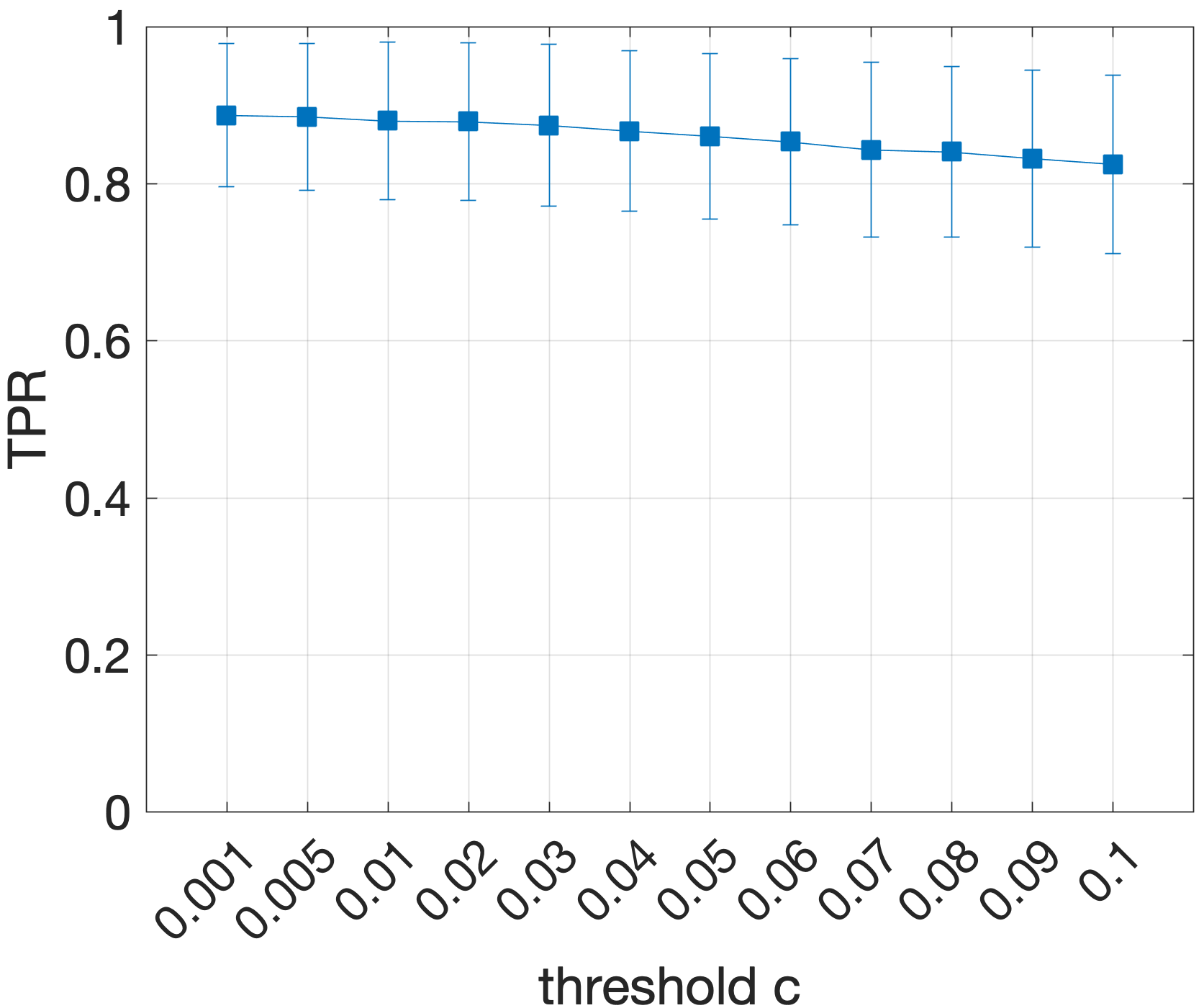}
	  }$
	\end{minipage}
	\hfill
    \begin{minipage}[b]{0.24\textwidth}
    \centering
    $\vcenter{\includegraphics[width=\textwidth]{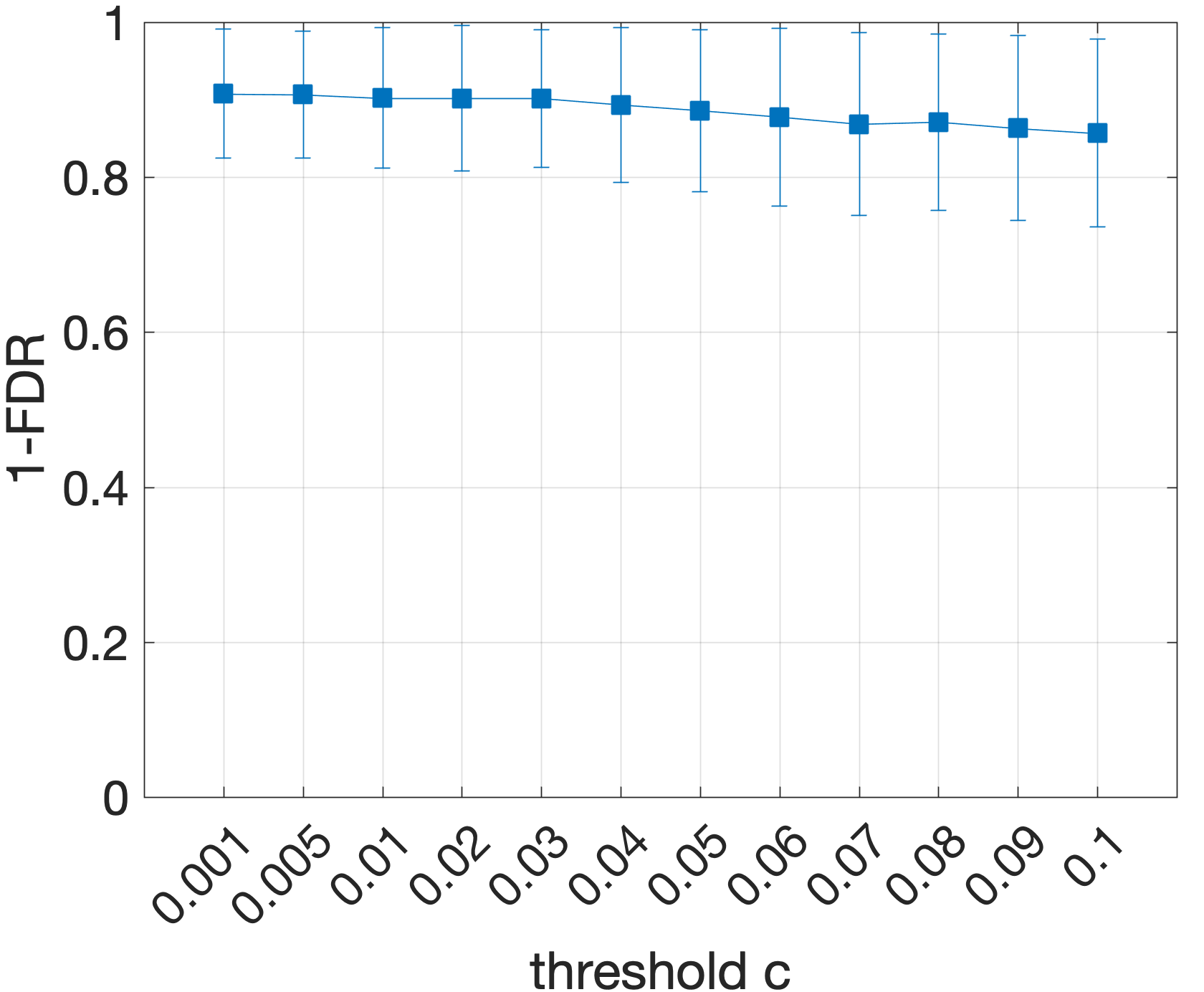}}$
	\end{minipage}
	\hfill
	\begin{minipage}[b]{0.24\textwidth}
	\centering
	  $\vcenter{\includegraphics[width=\textwidth]{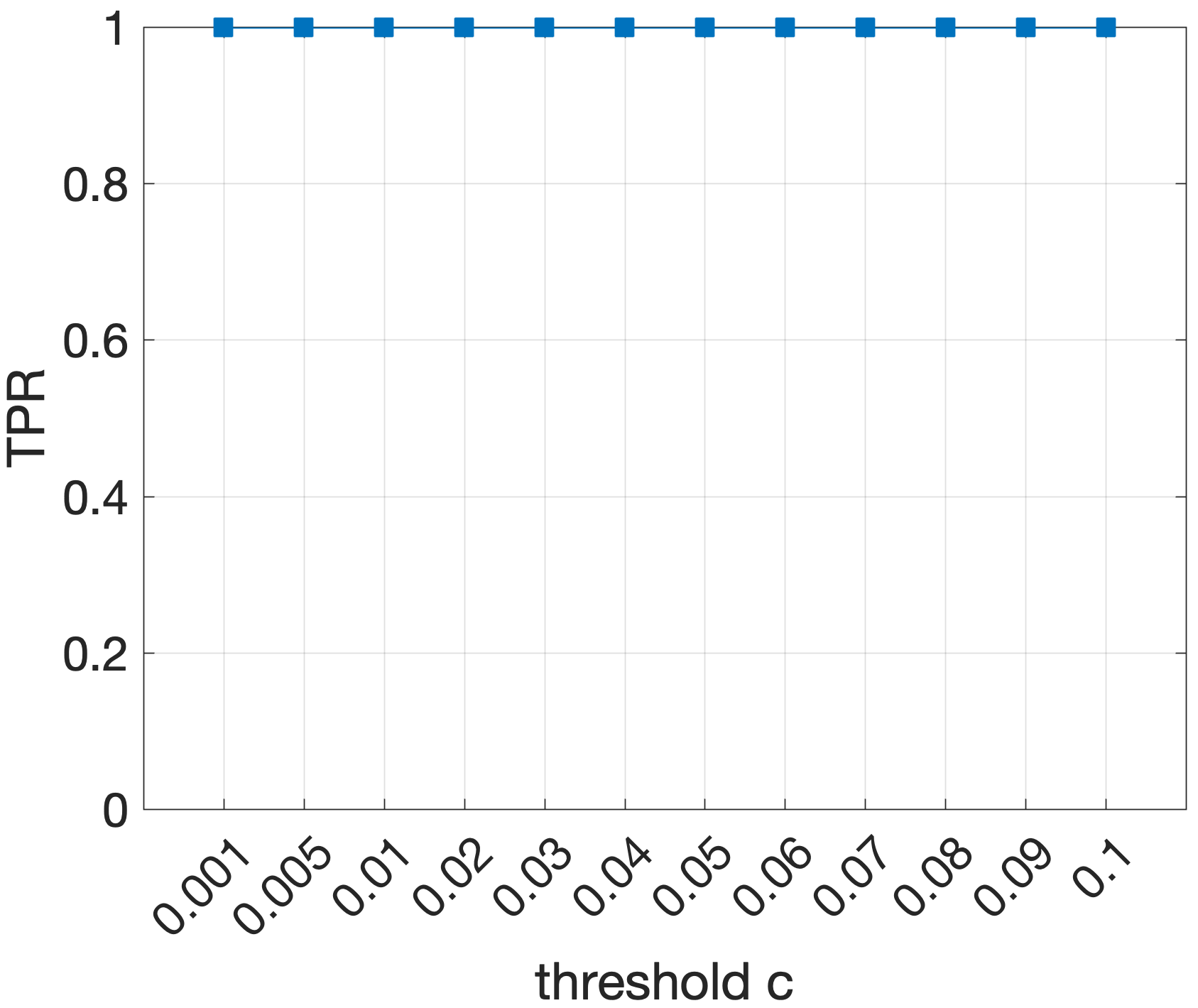}
	  }$
	\end{minipage}
	\hfill
    \begin{minipage}[b]{0.24\textwidth}
    \centering
    $\vcenter{\includegraphics[width=\textwidth]{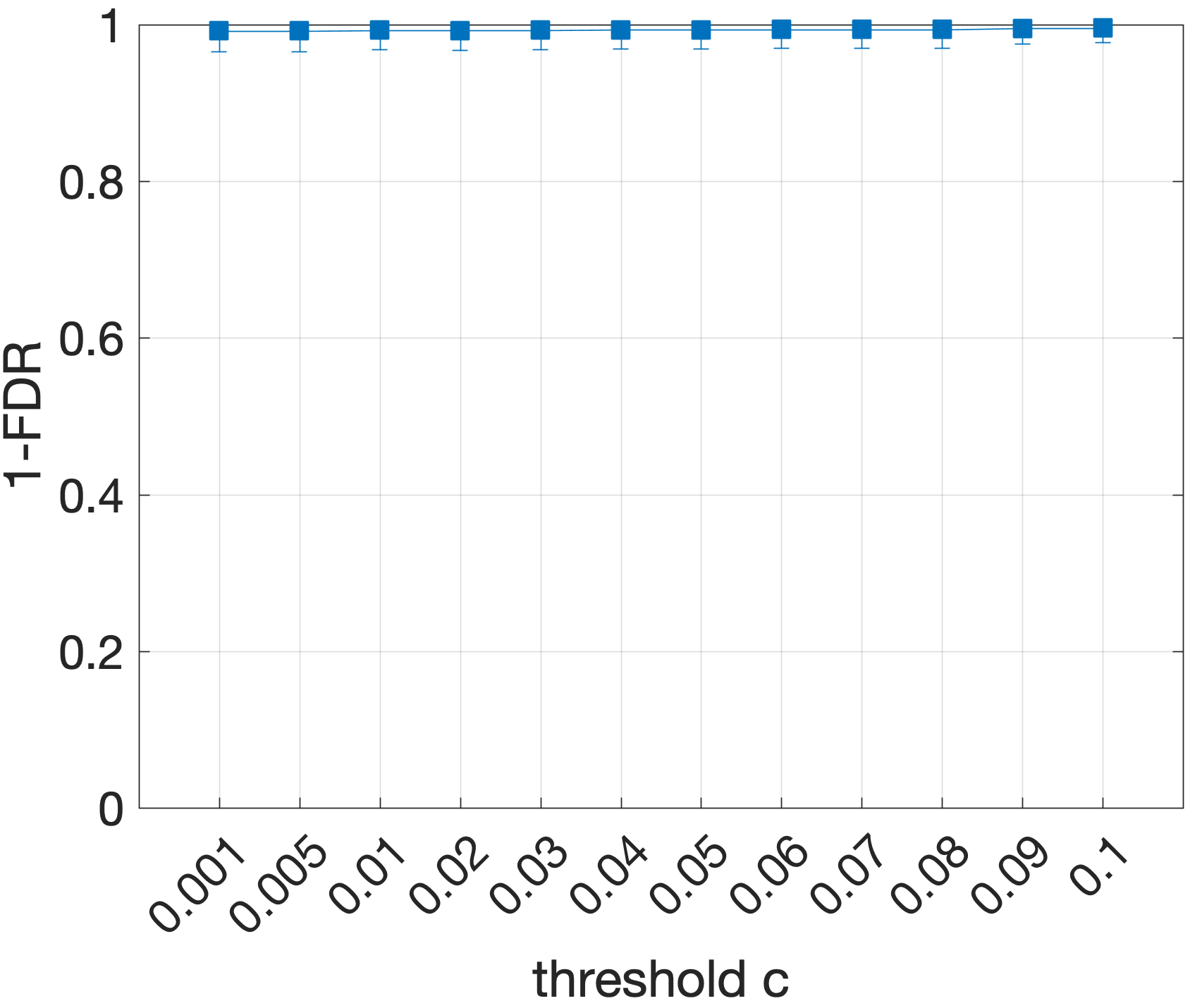}}$
	\end{minipage}
	
    \begin{minipage}[b]{0.24\textwidth}\centering
	\small{(a) TPR versus $c$ in Algo. \ref{algo-pem}, $N=150$}
	\end{minipage}
	\hfill
    \begin{minipage}[b]{0.24\textwidth}\centering
    \small{(b) $1-$FDR versus $c$ in Algo. \ref{algo-pem}, $N=150$}
	\end{minipage}
	\hfill
	\begin{minipage}[b]{0.24\textwidth}\centering
	\small{(c) TPR versus $c$ in Algo. \ref{algo-pem}, $N=500$}
	\end{minipage}
	\hfill
    \begin{minipage}[b]{0.24\textwidth}\centering
    \small{(d) $1-$FDR versus $c$ in Algo. \ref{algo-pem}, $N=500$}
	\end{minipage}

\caption{
{Performance  of Algorithm \ref{algo-pem} across various values for threshold $c$. Setting is $K=10$ and $1-\theta_j^+=\theta_j^-=0.2$. In each scenario 200 runs are carried out, and the error bar is within one standard deviation of the mean accuracy. }
}
\label{fig-thres-c}
\end{figure}

\paragraph{Algorithm \ref{algo-pem}'s performance on estimating the actual proportions of patterns.}
Other than the two accuracy measures for pattern selection presented in Table \ref{tab-dina}, we also evaluate how well the algorithms perform on estimating the actual proportions of the latent patterns. We use the simulation setting of the two-parameter SLAM with $K=10$, $|\mca_0|=10$, $Q=(Q_1^\top,Q_2^\top,Q_3^\top)^\top$, with parameters $p_{\aaa}=0.1$ for $\aaa\in\mca$ and $1-\theta_j^+=\theta_j^-=0.2$. This is the same setting as that of Example \ref{exp-path}.
We vary the sample size $N\in\{150, 300, 600, 900, 1200\}$ and compute the Root Mean Square Errors (RMSEs) of estimating the true proportions of latent patterns. The randomly generated 10 true patterns in $\mca_0$ are presented in Figure \ref{fig-rmse}(a), where each row represents a $K$-dimensional binary pattern. For each $N$, the RMSE of each proportion $p_{\aaa}$, $\aaa\in\mca_0$ is computed based on 200 runs; and in each run, we first perform pattern selection by using EBIC to choose $\lambda\in\{-0.2\times i:\,i=1,2,\ldots,20\}$ in Algorithm \ref{algo-pem} and then estimate the proportions based on the selected set of patterns.
 The  results of RMSEs are presented in Figure \ref{fig-rmse}(b).
As can be seen from the figure, under a small sample size $N=150$, the RMSEs of patterns are relatively diverse. In particular, the largest RMSE is around $0.06$ and corresponds to pattern 10, $\aaa_{10} = (0010000010)$, which is the pattern consisting of most ``0''s; while the smallest RMSE is less than half of the largest and corresponds to pattern 3, $\aaa_3 = (1110011111)$, which is the pattern consisting of most ``1''s. Interestingly, this observation implies for a very small sample size and a sparse $Q$-matrix (each row having at most three entries of ``1''s), those attribute patterns possessing fewer attributes are harder to estimate while those possessing more attributes are easier to estimate. While as $N$ increases, the RMSEs of all the proportions decrease and their difference become not discernible. For  $N=1200$, all the RMSEs are around 0.01.

\begin{figure}[h!]
\centering
\begin{minipage}[b]{0.43\textwidth}
	\centering
	  $\vcenter{\includegraphics[width=0.88\textwidth]{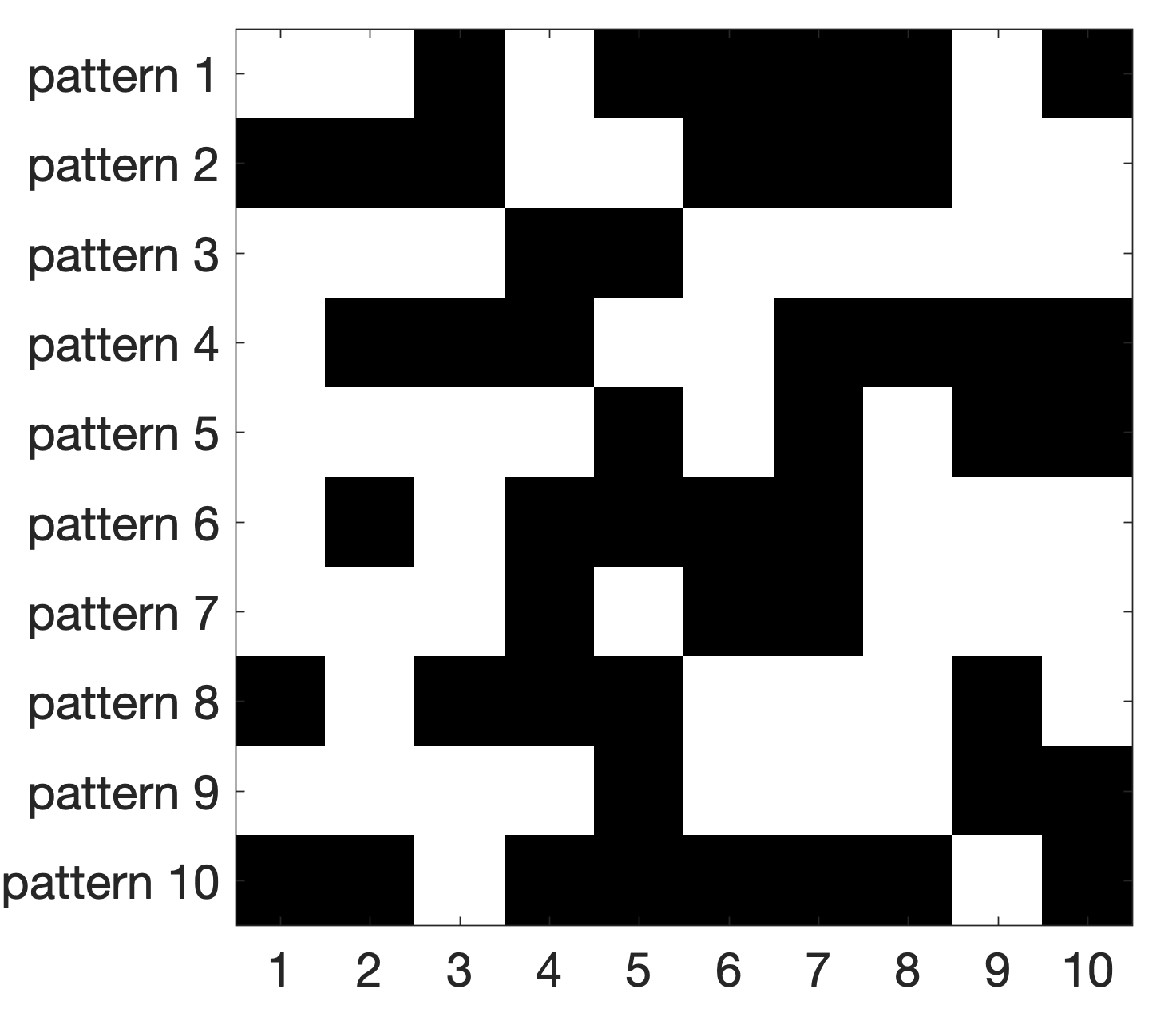}
	  }$
	\end{minipage}
	\hfill
    \begin{minipage}[b]{0.55\textwidth}
\centering
$\vcenter{\includegraphics[width=0.88\textwidth]{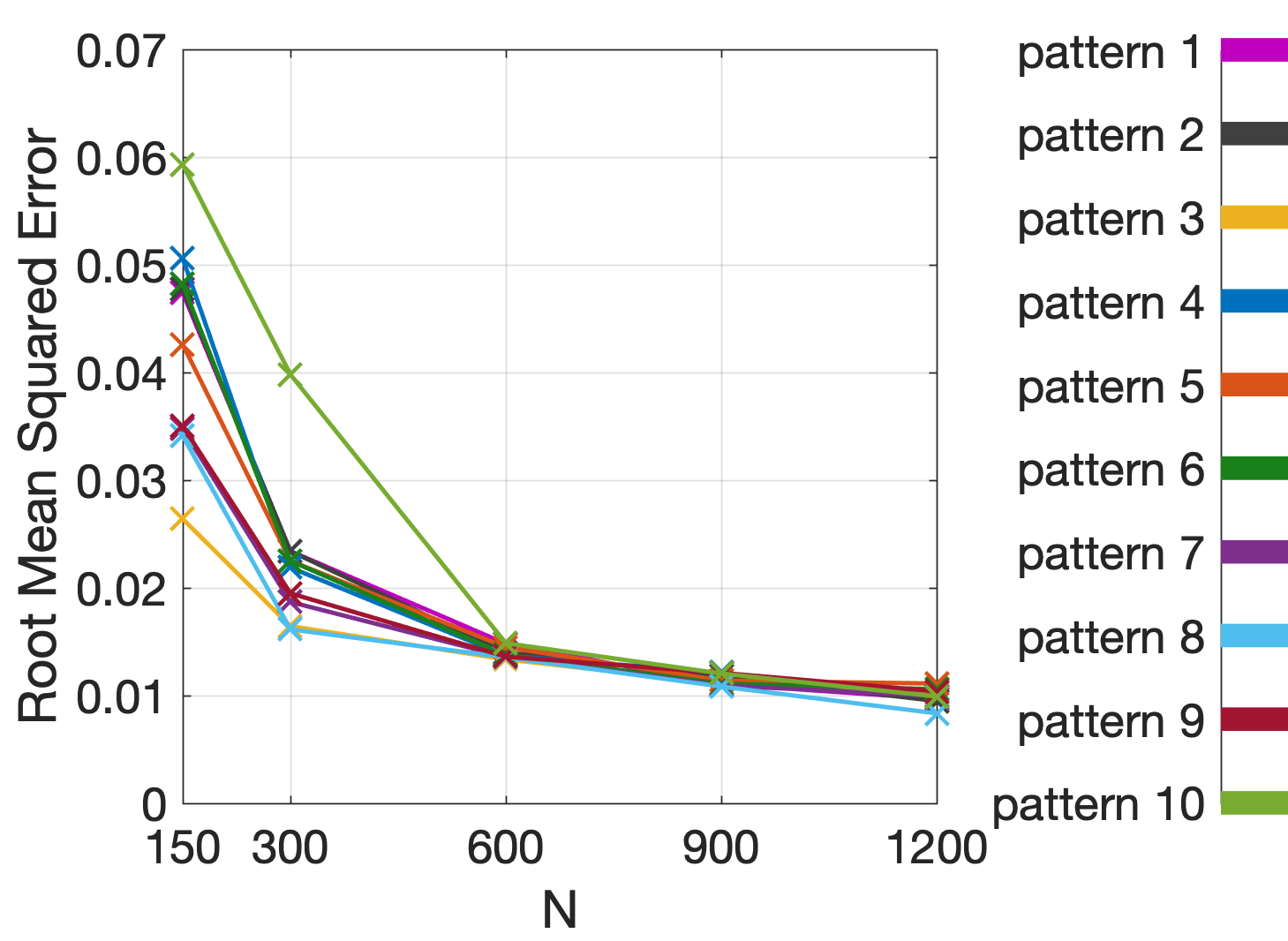}}$
	\end{minipage}
	
	\begin{minipage}[b]{0.43\textwidth}\centering
	\small{(a) patterns in $\mca_0$: white for 1, black for 0}
	\end{minipage}
	\hfill
    \begin{minipage}[b]{0.55\textwidth}\centering
    \small{(b) RMSE of $\{p_{\aaa}:\,\aaa\in\mca_0\}$ versus $N$}
	\end{minipage}

\caption{
{Root Mean Square Errors (RMSEs) for estimating the true proportions of patterns decrease as sample size $N$ increases. Results are based on 200 runs for each $N$.}
}
\label{fig-rmse}
\end{figure}

\paragraph{Evaluating the screening procedure under the multi-parameter SLAM.} In the multi-parameter setting, we also evaluate the performance of the approximate screening procedure that is developed based on the likelihood of the two-parameter model. 
 The results of the coverage probabilities are presented in Figure \ref{fig-screen-g}. The figure shows that despite being an approximate procedure, the screening Algorithm \ref{algo-screen} has excellent performance for the multi-parameter SLAM that covers the two-parameter model as a submodel. Specifically, Figure \ref{fig-screen-g} shows that for both $K=15$ and $K=20$, the approximate screening procedure almost always has a $100\%$ coverage probability for $N=500$ and $N=1000$.

\begin{figure}[h!]
\centering
\begin{minipage}[b]{0.48\textwidth}
	\centering
	  \includegraphics[width=0.7\textwidth]{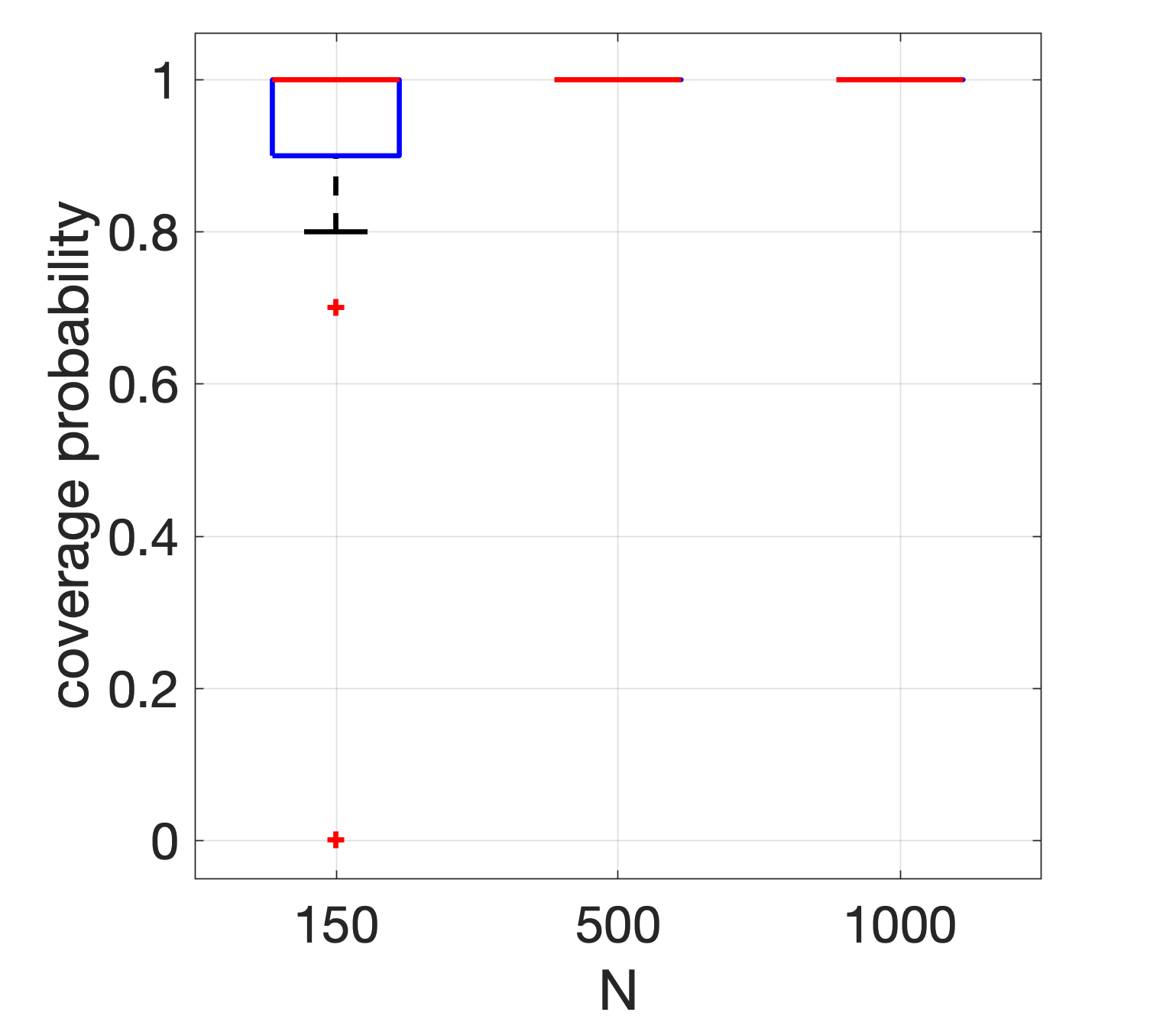}
	\end{minipage}
	%
    \begin{minipage}[b]{0.48\textwidth}
\centering
\includegraphics[width=0.7\textwidth]{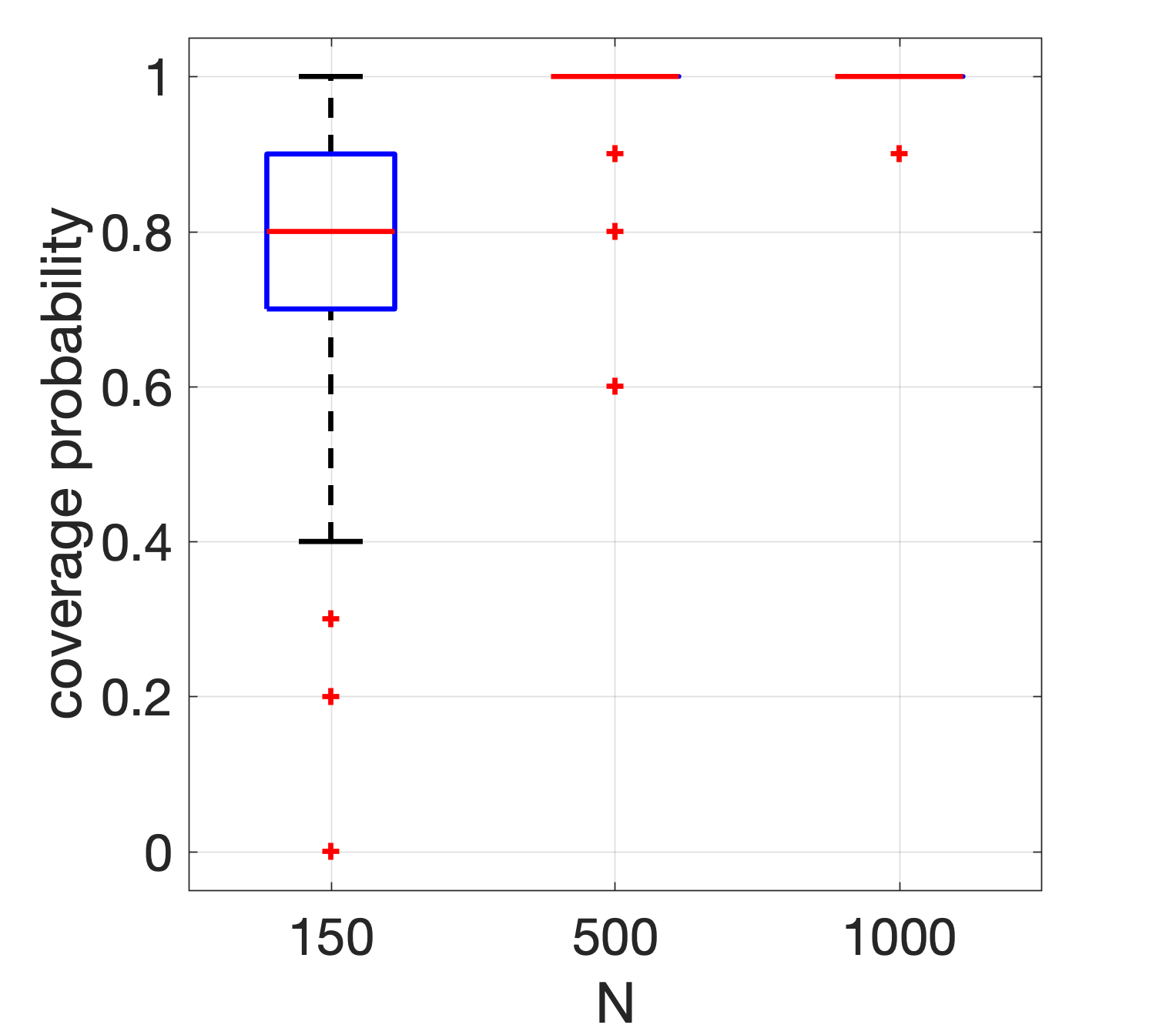}
	\end{minipage}

	\begin{minipage}[b]{0.48\textwidth}
	\centering
	  \small{(a) $K=15$, multi-parameter SLAM}
	\end{minipage}
	%
    \begin{minipage}[b]{0.48\textwidth}
\centering
    \small{(b) $K=20$, multi-parameter SLAM}
	\end{minipage}

\caption{
{Coverage probabilities of the true patterns, from the approximate screening procedure under the multi-parameter SLAM. Boxplots are from 200 runs in each scenario.}
}
\label{fig-screen-g}
\end{figure}


\paragraph{Sizes of the set of finally selected patterns under scenarios in Table \ref{tab-dina}.}
We   present the results of the number of patterns that are finally selected by the proposed methods, corresponding to simulation scenarios in Table \ref{tab-dina}. Denote the set of patterns selected by the PEM algorithm and that selected by the FP-VEM algorithm by $\widehat\mca_{\text{PEM}}$ and $\widehat\mca_{\text{FP-VEM}}$, respectively.
As shown in Figure \ref{fig-screen-size}, in the relatively strong signal setting with $1-\theta_j^+=\theta_j^-=10\%$, the sizes of $\widehat\mca_{\text{PEM}}$ and $\widehat\mca_{\text{FP-VEM}}$  almost always equal 10, the number of true patterns. Combined with the accuracy measures presented in Table \ref{tab-dina} in the main text, in most cases these selected 10 patterns are indeed the true ones in $\mca_0$. And in the relatively weak signal setting with $1-\theta_j^+=\theta_j^-=20\%$, the sizes of $\widehat\mca_{\text{PEM}}$ and $\widehat\mca_{\text{FP-VEM}}$ can be slightly larger than $|\mca_0|$ but still close to it.

\begin{figure}[h!]
\centering
\begin{minipage}[b]{0.24\textwidth}
	\centering
	  \includegraphics[width=\textwidth]{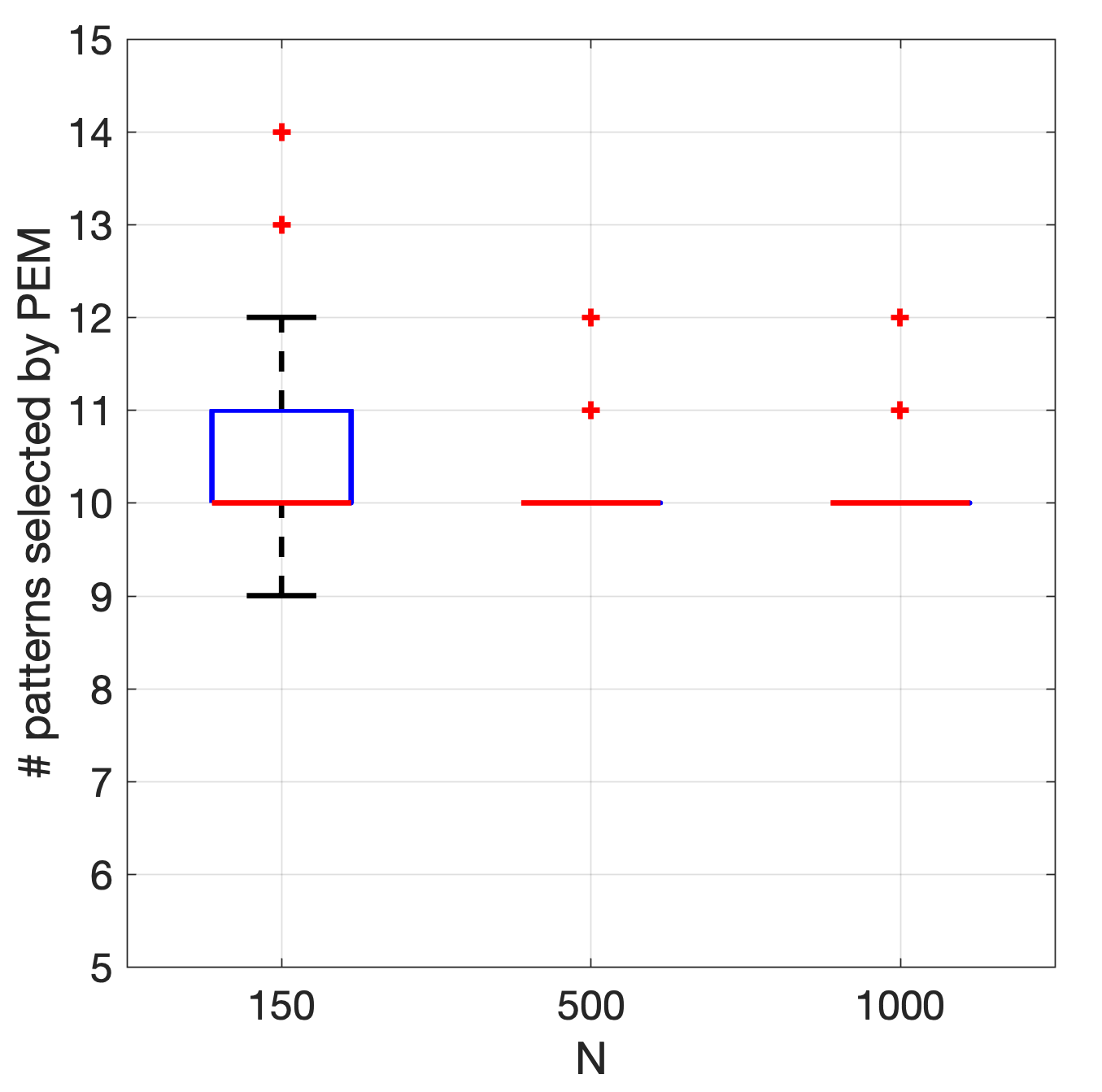}
	\end{minipage}
	\hfill
    \begin{minipage}[b]{0.24\textwidth}
\centering
\includegraphics[width=\textwidth]{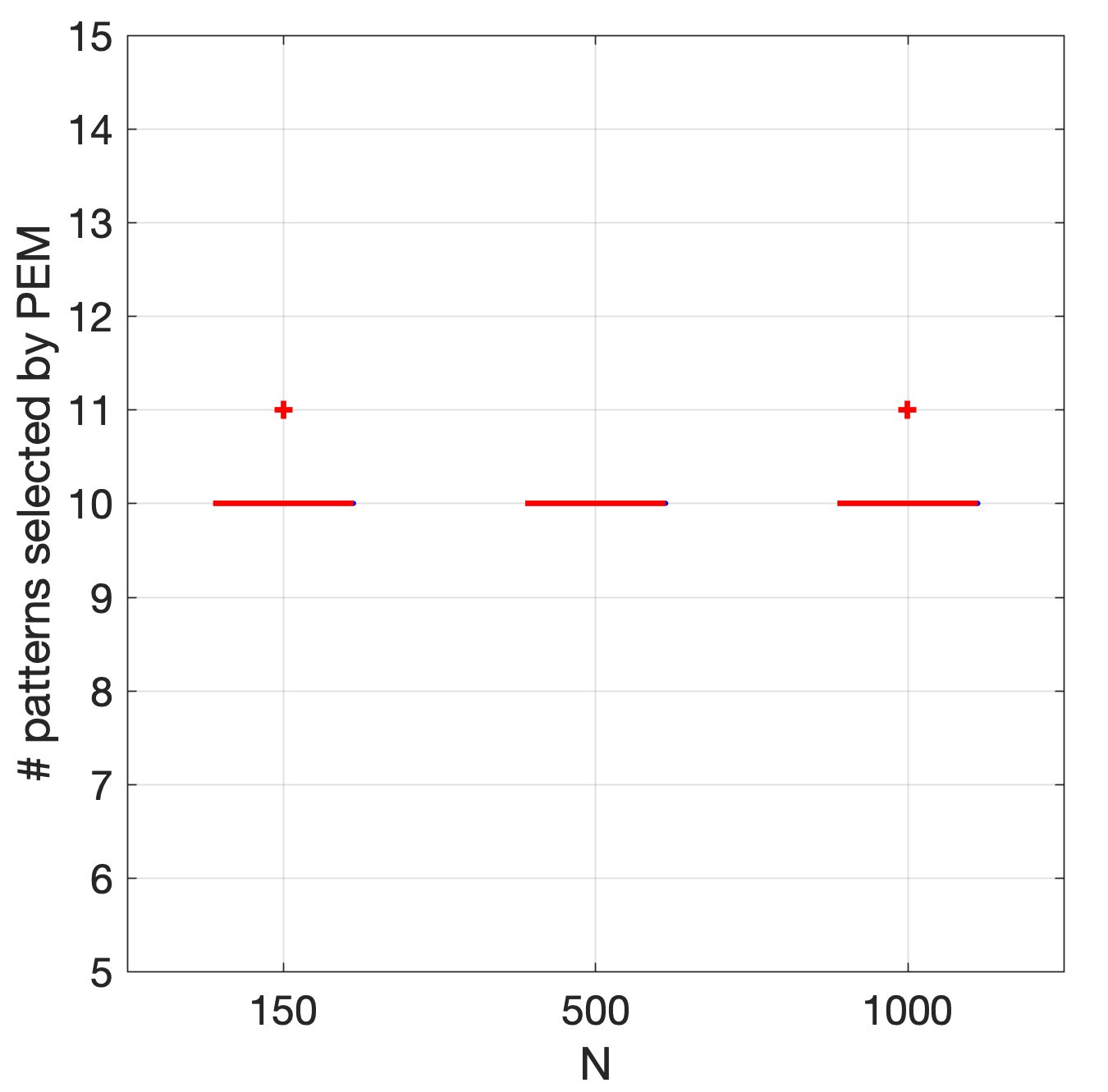}
	\end{minipage}
	\hfill
	\begin{minipage}[b]{0.24\textwidth}
	\centering
	  \includegraphics[width=\textwidth]{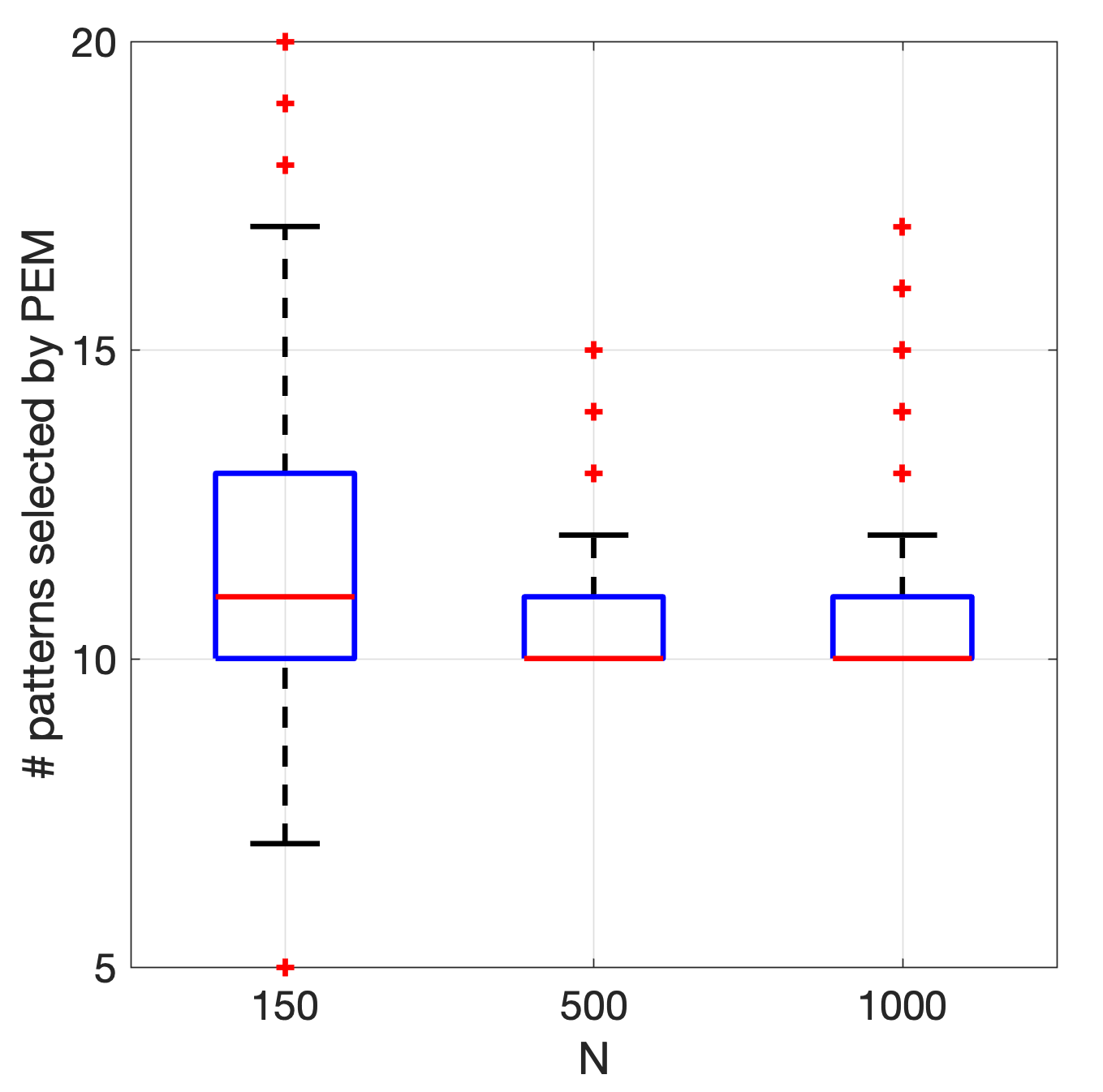}
	\end{minipage}
	\hfill
    \begin{minipage}[b]{0.24\textwidth}
\centering
\includegraphics[width=\textwidth]{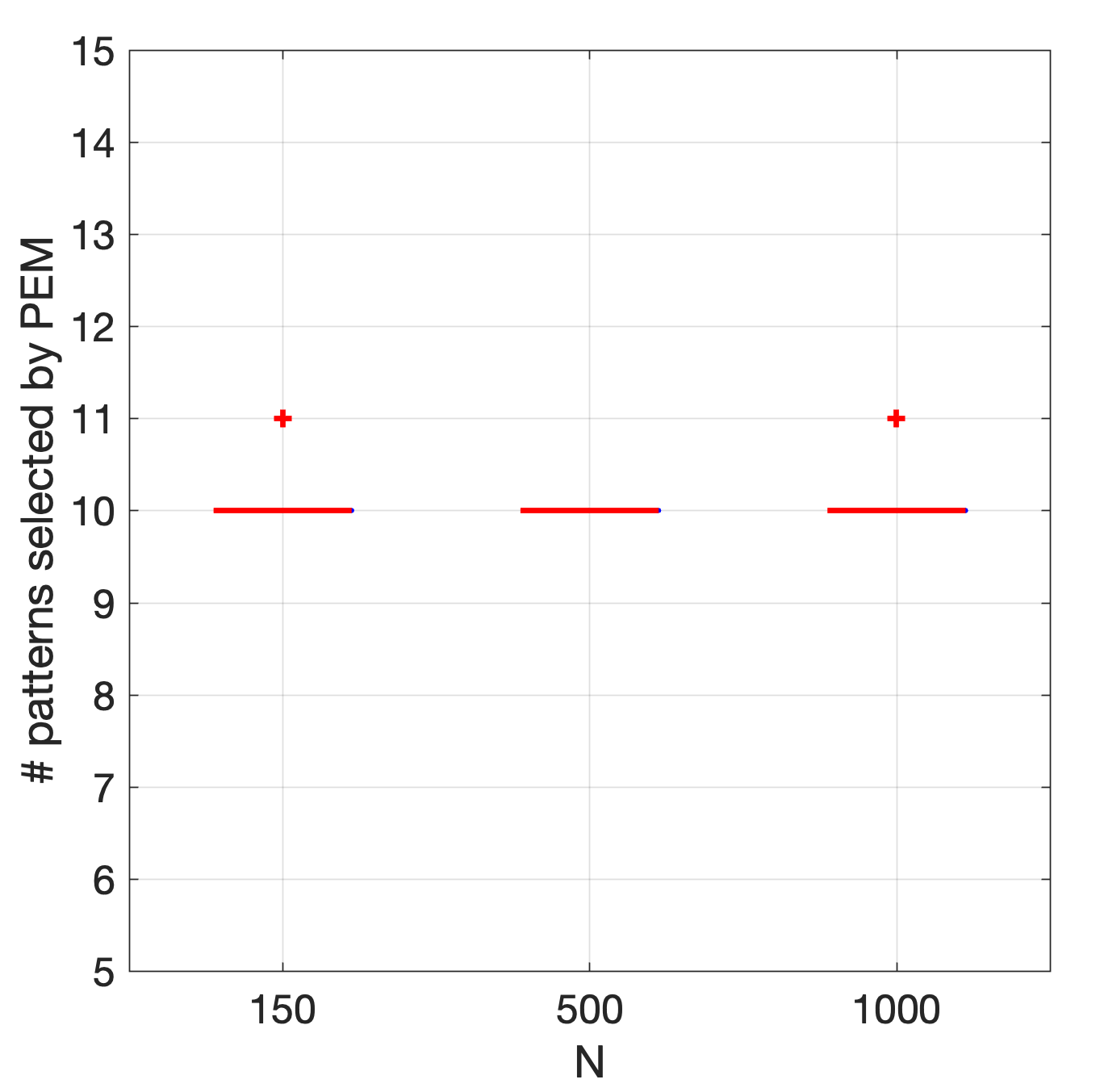}
	\end{minipage}
	
	\begin{minipage}[b]{0.24\textwidth}
	\centering
	  \small{(a) $K=15$, noise $20\%$; PEM}
	\end{minipage}
	\hfill
    \begin{minipage}[b]{0.24\textwidth}
\centering
    \small{(b) $K=15$, noise $10\%$; PEM}
	\end{minipage}
	\hfill
	\begin{minipage}[b]{0.24\textwidth}
	\centering
	\small{(c) $K=20$, noise $20\%$; PEM}
	\end{minipage}
	\hfill
    \begin{minipage}[b]{0.24\textwidth}
\centering
    \small{(d) $K=20$, noise $10\%$; PEM}
	\end{minipage}

	\begin{minipage}[b]{0.24\textwidth}
	\centering
	  \includegraphics[width=\textwidth]{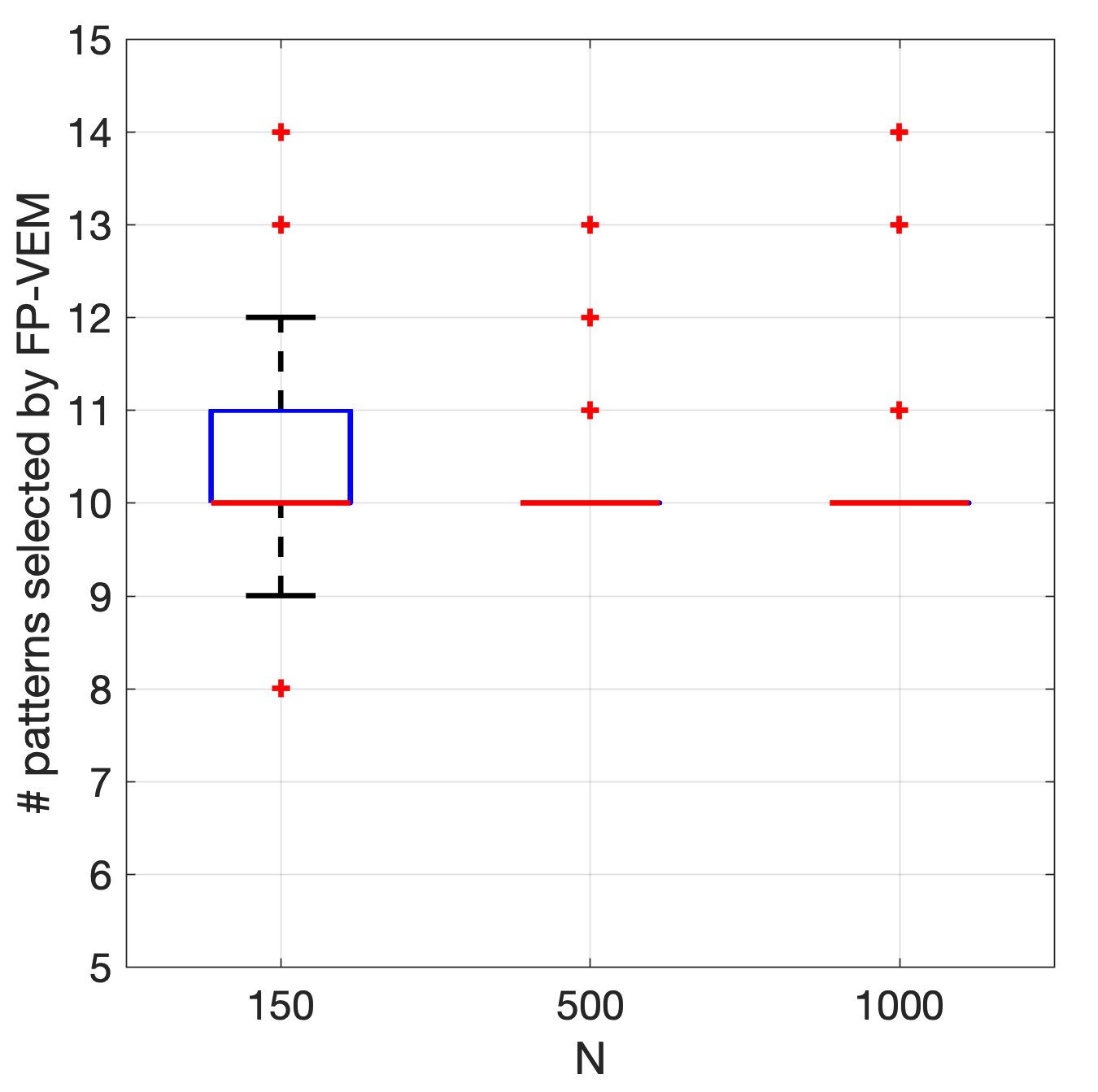}
	\end{minipage}
	\hfill
    \begin{minipage}[b]{0.24\textwidth}
\centering
\includegraphics[width=\textwidth]{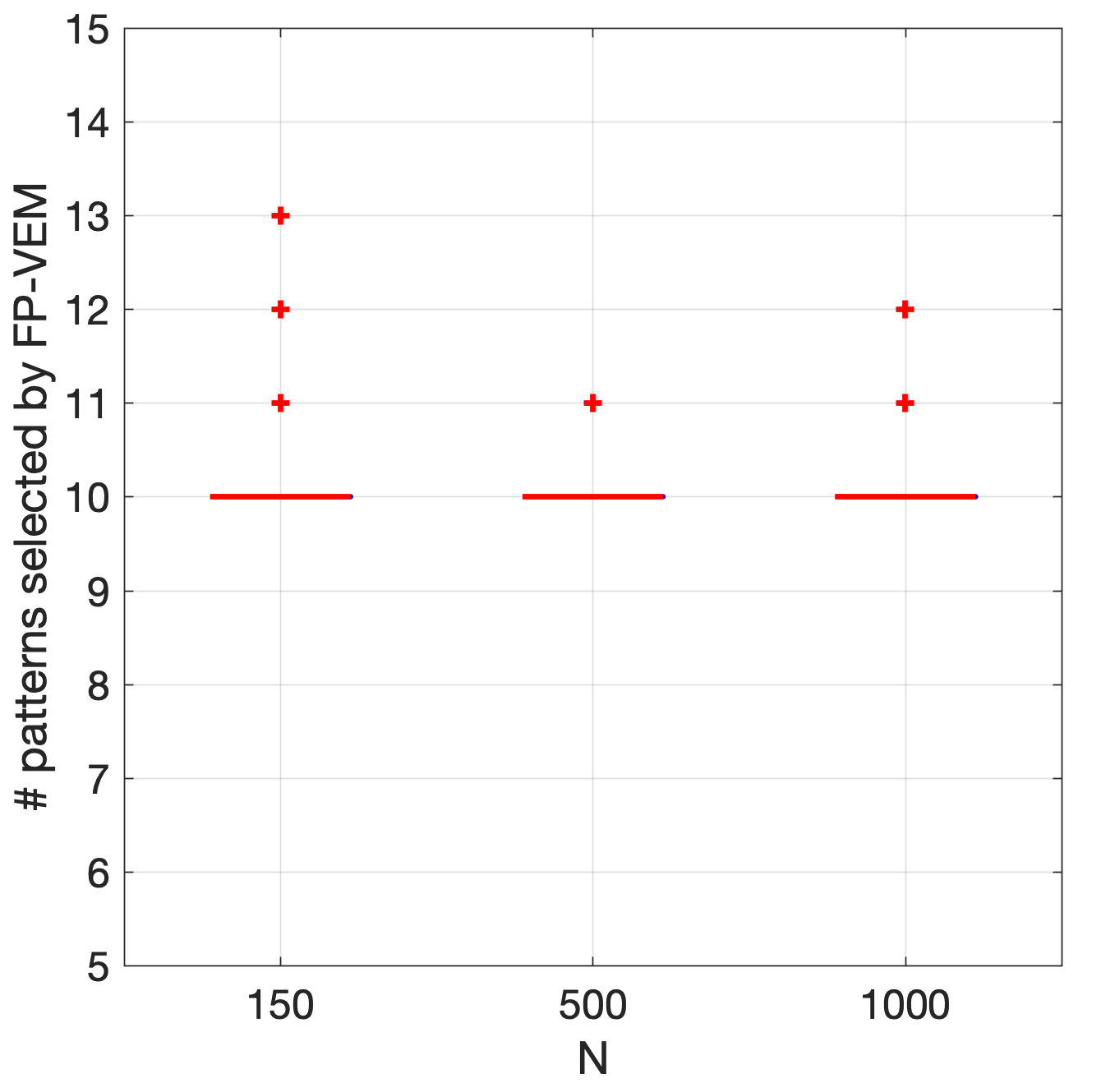}
	\end{minipage}
	\hfill
	\begin{minipage}[b]{0.24\textwidth}
	\centering
	  \includegraphics[width=\textwidth]{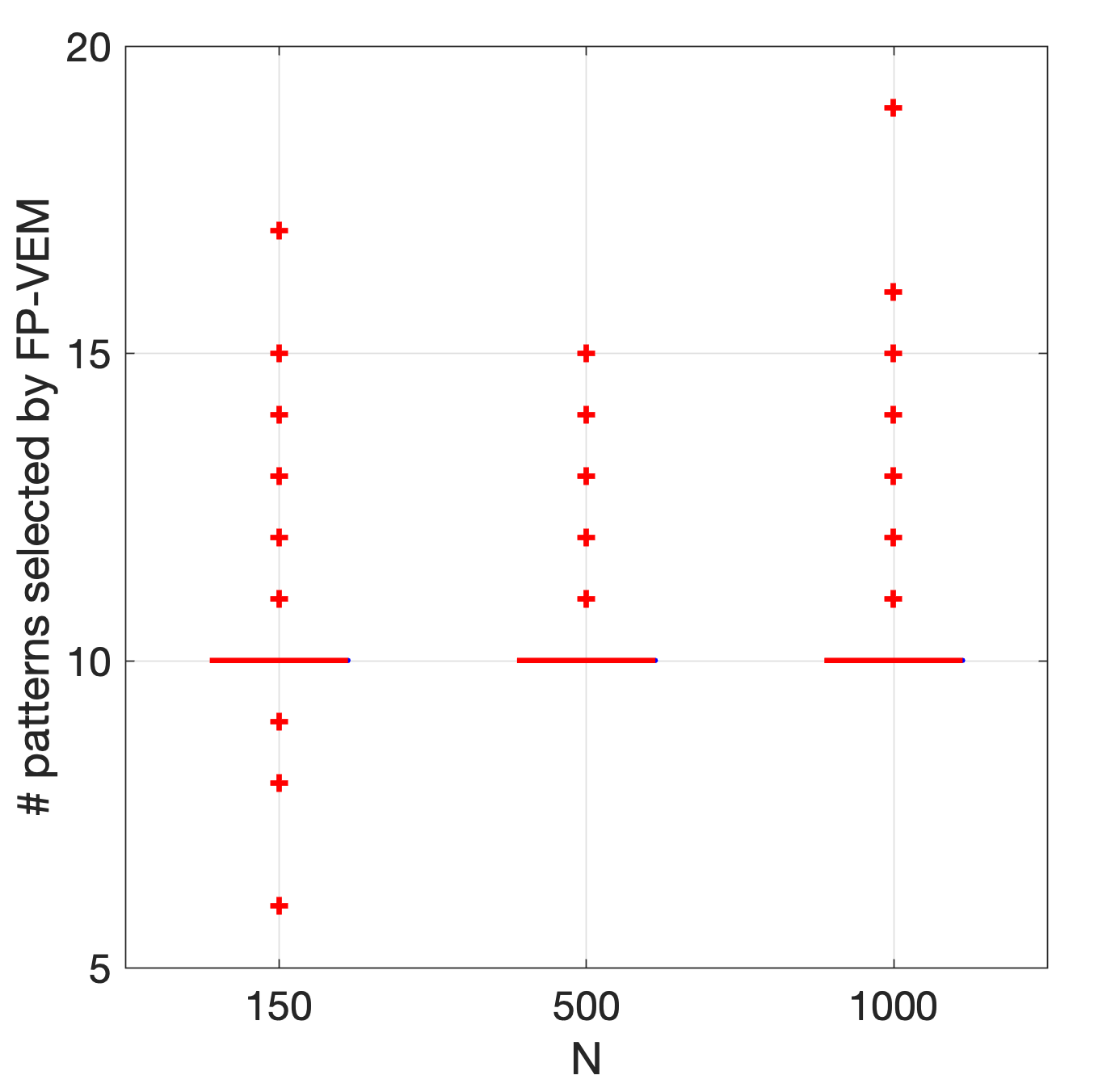}
	\end{minipage}
	\hfill
    \begin{minipage}[b]{0.24\textwidth}
\centering
\includegraphics[width=\textwidth]{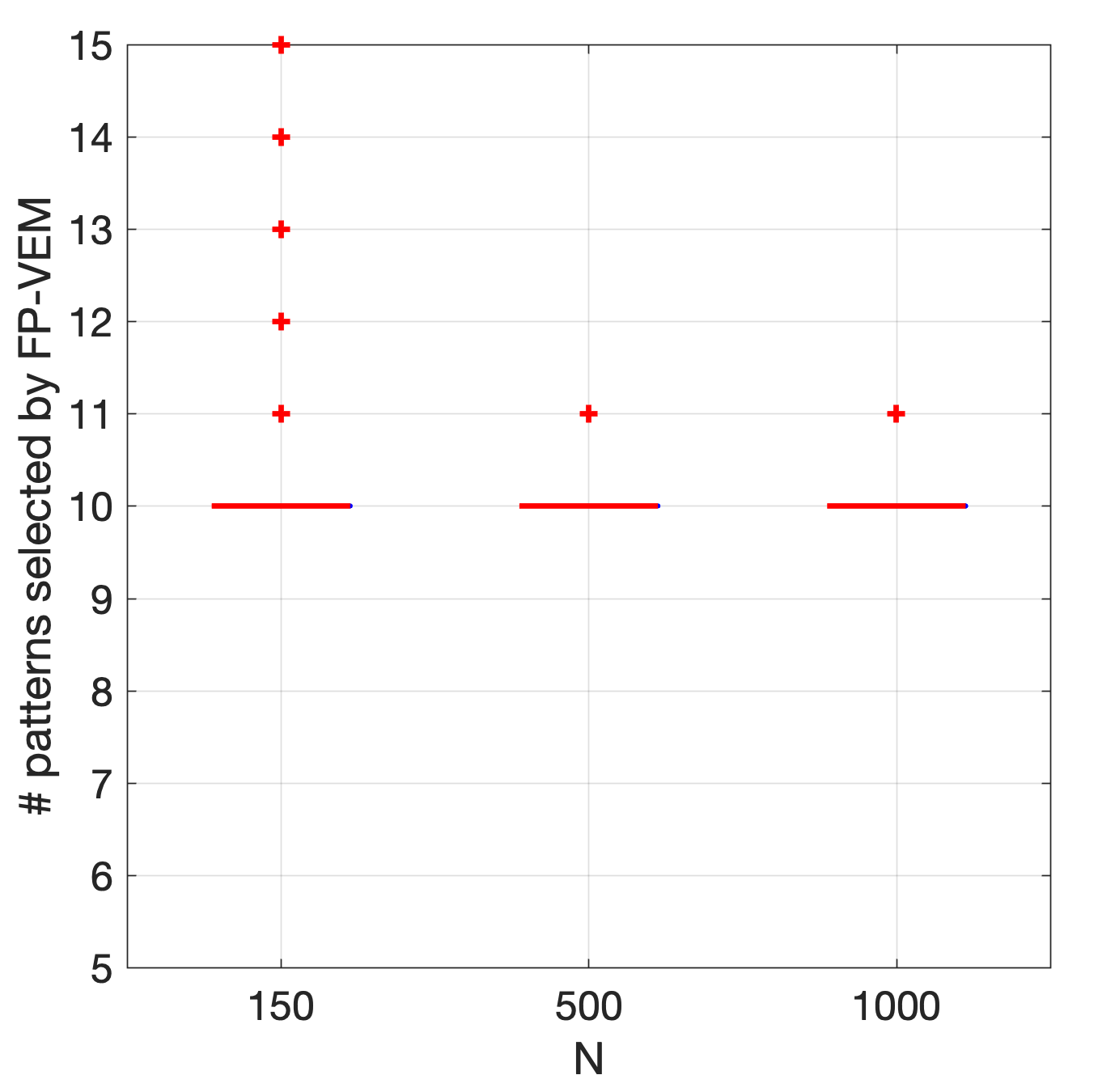}
	\end{minipage}
	
	\begin{minipage}[b]{0.24\textwidth}
	\centering
	  \small{(e) $K=15$, noise $20\%$; FP-VEM}
	\end{minipage}
	\hfill
    \begin{minipage}[b]{0.24\textwidth}
\centering
    \small{(f) $K=15$, noise $10\%$; FP-VEM}
	\end{minipage}
	\hfill
	\begin{minipage}[b]{0.24\textwidth}
	\centering
	\small{(g) $K=20$, noise $20\%$; FP-VEM}
	\end{minipage}
	\hfill
    \begin{minipage}[b]{0.24\textwidth}
\centering
    \small{(h) $K=20$, noise $10\%$; FP-VEM}
	\end{minipage}	

\caption{
{Sizes of the finally selected patterns $\widehat\mca_{\text{PEM}}$ and $\widehat\mca_{\text{FP-VEM}}$ under the two-parameter SLAM. The ``noise'' refers to the value of $1-\theta_j^+=\theta_j^-$. The number of true patterns is $|\mca_0|=10$.}
}
\label{fig-screen-size}
\end{figure}

\paragraph{TIMSS Data: Attribute structures corresponding to different $\Upsilon$'s.} 
For the TIMSS data, we  obtain  those different attribute structures corresponding to different $\Upsilon$'s in the FP-VEM algorithm. The results are presented in Figure \ref{fig-timss-evolve}.
Apart from the five structures shown in Figure \ref{fig-timss-evolve}(a)--(e),    the two patterns selected when $\Upsilon\in[0.70,0.74]$ are the all-zero and the all-one patterns, which do not result in any structure among the 13 attributes. Note that the structure in Figure \ref{fig-timss-evolve}(d) is equivalent to the structure selected by EBIC in Figure \ref{fig-tim}(b).

\begin{figure}[h!]
\centering
\begin{minipage}[t]{.19\linewidth}
\centering
\includegraphics[width=\textwidth]{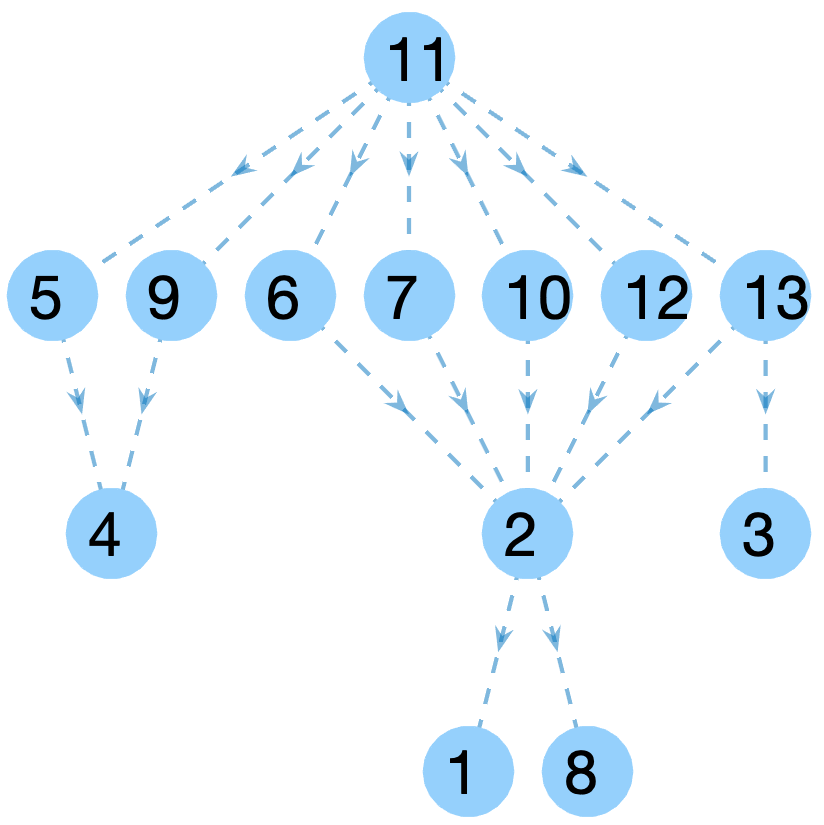}
\end{minipage}%
\hfill
\begin{minipage}[t]{.19\linewidth}
\includegraphics[width=0.95\textwidth]{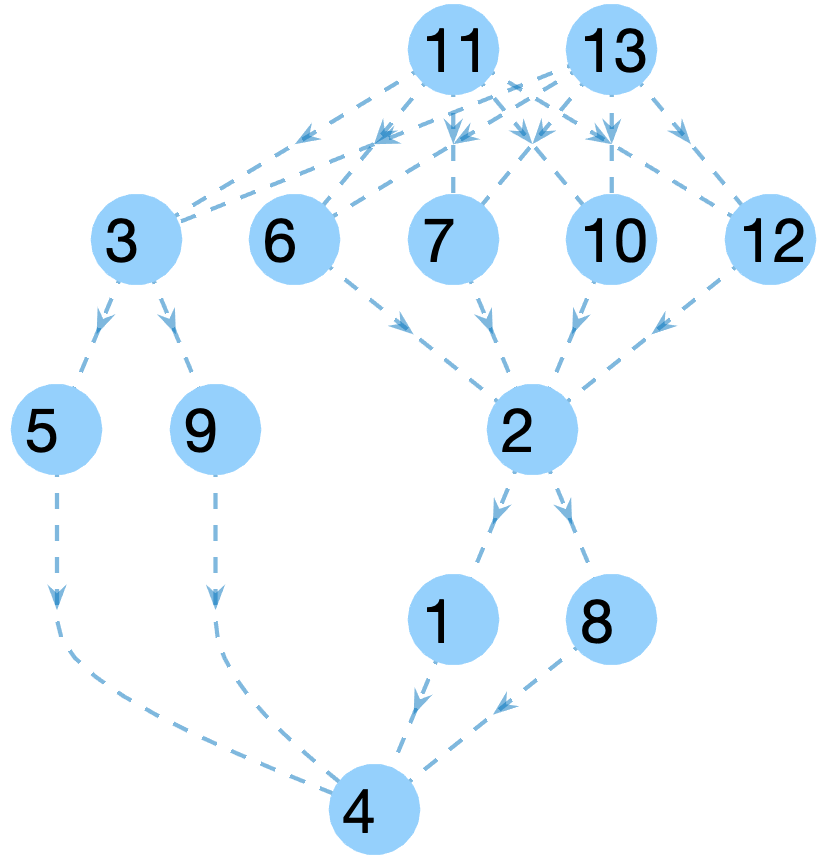}
\end{minipage}%
\hfill
\begin{minipage}[t]{.19\linewidth}
\includegraphics[width=\textwidth]{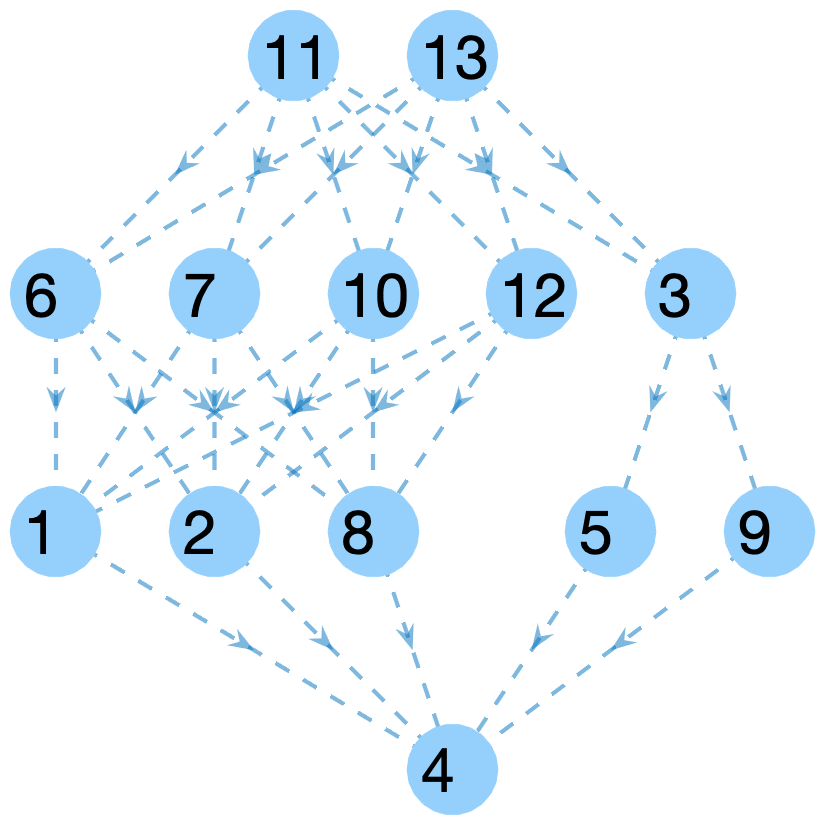}
\end{minipage}%
\hfill
\begin{minipage}[t]{.19\linewidth}
\includegraphics[width=\textwidth]{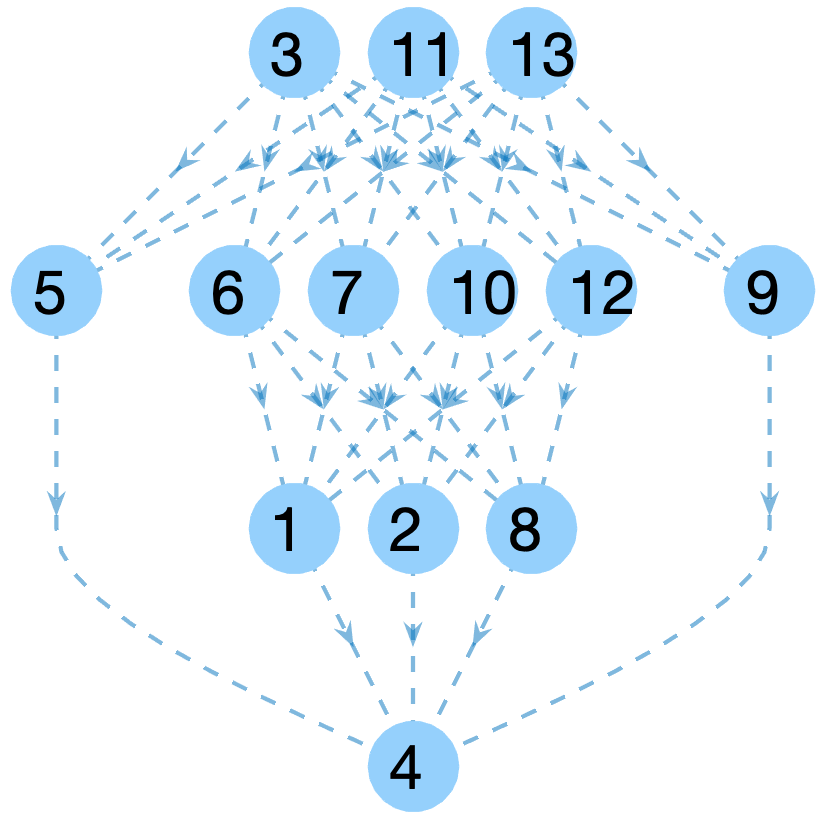}
\end{minipage}%
\hfill
\begin{minipage}[t]{.22\linewidth}
\includegraphics[width=\textwidth]{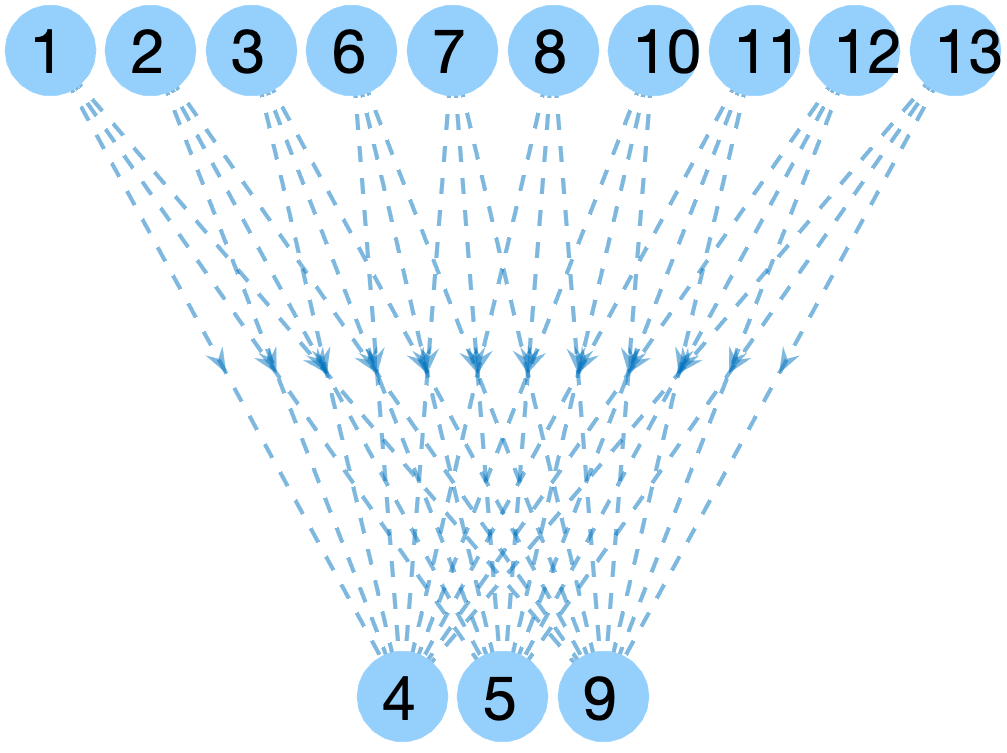}
\end{minipage}%

\begin{minipage}[t]{.19\linewidth}
\centering
\small{(a) $\Upsilon=0.90$}
\end{minipage}%
\hfill
\begin{minipage}[t]{.19\linewidth}\centering
\small{(b) $\Upsilon=0.88$}
\end{minipage}%
\hfill
\begin{minipage}[t]{.19\linewidth}\centering
\small{(c) $\Upsilon=0.86$}
\end{minipage}%
\hfill
\begin{minipage}[t]{.19\linewidth}\centering
\small{(d) $\Upsilon\in[0.80,0.84]$}
\end{minipage}%
\hfill
\begin{minipage}[t]{.22\linewidth}\centering
\small{(e) $\Upsilon\in[0.76,0.78]$}
\end{minipage}%

\caption{
{
Different attribute structures corresponding to various $\Upsilon$'s in Algorithm \ref{algo-fpvem}. Plot (d) here is equivalent to Figure \ref{fig-tim}(b), the attribute structure selected by EBIC.}
}
\label{fig-timss-evolve}
\end{figure}

\color{black}

\clearpage

\paragraph{Acknowledgement}
This research is partially supported by National Science Foundation grants SES1659328 and DMS-1712717, and Institute of Education Sciences grant R305D160010.

\bibliography{ref}

\end{document}